\documentclass[twocolumn,twocolappendix]{aastex63}

\usepackage{amsmath}
\usepackage{bbm,bm}
\usepackage{multirow}
\usepackage{colortbl}
\usepackage{preamble}
\usepackage{hyperref}

\turnoffedit

\graphicspath{{./}{paper-figures/}}

\shorttitle{TIGRESS-NCR Metallicity Suite}
\shortauthors{Kim et al.}

\begin{document}

\title{Metallicity Dependence of Pressure-Regulated Feedback-Modulated Star Formation in the TIGRESS-NCR Simulation Suite}

\author[0000-0003-2896-3725]{Chang-Goo Kim}
\affiliation{Department of Astrophysical Sciences, Princeton University, 4 Ivy Lane, Princeton, NJ 08544, USA}
\email{cgkim@astro.princeton.edu}

\author[0000-0002-0509-9113]{Eve C. Ostriker}
\affiliation{Department of Astrophysical Sciences, Princeton University, 4 Ivy Lane, Princeton, NJ 08544, USA}
\affiliation{Institute for Advanced Study, 1 Einstein Drive, Princeton, NJ 08540, USA}

\author[0000-0001-6228-8634]{Jeong-Gyu Kim}
\affiliation{Division of Science, National Astronomical Observatory of Japan, Mitaka, Tokyo 181-0015, Japan}
\affiliation{Korea Astronomy and Space Science Institute, Daejeon 34055, Republic Of Korea}

\author[0000-0003-1613-6263]{Munan Gong}
\affiliation{Max-Planck Institute for Extraterrestrial Physics, Garching near Munich, D-85748, Germany}

\author[0000-0003-2630-9228]{Greg L. Bryan}
\affiliation{Department of Astronomy, Columbia University, 550 West 120th Street, New York, NY 10027, USA}

\author[0000-0003-3806-8548]{Drummond B. Fielding}
\affiliation{Center for Computational Astrophysics, Flatiron Institute, 162 5th Ave, New York, NY 10010, USA}
\affiliation{Department of Astronomy, Cornell University, Ithaca, NY 14853, USA}

\author[0000-0002-1050-7572]{Sultan Hassan}
\altaffiliation{NASA Hubble fellow}
\affiliation{Center for Cosmology and Particle Physics, Department of Physics, New York University, 726 Broadway, New York, NY 10003, USA}
\affiliation{New York University Abu Dhabi, PO Box 129188, Abu Dhabi, United Arab Emirates}
\affiliation{Center for Computational Astrophysics, Flatiron Institute, 162 5th Ave, New York, NY 10010, USA}
\affiliation{Department of Physics \& Astronomy, University of the Western Cape, Cape Town 7535,
South Africa}

\author[0000-0003-3207-8868]{Matthew Ho}
\affiliation{CNRS \& Sorbonne Universit\'e, Institut d’Astrophysique de Paris (IAP), UMR 7095, 98 bis bd Arago, F-75014 Paris, France}

\author[0000-0002-4232-0200]{Sarah M.~R.~Jeffreson}
\affiliation{Center for Astrophysics | Harvard \& Smithsonian, 60 Garden St, Cambridge, MA 02138, USA}

\author[0000-0002-6748-6821]{Rachel S. Somerville}
\affiliation{Center for Computational Astrophysics, Flatiron Institute, 162 5th Ave, New York, NY 10010, USA}

\author[0000-0001-8867-5026]{Ulrich P. Steinwandel}
\affiliation{Center for Computational Astrophysics, Flatiron Institute, 162 5th Ave, New York, NY 10010, USA}

\begin{abstract}
We present a new suite of numerical simulations of the star-forming interstellar medium (ISM) in galactic disks using the TIGRESS-NCR framework. 
Distinctive aspects of our simulation suite are: (1) sophisticated and comprehensive numerical treatments of essential physical processes including magnetohydrodynamics, self-gravity, and galactic differential rotation, as well as photochemistry, cooling, and heating coupled with direct ray-tracing UV radiation transfer and resolved supernova feedback and 
(2) wide parameter coverage including variation in metallicity over $Z'\equiv Z/Z_\odot\sim 0.1-3$, gas surface density $\Sgas\sim5-150\Surf$, and stellar surface density $\Sstar\sim 1-50\Surf$. 
The range of emergent star formation rate (SFR) surface density range is $\Ssfr\sim 10^{-4}-0.5\sfrunit$ and  ISM total midplane pressure is $\Ptot/k_B= 10^3 - 10^6 \Punit$, with $\Ptot$ equal to the ISM weight $\W$.
For given $\Sgas$ and $\Sstar$, we find $\Ssfr \propto {Z'}^{0.3}$.
We provide an interpretation based on the pressure-regulated feedback-modulated (PRFM) star formation theory.
The total midplane pressure consists of thermal, turbulent, and magnetic stresses. 
We characterize feedback modulation in terms of the yield $\Upsilon$, defined as the ratio of each stress to $\Ssfr$. 
The thermal feedback yield varies sensitively with both weight and metallicity as $\Yth\propto\W^{-0.46}Z'^{-0.53}$, while the combined turbulent and magnetic feedback yield shows weaker dependence $\Ynonth\propto\W^{-0.22}Z'^{-0.18}$.
The reduction in $\Ssfr$ at low metallicity is due mainly to enhanced thermal feedback yield, resulting from reduced attenuation of UV radiation.
With the metallicity-dependent calibrations we provide, PRFM theory can be used for a new subgrid star formation prescription in cosmological simulations where the ISM is unresolved.
\end{abstract}

\section{Introduction}\label{sec:intro}

Galactic star formation rates (SFRs) and the physical state of the interstellar medium (ISM) are observed to be tightly connected \citep[e.g.,][]{2012ARA&A..50..531K,2020ApJ...892..148S,2023ApJ...945L..19S,2021MNRAS.503.3643B}.
This connection can be understood theoretically based on ISM dynamics and thermodynamics and the physics of stellar feedback.
On the one hand, the loss of energy in the ISM occurs on relatively short time scales via radiative cooling and turbulence dissipation. \edit1{Gas with locally reduced pressure support in turn collapses by gravity and forms stars.}
On the other hand, newborn stars return energy (sourced by nuclear fusion) that pervades the surrounding ISM, offsetting losses and recovering the balance between pressure (provided by thermal, turbulent, and magnetic components) and gravity \citep[e.g.,][]{2010ApJ...721..975O,2011ApJ...731...41O,2011ApJ...743...25K,2022ApJ...936..137O}.
Stellar feedback is thus key in controlling future star formation and maintaining the physical state of ISM disks.
Because ISM evolution, star formation, and feedback are inherently cyclic, realistic understanding of the galactic ecosystem necessitates a holistic approach to these tightly coupled physical processes.

Representing ISM physics and stellar feedback in numerical simulations requires treatments of the thermodynamic properties of gas in different phases and of localized injection of energy from stellar feedback.
With varying degrees of accuracy, there exist many numerical frameworks that solve (magneto)hydrodynamics equations including losses -- from shocks, turbulent cascades, and phase mixing followed by radiative cooling -- and modeling gains -- energy returns mainly from massive young stars -- in the context of the galactic ecosystem.
Such efforts can be categorized into three different types based on their outer scales: cosmological zoom-in simulations \citep[e.g.,][]{2014MNRAS.445..581H,2018MNRAS.480..800H,2023MNRAS.519.3154H,2020MNRAS.491.3461B}, isolated global galaxy simulations \citep[e.g.,][]{2020MNRAS.499.5732K,2021MNRAS.505.3470J,2023MNRAS.523.6336B,2017MNRAS.471.2151H,2023ApJ...952..140H,2023ApJ...950..132H,2021MNRAS.506.3882S,2023MNRAS.526.1408S,2024ApJ...960..100S,2020MNRAS.499.5862L,2021MNRAS.505.5438T,2020MNRAS.492.1594S,2024MNRAS.tmp..833L}, and \edit1{vertically-stratified local} simulations of galactic disks \citep[e.g.,][]{2017ApJ...846..133K,2020ApJ...900...61K,2023ApJ...946....3K,2017MNRAS.466.1903G,2021MNRAS.504.1039R,2023MNRAS.522.1843R,2020MNRAS.491.2088K,2021ApJ...920...44H,2023ApJ...952..140H}.
Given the limited dynamic range of any numerical simulation, larger outer scales also imply coarser resolution of the ISM, which makes it difficult to follow multiphase physics explicitly.
In particular, following the creation and evolution of the hot ISM is challenging for pseudo-Lagrangian approaches because it is so diffuse; at realistic hot densities $n_{\rm H} < 10^{-2}\pcc$, even a $(10\pc)^3$ volume contains $<0.3 M_\odot$.
Limited resolution also means that when gravitational collapse occurs, the mass involved may be more strongly clustered than is realistic, leading to an excessive spatio-temporal correlation of feedback.
To date, larger-scale simulations \edit1{of cosmological volumes typically have not attempted to explicitly resolve the multiphase ISM due to their insufficient resolution, instead adopting subgrid models such as those of \citet{2003MNRAS.339..289S} or \citet{2008MNRAS.383.1210S} to model star formation \citep[e.g.][]{2020NatRP...2...42V}. However, efforts to move towards more explicit treatments of key physics in cosmological simulations are underway \citep[e.g.][]{2021A&A...651A.109D,2023MNRAS.522.3831F}.}

Localized energy injection from stellar feedback results in expanding bubbles of different astronomical types, which have been studied using targeted numerical simulations.
These include supernova remnants and superbubbles  \citep{2015ApJ...802...99K,2017ApJ...834...25K,2015MNRAS.450..504M,2015A&A...576A..95I,2015MNRAS.451.2757W,2018MNRAS.481.3325F,2019MNRAS.483.3647G,2019MNRAS.490.1961E,2020MNRAS.495.1035S}, \ion{H}{2} regions \citep{2018ApJ...859...68K,2021ApJ...911..128K,2016MNRAS.463.3129G,2020MNRAS.492..915G,2024MNRAS.527..478D}, and stellar wind blown bubbles \citep{2018MNRAS.478.4799H,2021MNRAS.501.1352G,2021ApJ...914...90L,2021ApJ...922L...3L};  regions with strong feedback interact with each other and the surrounding inhomogeneous ISM.
Expanding feedback-driven bubbles inject a significant amount of radial momentum in the ISM and cause phase transitions to hotter phases by photoionization and shocks in the gas \citep[see e.g. reviews of][]{1988RvMP...60....1O,2019ARA&A..57..227K,2023ASPC..534....1C}.
It is also important to note that the motions driven by feedback are coupled with other large-scale flows in galactic disks, e.g., shear and epicyclic motions induced by galactic differential rotation, and other externally driven gas flows, e.g., cosmic accretion.

Each of the different types of holistic numerical simulations has advantages and disadvantages.
Cosmological zoom-in simulations can realistically capture the cosmic flows that build a given galaxy and interactions with other galaxies, while sacrificing accuracy in modeling the multiphase gas.
In order to model feedback at low resolution in cosmological zoom simulations, the current best practice is to inject the terminal radial momentum that feedback bubbles should have (mainly due to SNe) as calibrated from higher resolution simulations \citep[e.g.,][]{2014MNRAS.445..581H,2014ApJ...788..121K,2022ApJS..262....9O}.
While this approach can drive turbulence in warm and cold ISM phases, it misses transitions to the hot phase. Doing so would require much higher resolution to follow the energy-conserving stage of supernova remnants and shocks that create hot gas  \citep[e.g.,][]{2015ApJ...802...99K,2020MNRAS.495.1035S}.
In cosmological zoom and global galaxy simulations, treatments of thermodynamic, chemical, and radiative processes in the ISM also involve many approximations.
Local tall-box simulations as cited above are in the opposite limit, in which the ISM physics are treated with more accurate and explicit methods and supernova shock heating can be directly resolved.\footnote{We note that there is an additional class of local simulations that do not follow self-consistent cycles of star formation and feedback \citep[e.g.,][]{2006ApJ...653.1266J,2009ApJ...704..137J,2015MNRAS.454..238W,2016MNRAS.456.3432G,2018MNRAS.479.3042G,2016MNRAS.459.2311M,2018MNRAS.481.3325F,2023MNRAS.tmp.3640T}.
Rather, these models focus on the response of the ISM to prescribed injection of energy mimicking stellar feedback and use the local box set-up in the interest of more control and higher resolution (see \citealt{2020ApJ...895...43S,2024arXiv240212474S} for a similar example of global galaxy models.).
These numerical experiments have specific goals and merits, but work of this kind should not be confused with self-consistent simulations of the star-forming ISM like those presented in this paper, which aim to provide a holistic view of co-regulation of the ISM and star formation.}
Some local (or semi-global) models include galactic differential rotation using a shearing-box approximation \citep{2017ApJ...846..133K,2018A&A...620A..21C} and galactic structures like spiral arms \citep{2020ApJ...898...35K} and bar driven inflows \citep{2021ApJ...914....9M,2023ApJ...946..114M}.
However, the effects of global geometry and cosmic inflows cannot be captured directly.
Isolated global galaxy simulations are in between; models of more massive galaxies (Milky Way-like) tend to have numerical approaches closer to cosmological zooms \citep[e.g.,][which treats feedback via prescribed momentum injection]{2021MNRAS.505.3470J,2022MNRAS.515.1663J,2022ApJ...928..144L}, while models of less massive galaxies (dwarf galaxies) include more explicit ISM physics \citep{2020MNRAS.499.5732K,2022MNRAS.512..348K,2022arXiv221104626K} and directly resolve supernova feedback \citep{2023MNRAS.526.1408S,2024ApJ...960..100S,2023arXiv231011495S,2023ApJ...952..140H}.

In a recent publication \citep{2023ApJ...946....3K}, we presented the first results from simulations employing the new TIGRESS-NCR\footnote{``TIGRESS'' stands for ``Three-phase ISM in Galaxies Resolving Evolution with Star formation and Stellar feedback,'' and ``NCR'' stands for ``Non-equilibrium Cooling and Radiation''.} framework to study the star-forming ISM in conditions similar to those in the solar neighborhood and in local-universe galaxies such as those observed by PHANGS \citep{2019Msngr.177...36S}.
Here, we shall present results from a much larger simulation survey conducted using TIGRESS-NCR.
The range of gas and stellar surface density in our new parameter survey is similar to that explored in \citet{2020ApJ...900...61K} and \citet{2022ApJ...936..137O} using the original TIGRESS framework \citep[][we shall refer to this framework as ``TIGRESS-classic'' henceforth]{2017ApJ...846..133K}.
The TIGRESS-classic framework solves ideal MHD equations within a local shearing box, utilizing uniformly high resolution ($\sim 2$--$8\pc$) which enables gravitational collapse in dense regions -- producing sink particles that act as feedback sources, while also being sufficient to resolve the energy-conserving stage of SNRs.

TIGRESS-NCR extends the TIGRESS-classic framework by including explicit UV radiation transfer using an adaptive ray tracing (ART) method \citep{2017ApJ...851...93K} coupled with non-equilibrium photochemistry, cooling, and heating, as detailed in \citet{2023ApJS..264...10K}.
This comprehensive ``NCR'' treatment of microphysics replaces the simplified treatment in TIGRESS-classic, which adopted approximate formulae for heating and cooling in warm-cold gas and was restricted to solar metallicity.
Taking advantage of our expanded ISM modeling capabilities, the new suite of TIGRESS-NCR simulations covers \edit1{from super-solar to} low metallicity regimes.
\edit1{
Combining more than a decade variation in metallicity with a wide range of galactic conditions (gas surface density and gravitational potential),
the simulation parameter study presented here allows us to broadly characterize scaling relations of galactic SFRs and turbulence. 
We note that effects of varying metallicity were previously studied in local-box star-forming ISM simulations by \citet{2021ApJ...920...44H}.
They explored the same metallicity range as we do, for solar neighborhood conditions, but the main scientific focus of that paper was on the metallicity dependence of the atomic-to-molecular transition and C$^{+}$/C/CO distributions.}

The exploration of metallicity dependence is critical as the thermal balance in the ISM is sensitive to the abundances of metals and dust\footnote{Throughout this paper, we will use ``metallicity'' as a collective term for the total elemental abundance of heavy metals in gas and dust.
When a distinction is needed, we will use ``gas metallicity'' and ``dust abundance'' to denote the metal abundance in the gas phase and dust, respectively.}.
A variety of processes are responsible for radiative cooling and heating in different gas phases \citep{2011piim.book.....D,2017RMxAA..53..385F,2022ARA&A..60..247W,2023ApJS..264...10K}.
In the cold neutral ISM, fine-structure metal lines from C$^+$ and O are the main cooling channels in atomic and diffuse molecular gas, with rotational transitions of CO becoming the dominant coolant in denser and more shielded gas; FUV producing the photoelectric (PE) effect in small grains and polycyclic aromatic hydrocarbons (PAHs) and cosmic ray (CR) ionization are the main heating channels \citep{1969ApJ...155L.149F,1972ApJ...176..103W,1994ApJ...427..822B,2001ApJS..134..263W,1995ApJ...443..152W,2003ApJ...587..278W,2019ApJ...881..160B}.
In the warm neutral ISM, cooling is dominated by the hydrogen Ly$\alpha$ line, while PE heating still dominates when grain abundances are sufficiently high; CR heating begins to dominate at low dust abundance.
In the warm ionized ISM \citep{2009RvMP...81..969H}, EUV photons ionize hydrogen, helium, and heavier atoms, making photoionization heating the dominant heating channel, while photoionized metals (e.g., O$^+$, O$^{+2}$, N$^+$) are the major coolants (we collectively refer to this as nebular line cooling; see \citealt{2017RMxAA..53..385F} for comprehensive nebular modeling).
Finally, hot gas ($T>10^6\Kel$) is created by very high-speed shocks generated by supernovae \citep{1972ApJ...178..159C,1977ApJ...218..148M} and cools very inefficiently.
Shocks at somewhat lower speed and mixing of the hot gas with denser material lead to temperatures $T\sim 10^{5-6}\Kel$, at which highly ionized metals become the dominant coolants \citep{1993ApJS...88..253S,2007ApJS..168..213G,2012ApJS..199...20G}.

Metallicity affects ISM thermodynamics not only through direct agents of cooling (with lower metals reducing cooling) and heating (with lower dust reducing heating), but also through effects on UV radiative transfer \citep[see][for a review]{2022ARA&A..60..247W}.
Since grains absorb both FUV and EUV photons as they propagate through the ISM, lower abundances of dust reduce the attenuation of radiation and thereby tend to increase the gas heating for a given rate of radiation production.
Because of the complex interplay among the different effects involved, a quantitative understanding of ISM thermodynamics at varying metallicity requires numerical simulations.

In the pressure-regulated, feedback-modulated (PRFM) theory of the star-forming ISM \citep[see][and references therein]{2022ApJ...936..137O}, the ISM pressure varies directly with the star formation rate per unit area because feedback is responsible for heating and driving of turbulence.
This relationship is quantified in terms of the feedback yield $\Upsilon$ (see \autoref{eq:yield} below for definition), which has previously been measured using TIGRESS-classic simulations in \citet{2022ApJ...936..137O} and TIGRESS-NCR simulations in \citet{2023ApJ...946....3K}.
Since midplane pressure is regulated to match the ISM weight under vertical equilibrium in disk galaxies, the feedback yield can be used to predict the mean star formation rate given large-scale galactic properties (primarily gas and stellar surface density); this prediction, as well as predictions for the relations between component pressures and star formation, have been validated in nearby galaxies \citep{2017ApJ...835..201H,2020ApJ...892..148S,2023ApJ...945L..19S,2021MNRAS.503.3643B}.
A question of considerable interest is how the feedback yield depends on metallicity.
Because star formation is expected to vary inversely with feedback yield, this has ramifications for predictions of star formation rates in low-metallicity dwarfs in the local universe, as well as low-metallicity galaxies at high redshift.
With the numerical implementation in TIGRESS-NCR, we can address this question; quantifying feedback yields over a wide range of metallicity and pressure is a key motivation for the present study.

Simulations that evolve the ISM with explicit treatments of physics and uniformly high resolution are valuable as laboratories where the interactions behind emergent properties of galaxies may be scrutinized in great detail.
In addition, suites of such simulations offer a means to develop realistic subgrid treatments for deployment in large-scale, low-resolution galaxy formation models.
While it is not possible to resolve the ISM directly in simulations of this kind, galactic-scale baryonic evolution depends entirely on the choices adopted for SFRs and galactic winds.
An effort to develop a new generation of physically-motivated, numerically-calibrated subgrid models has been recently launched, under the umbrella of the SMAUG\footnote{Simulating Multiscale Astrophysics to Understand Galaxies; https://www.simonsfoundation.org/flatiron/center-for-computational-astrophysics/galaxy-formation/smaug/} and Learning the Universe collaborations (see \citealt{2020ApJ...903L..34K,2024MNRAS.527.1216S} as examples of subgrid wind modeling)
In S. Hassan et al (2024, submitted), a comparison was made between the SFR in \edit1{galaxies in the Illustris-TNG50 simulation} \citep{2019MNRAS.490.3196P,2019MNRAS.490.3234N}, \edit1{calculated with the \citet{2003MNRAS.339..289S} subgrid model}, and what would have been predicted for the SFR using the PRFM theory and yield calibration from the TIGRESS-classic simulation suite \citep{2022ApJ...936..137O}.
This comparison shows intriguing differences: higher SFRs and shorter gas depletion times would be predicted from PRFM at high redshifts where gas is denser and pressure is higher, compared to the native TNG SFR from the \citet{2003MNRAS.339..289S} subgrid model.
However, for a fully quantitative and self-consistent prediction of the SFR (especially, at higher redshifts), it is critical to include a dependence of the yield on metallicity.
Calibration of the feedback yield from TIGRESS-NCR simulations, which account for the dependence of both heating/cooling and radiative transfer on metallicity, is thus an important goal of this paper.

The remainder of this paper is organized as follows.
We first summarize key methods in \S~\ref{sec:methods_ncr} and model parameters in \S~\ref{sec:models}.
In \S~\ref{sec:results}, we then provide an overview for two chosen galactic conditions (solar neighborhood and \edit1{conditions similar to the SFR-weighted mean in nearby star-forming galaxies from the}  PHANGS sample) with varying metallicities.
We analyze the maps of gas and radiation properties and quantify the emergent SFRs as a function of metallicity.
In \S~\ref{sec:prfm_lowz}, we analyze the full simulation suite in the context of the PRFM theory, and provide a new calibration to the feedback yield including metallicity dependence.
We also introduce an effective equation of state for multiphase, star-forming gas in \S~\ref{sec:eEoS}.
We discuss and summarize our results in \S~\ref{sec:discussion} and \S~\ref{sec:summary}.

\section{Methods \& Models}\label{sec:methods}

\subsection{TIGRESS-NCR}\label{sec:methods_ncr}

We use the TIGRESS-NCR framework to run a suite of numerical simulations under widely varying galactic conditions, where the novel feature compared to our previous work is the range of gas metallicity and dust abundance.
The TIGRESS-NCR framework models the star-forming ISM in a patch of a galactic disk including the effects of magnetic fields, gravity, galactic differential rotation, stellar feedback including UV radiation and supernovae, cooling, heating, and chemistry.

We aim to treat most physical processes in the star-forming ISM as \emph{explicitly} as possible (to the extent the resolution allows), avoiding \emph{ad hoc} approximations.
The numerical framework we have built has enabled many previous scientific studies, with technical details of the physical elements we have implemented described in several published papers.
Interested readers should consult the TIGRESS-classic method paper \citep{2017ApJ...846..133K} for summaries of methods for MHD, shearing-box, gravity, sink particles, and supernova feedback \citep[see also][for an update in gas accretion to sinks]{2020ApJ...900...61K}, and the TIGRESS-NCR method paper \citep{2023ApJ...946....3K} for an explanation of UV radiation transfer and review of selected photochemistry, cooling, and heating processes.
Comprehensive descriptions of our formulations for photochemistry, cooling, and heating processes (separating neutral, photoionized, and collisionally ionized regimes), including detailed rate coefficients, are presented in  \citet{2023ApJS..264...10K}, along with tests of our implementations.

Here, we briefly summarize the included chemical and thermodynamic processes that directly depend on gas metallicity and dust abundance.
\begin{itemize}
  \item {\bf Metal cooling.} The metal cooling is directly proportional to gas metallicity. This includes fine structure line cooling by C, C$^+$, O, rotational line cooling by CO, combined nebular line cooling in the warm ionized gas \citep{1995ApJ...443..152W,2011piim.book.....D,2018A&C....23...40V}, and CIE metal cooling in the hot gas \citep{1993ApJS...88..253S,2009MNRAS.393...99W,2012ApJS..199...20G}.
  \item {\bf Photoelectric heating on small grains and PAHs.}
  Incident FUV photons cause electrons to be dislodged via the PE effect from surfaces of small grains and PAHs, sharing the excess energy with the surrounding gas \citep{1972ApJ...176..103W}.
  The PE heating rate \edit1{per hydrogen ($\Gamma_{\rm PE}$)} thus scales directly with the abundance of small grains, which we take as proportional to the total dust abundance ($\Zd$); we do not explicitly consider possible variations of the PAH fraction \citep[but see][for observational evidence of decreasing PAH fraction at low metallicities]{2007ApJ...663..866D,2020ApJ...889..150A}.
  The PE heating efficiency ($\epsilon_{\rm PE}$) depends on grain charging, which depends on both the local radiation field ($J_{\rm FUV})$ and the CR ionization rate ($\xi_{\rm cr}$) via the free electron abundance $x_{\rm e}$ \citep{1994ApJ...427..822B,1995ApJ...443..152W,2001ApJS..134..263W}.
  \item {\bf Grain-assisted processes.} We include grain-assisted recombination and $\HH$ formation that scale with dust abundance \citep{1979ApJS...41..555H,2001ApJS..134..263W}.
  The former also contributes to cooling.
  Heating and cooling related to $\HH$ formation and dissociation are included but play a minor role in overall energetics.
\end{itemize}

In our ART module, UV radiation in all three bands -- photoelectric (PE; 6 eV $<h\nu <$ 10.2 eV), Lyman-Werner (LW; 10.2 eV $<h\nu<$ 13.6 eV), and Lyman continuum (LyC; 13.6 eV $<h\nu$) -- are attenuated by dust along each ray. \edit1{In our adopted nomenclature, FUV means both PE and LW bands while EUV means the LyC band.}
We linearly scale the spectrum-averaged dust cross sections in Appendix B of \citet{2023ApJS..264...10K} with dust abundance.

We also include heating by CR ionization that is independent of the metallicity.
We adopt a simple prescription for the CR ionization rate $\xi_{\rm cr}\propto \Sigma_{\rm SFR}/\Sigma_{\rm gas}$.
The linear scaling with $\Sigma_{\rm SFR}$ is motivated by the source of CRs in SNR shocks, while the inverse dependence on gas column density nominally represents collisional losses on large scales \citep[as adopted by][]{2003ApJ...587..278W}.
Strictly speaking, losses by transport out of the galactic disk likely exceed collisional losses at moderate ISM surface density (similar to the solar neighborhood), while following the above relation at higher surface density (most of our parameter space for the current simulation suite).
We normalize based on solar neighborhood conditions, i.e.  $\xi_{\rm cr,0} = 2\times10^{-16}\,{\rm s^{-1}}$ is adopted for  $\Ssfr=2.5\times10^{-3}\sfrunit$ and $\Sgas = 10\Surf$ \citep{2015ApJ...800...40I}.
We apply an additional local attenuation recipe $\xi_{\rm cr}\propto N_{\rm eff}^{-1}$ if a local column density estimator $N_{\rm eff}= 1.5\times 10^{21}\cm^{-2}(\nH/100\pcc)^{0.3}$ exceeds $N_0 = 9.35\times10^{20}\cm^{-2}$ \citep{2017ApJ...845..163N}.
We note that this column density estimator is different from what we adopted in \citet{2023ApJS..264...10K} based on the ratio between the attenuated and unattenuated PE radiation fields from RT solutions.
The new form is adopted for simplicity and robustness since the previous prescription using the RT solutions does not converge with the ray truncation parameters.

The ART method we use in the TIGRESS-NCR framework is a direct but expensive method to follow the UV radiation fields.
Even with highly optimized performance and parallel efficiency \citep{2017ApJ...851...93K}, there are a few additional assumptions we adopt to reduce the overall cost: sparse ART calculation at every hydro time step of the warm and cold gas, and ray termination for FUV at $|\zpp|=300\pc$ above/below which we transition to a horizontally uniform FUV field from a plane-parallel RT solution\footnote{There can be radiation source particles located above/below $\zpp$. In such (very rare) cases, we neglect their FUV photons propagating upwards/downwards, but rays propagating downwards/upwards are followed consistently until the FUV photons are collected at the other side of $\zpp$.} \citep{2023ApJ...946....3K}.
In addition, two more parameters determine the ray termination conditions: (1) the maximum travel distance in the horizontal direction ($\dxymax$) and (2) the ratio of the luminosity of the photon packet to the total luminosity of all sources in the domain ($\epp$; FUV only).
As we showed in Appendix of \citet{2023ApJ...946....3K}, the impact of these RT termination parameters on pressures, SFR surface density, and feedback yields are minimal, although the radiation fields at large distances from the midplane are more sensitive to these parameters.
To save the computational cost, therefore, we first run the early stage of simulations (typically $<1-2\torb$) using smaller $\dxymax$ and larger $\epp$ parameters.
We then restart the simulation for a longer period (typically $>2-4\torb$) with the values that give reasonable convergence for radiation fields in each model.

\begin{deluxetable*}{lcccccccccc}
    \tablecaption{Input Physical Parameters\label{tbl:model}}
    \tablehead{
      \colhead{Model} &
      \dcolhead{\Zd} &
      \dcolhead{{\Sgas}_{,0}} &
      \dcolhead{\Sstar} &
      \dcolhead{z_*} &
      \dcolhead{\rho_{\rm dm}} &
      \dcolhead{\Omega} &
      \dcolhead{R_0} &
      \dcolhead{L_{x,y}} &
      \dcolhead{L_z} &
      \dcolhead{\Delta x}
      \\
      \colhead{Series} &
      \colhead{} &
      \colhead{$\mathrm{M_{\odot}\,pc^{-2}}$} &
      \colhead{$\mathrm{M_{\odot}\,pc^{-3}}$} &
      \colhead{$\mathrm{pc}$} &
      \colhead{$\mathrm{M_{\odot}\,pc^{-3}}$} &
      \colhead{$\mathrm{km\,s^{-1}\,kpc^{-1}}$} &
      \colhead{$\mathrm{kpc}$} &
      \colhead{$\mathrm{pc}$} &
      \colhead{$\mathrm{pc}$} &
      \colhead{$\mathrm{pc}$}
    }
    \colnumbers
    \startdata
    {\tt S05}        &  1, 0.1 & 5 & 1 & 500 & 0.002 & 15 & 8 & 2048 & 6144 & 8 \\
    {\tt R8}   &  3, 1, 0.3, 0.1, 0.025 & 12 & 42 & 245 & 0.0064 & 28 & 8 & 1024 & 6144 & 8\\
    {\tt S30}        &  1, 0.1  & 30 & 42 & 245 & 0.0064 & 28 & 8 & 1024 & 6144 & 8\\
    {\tt LGR4} &  3, 1, 0.3, 0.1, 0.025 & 50 & 50 & 500 & 0.005 & 60 & 4 & 512 & 3072 & 4\\
    {\tt S100}       &  1, 0.1  & 100 & 50 & 500 & 0.005 & 60 & 4 & 512 & 3072 & 4\\
    {\tt S150}       &  1, 0.1  & 150 & 50 & 500 & 0.005 & 100, 200 & 2 & 512 & 3072 & 4\\
    \enddata
    \tablecomments{
      We assume $\Zg=\Zd$ except for $\Zd=0.025$ for which we adopt $\Zg=0.1$.
      In each model series, a suffix in the model name is used to denote the metallicity parameters, i.e., ``{\tt -ZXX}'' means $\Zg=\Zd=$XX or ``{\tt -ZgXX.ZdYY}'' stands for $\Zg=$XX and $\Zd=$YY. 
      For {\tt R8} and {\tt LGR4}, the additional suffix ``{\tt b10}'' is used to denote models with weaker initial magnetic fields (initial plasma beta $\beta_0=10$ instead of $\beta_0=1$).
      For {\tt S150}, an additional suffix for the galactic rotation parameters is added; ``{\tt Om100q0}'' and ``{\tt Om200}'' for $\Omega=100\kms\kpc^{-1}$ with $q=0.01$ and $\Omega=200\kms\kpc^{-1}$, respectively.
      See \autoref{tbl:result} and \autoref{tbl:result2} for the expanded model names.
      The solar metallicity {\tt R8} and {\tt LGR4} models ({\tt R8-Z1} and {\tt LGR4-Z1}) are identical to the {\tt R8} and {\tt LGR4} models presented in \citet{2023ApJ...946....3K}.
    }
  \end{deluxetable*}

\subsection{Simulation initialization}\label{sec:init}

For initial conditions, we adopt horizontally uniform, vertically stratified gas profiles for density, temperature, and magnetic field strength following double Gaussians representing warm and hot components \citep{2017ApJ...846..133K,2023ApJ...946....3K}.
Following our previous practice, we also apply initial velocity perturbations with amplitude of $10-30\kms$ (higher values for higher surface density models) and create initial star clusters that provide initial UV radiation and supernovae during the early evolution, prior to the formation of the first generation star clusters.
The magnetic field is initialized along the local azimuthal direction ($\yhat$) with a vertically constant ratio of the initial magnetic pressure to the thermal pressure  $\beta_0\equiv 8\pi \Pth/B_0^2$.

Galactic rotation is modeled in the local shearing-box approximation \citep{2010ApJS..189..142S}.
The simulation box is centered on galactocentric radius $R_0$ and co-rotates with this point in the disk at an angular frequency $\Omega(R_0) =2\pi/\torb$. Galactic differential rotation is characterized by the shear parameter $q\equiv -d\ln\Omega/d\ln R|_{R_0}$.
The resulting background flow along the local azimuthal direction ($\yhat$) is $\bm{v}_{0} = -q\Omega x\yhat$, sheared in the local radial coordinate ($x$); this flow is imposed in the initial conditions and maintained through shearing periodic boundary conditions at the $x$-faces of the box.
The additional forces arising in this rotating frame are the Coriolis force $-2\bm{\Omega}\times\bm{v}$ and the tidal force $2q\Omega^2 x\xhat$.
We assume a flat rotation curve ($q=1$) unless stated otherwise.

As we shall show, the final self-regulated state is insensitive to the initial conditions in most simulations.
The initial gas profiles matter for the initial transient phase that typically lasts less than one orbit time ($\torb$).
Still, there remains a longer-term memory of initial magnetic fields \citep[e.g.,][]{2015ApJ...815...67K}, which converge to a self-consistent saturation state after a few orbit times unless the chosen initial field strength is widely different from the converged value.
We defer to future investigations of the effects of initial magnetic fields and their growth due to galactic dynamo at varying rotation parameters and feedback.

\subsection{Model parameters}\label{sec:models}

In the current simulation suite, the main gas parameters that matter the most are the initial gas surface density $\Sgasinit$, gas metallicity $\Zg$, and dust abundance $\Zd$.
Here, the prime means that values for the metal and dust ratios to gas mass are defined relative to solar neighborhood values; we adopt metal-to-gas $Z_{g,\odot}=0.014$ \citep{2009ARA&A..47..481A} and dust-to-gas $Z_{ d,\odot} = 0.0081$ \citep{2001ApJ...548..296W}.

We assume $\Zg=\Zd$ for $\Zg=3,$ 1, 0.3, and 0.1 except for one model with $\Zg=0.1$ and $\Zd=0.025$.
\edit1{This choice is motivated by the observational evidence \citep[e.g.,][]{2014A&A...563A..31R,2019A&A...623A...5D,2022ApJ...928...90R} showing the linear relation between $\Zg$ and $\Zd$ (or a constant dust-to-metal ratio) at near solar metallicities and a significant drop of the dust-to-metal ratio below $\Zg=0.1$.}
Although we trace the gas metallicity field locally using passive scalars with additional metal injection from SN ejecta, we do not use this cell-by-cell gas metallicity information for cooling and heating.
We instead simply assume globally constant $\Zg$ and $\Zd$ for each simulation.
Investigation of the effect of locally varying gas metallicity and dust abundance is deferred to future models with an explicit dust evolution \citep[e.g.,][]{2019MNRAS.487.3252H,2023ApJ...952..140H}.

In addition to the self-gravity of gas, the vertical stratification of the gas disk is controlled by the vertical gravity of the old stellar disk and dark matter halo.
We use a fixed potential for this ``external'' gravity adopting a functional form similar to \citet{1989MNRAS.239..571K} with three main parameters \citep[see][]{2017ApJ...846..133K}: stellar surface density $\Sstar$, stellar disk scale height $z_*$, and dark matter volume density $\rho_{\rm dm}$.

We summarize the model parameters in Table~\ref{tbl:model}.
For the model name, we follow the naming convention used in \citet{2023ApJ...946....3K} for galactic conditions with $\Sigma_{\rm gas}$ and background gravity similar to solar neighborhood ({\tt R8}),  and \edit1{similar to PHANGS galaxies} ({\tt LGR4}).
For other models, we introduce a new naming convention simply representing initial gas surface density, e.g., {\tt S100} for $\Sgasinit=100\Surf$.
For these models, $\Sigma_{\rm gas}$ is similar to the TIGRESS-classic suite \citep{2020ApJ...900...61K,2022ApJ...936..137O} but external gravity parameters are different.

Galactic rotation is parameterized by $\Omega$ and $q$.
For most model series, we adopt a single value of $\Omega$, increasing with $\Sigma_{\rm gas}$. We only vary the galactic rotation parameters for high gas surface density models where they make notable differences.

The parameters for different model series are chosen to cover a range of conditions in typical star-forming galactic disks.
We then explore full metallicity variation for {\tt R8} and {\tt LGR4} models and only run two metallicities for other models.
We use a single suffix with `$Z$' to denote the metallicity parameter models if $\Zg=\Zd$.
The models with $\Zg=0.1$ and $\Zd=0.025$ use {\tt Zg0.1Zd0.025} as a suffix.
Finally, we also explore the effect of initial magnetic field strength by running the {\tt R8} and {\tt LGR4} models with weaker initial fields (`{\tt b10}' to denote $\beta_0=10$ while $\beta_0=1$ in other models) and the effect of galactic rotation in the {\tt S150} models (`{\tt Om100q0}' for $\Omega=100\kms\kpc$ and $q=0.01$, close to a near rigid body rotation).
For example, {\tt R8-Z1.0} is identical to {\tt R8-8pc} in \citet{2023ApJ...946....3K}, while {\tt R8-Zg0.1Zd0.025-b10} denotes the same model with low metallicity $\Zg=0.1$ and dust abundance $\Zd=0.025$ with a weaker initial magnetic field.

The simulation domain is a vertically-extended cuboid with the horizontal dimension of $L_x=L_y=512\pc$ to $2048\pc$ and the vertical dimension of $L_z=3072\pc$ to $6144\pc$ (see parameters in \autoref{tbl:model}).
We adopt a cubic resolution element with the side length of $\Delta x = 4\pc$ or $8\pc$, depending on the models.
The resolution convergence is demonstrated in \citet{2023ApJ...946....3K} for {\tt R8} and {\tt LGR4} with $Z'=1$.

\section{Simulation Overview}\label{sec:results}

\begin{figure*}
  \includegraphics[width=\linewidth]{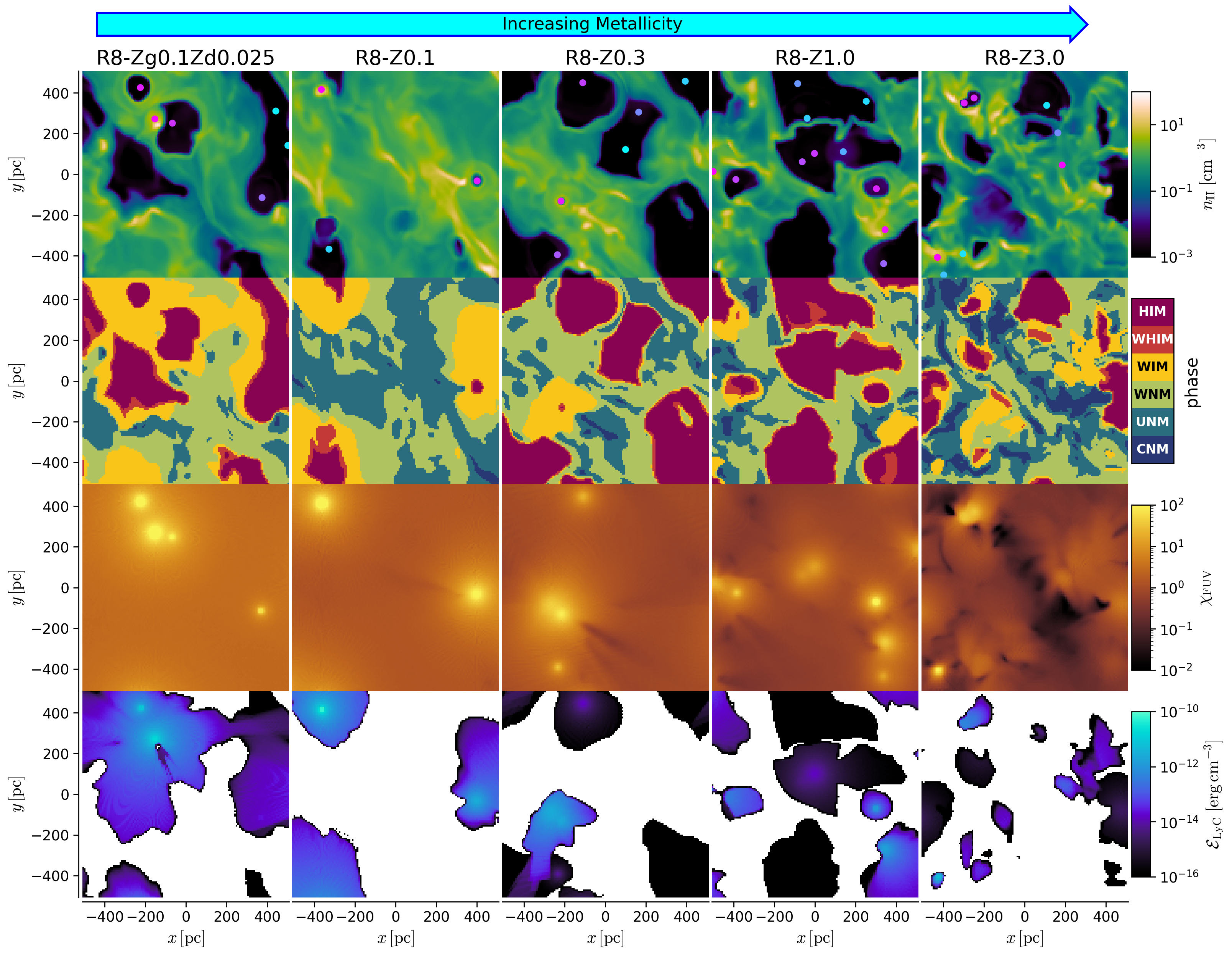}
  \caption{Visualization of the metallicity dependence of gas and radiation properties in the {\tt R8} model series (with $\Sigma_{\rm gas,0}=12 \Surf$ and $\Sstar=42\Surf$ comparable to the solar neighborhood); $\Zg$ and $\Zd$ increase from left to right.
  From top to bottom, we show midplane slices of number density, gas phase, FUV mean intensity normalized to the Draine field, and EUV radiation energy density.
  Young star clusters (color coded from age=0 (magenta) to 40~Myr (cyan)) within $|z|<50\pc$ are overplotted as circles in the top row.
  We select snapshots from times near local star formation peaks with similar total luminosities from sources.
  From left to right, the snapshots are at $t=453$, 378, 420, 438, and 428 Myr with $\Ssfr=6.1$, 3.4, 4.0, 3.0, and $4.2\times10^{-3}\sfrunit$, $L_{\rm LyC}/(L_xL_y)=6.3$, 5.4, 7.5, 6.6, and 6.5 $L_\odot\,\pc^{-2}$ , and $L_{\rm FUV}/(L_xL_y)=19.2$, 12.5, 13.7, 13.2, and 15.5 $L_\odot\,\pc^{-2}$.
  \label{fig:R8-map}}
\end{figure*}

\begin{figure*}
  \includegraphics[width=\linewidth]{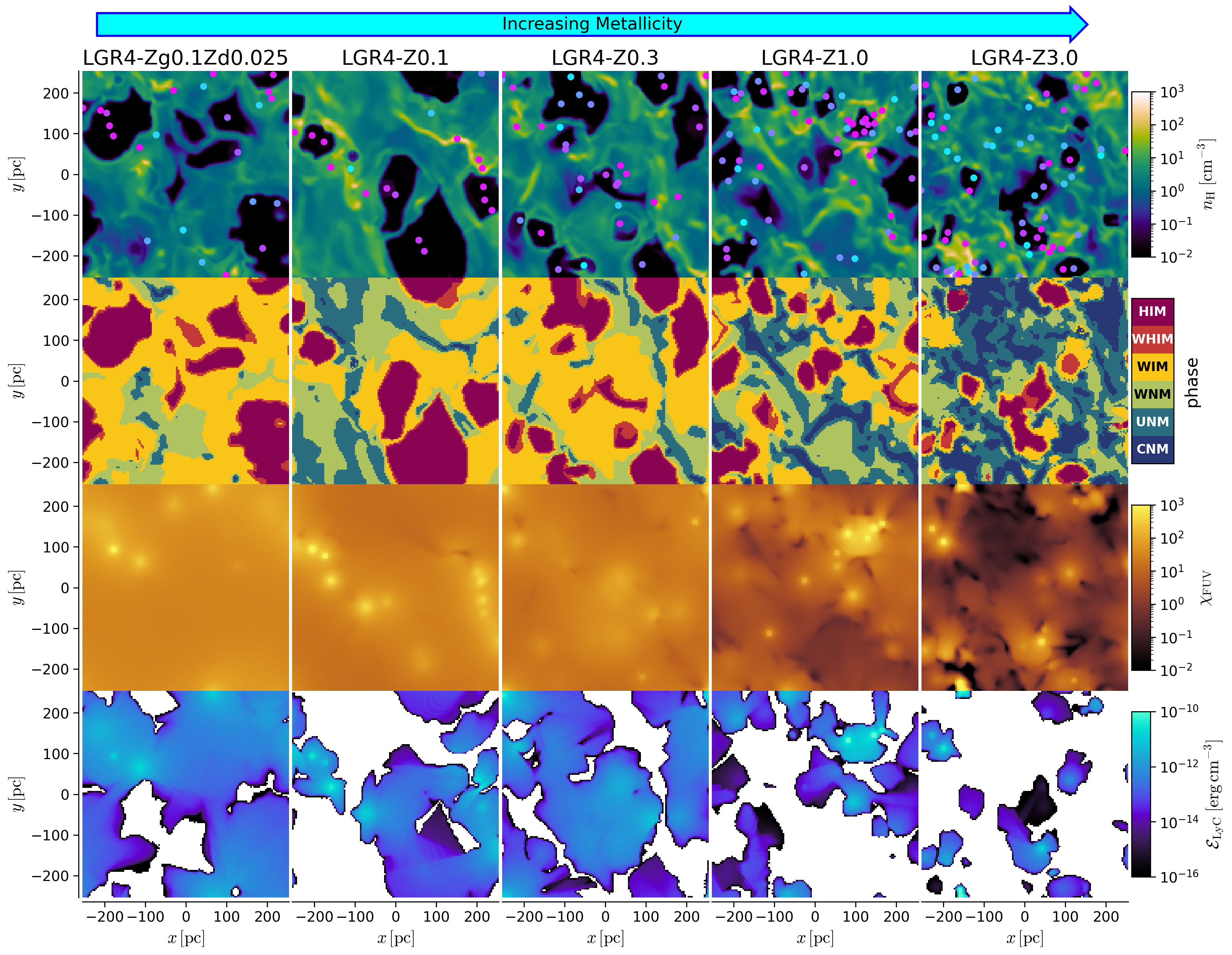}
  \caption{Same as \autoref{fig:R8-map} but for {\tt LGR4} model series (with $\Sigma_{\rm gas,0}=50 \Surf$ and $\Sstar=50\Surf$, comparable to mean PHANGS galaxies).
  From left to right, the snapshots are at $t=386$, 233, 293, 333, and 277~Myr with $\Ssfr=5.7$, 5.5, 5.4, 5.8, and $4.3\times10^{-2}\sfrunit$, $L_{\rm LyC}/(L_xL_y)=110$, 95, 97, 79, and 69 $L_\odot\,\pc^{-2}$ , and $L_{\rm FUV}/(L_xL_y)=272$, 228, 242, 192, and 191 $L_\odot\,\pc^{-2}$.
  \label{fig:LGR4-map}}
\end{figure*}

\begin{figure*}
  \includegraphics[width=\linewidth]{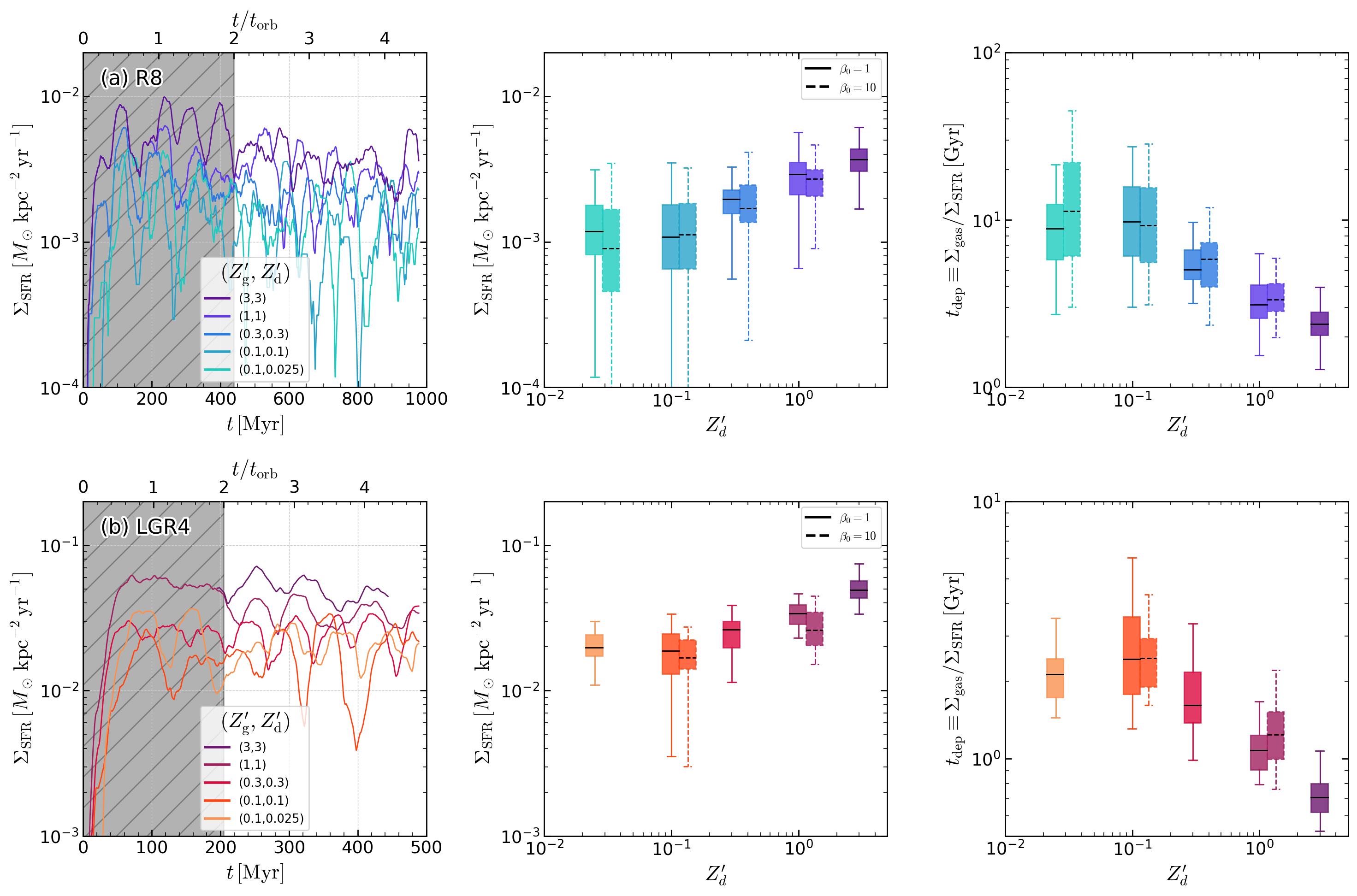}
  \caption{
    Summary of the emergent SFRs in the (a) {\tt R8} (top row) and (b) {\tt LGR4} (bottom row) model series which have gas surface density comparable to local disk galaxies, showing lower SFRs at lower metallicities.
    {\bf Left:} Time evolution of $\Ssfr$ from star cluster particles with age younger than 40~Myr.
    Early evolution for $t < 2\torb$ (shaded area) is excluded in the analysis of the saturated state.
    {\bf Middle:} Box-and-whisker plots summarizing $\Ssfr$ for the late stage evolution ($t>2\torb$) as a function of $\Zd$.
    For visual clarity, the {\tt b10} models with initially weaker magnetic fields are shifted by 30\% to the right.
    {\bf Right:} Box-and-whisker plots summarizing $\tdep\equiv\Sgas/\Ssfr$ for the late stage evolution ($t>2\torb$) as a function of $\Zd$.
    Note that $\Sgas$ at late stages are lower in higher metallicity models due to the prolonged high SFRs (see \autoref{tbl:result}), making the difference in $\tdep$ larger than in $\Ssfr$.
    Initially, reduced magnetic fields (shown in the box-and-whisker plots with dashed lines) do not affect the late time evolution as all models reach similar saturated magnetic field strengths.
    At $\Zg=0.1$, further reduction in the dust abundance ($\Zd=0.1$ vs $\Zd=0.025$) does not make a difference in $\Ssfr$.
    \label{fig:sfr_history}
  }
\end{figure*}

\subsection{Metallicity dependence of the ISM phases and radiation fields}\label{sec:result_map}

\autoref{fig:R8-map} and \autoref{fig:LGR4-map} qualitatively depict the effect of metallicity on the properties of gas and radiation fields for the {\tt R8} and {\tt LGR4} model series, respectively.
From top to bottom, we present the maps from midplane ($z=0$) slices of hydrogen number density, gas phase, far ultraviolet (FUV=PE+LW; $6\eV<h\nu<13.6\eV$) radiation intensity (normalized relative to the mean intensity of the Draine field \citep{1978ApJS...36..595D}, $J_{\rm FUV, Draine}=2.1\times10^{-4}\junits$), and radiation energy density of the LyC band.
The gas metallicity and dust abundance decrease from left to right (from $(\Zg,\Zd)=(3,3)$ to $(0.1,0.025)$).

For the phase separation shown in the second row, we use a reduced definition from the original phase definition presented in \citet{2023ApJ...946....3K}.
We use the temperature boundaries of $T=500\Kel$, 6000~K, 35000~K, and $5\times10^5\Kel$ to divide the gas into the cold neutral medium (\CNM), unstable neutral medium (\UNM), warm medium (\WNM+\WIM), warm-hot ionized medium (\WHIM), and hot ionized medium (\HIM).
For the warm medium, an additional hydrogen abundance cut is used to separate the warm neutral medium (\WNM; $\xHI>0.5$) and warm ionized medium (\WIM; $\xHI<0.5$).

The snapshots within each model series are chosen to have similar total luminosity emitted from sources, near a local star formation peak for each model.
Yet, the FUV radiation field is much lower in the high metallicity models (right two columns) with clear shadows cast by dense gas near sources.
Only very young clusters ($t_{\rm age}<5\Myr$) are significant LyC sources.
The LyC radiation field shows much sharper cutoffs due to the strong absorption by the neutral hydrogen.
The fraction of LyC photons absorbed by dust also decreases as the dust abundance decreases allowing for more mass/volume to be ionized at a given ionizing luminosity (visually clearer in \autoref{fig:LGR4-map}).
Although the gas density distribution is not very different, more pervasive UV radiation at lower metallicities allows more gas in the warm and hot phases than in the cold phase.

\subsection{Star formation rates and gas depletion times}\label{sec:sfr}

\autoref{fig:sfr_history} shows the time evolution of SFR surface density for all members of the {\tt R8} (top) and {\tt LGR4} (bottom) model series, representative of galactic conditions with gas and stellar surface density comparable to the Milky Way and nearby disk galaxies.
The SFR surface density is measured from the total star cluster mass with age less than $t_{\rm bin}$, i.e.,
\begin{equation}
  \Sigma_{\rm SFR} = \frac{\sum_{\rm i} m_{*,i}(t_{\rm age}<t_{\rm bin})}{L_xL_yt_{\rm bin}}.
\end{equation}
Here, we use $t_{\rm bin}=40 \Myr$.
From these temporal histories, it is evident that SFRs are lower and also more bursty at lower metallicity.

The second and third columns compare $\Ssfr$ and the gas depletion time
\begin{equation}\label{eq:tdep}
 \tdep \equiv \frac{\Sgas}{\Ssfr}
\end{equation}
as a function of metallicity (metallicity increases from left to right) during the late-stage evolution $t>2\torb$.
It is immediately noticeable that SFR decreases as gas metallicity decreases, while at $\Zg=0.1$, the further reduction of dust abundance does not make a difference in the SFR.
Since long-term evolution with different SFR surface densities leads to different gas surface densities (see \autoref{tbl:result}), gas depletion time provides a better quantitative comparison, with $\tdep$ increasing from $\sim 2 {\rm Gyr}$ at $\Zg=3$ to $\sim 10 {\rm Gyr}$ at $\Zg=0.1$ in the {\tt R8} series, and a factor $\sim 3-4$ lower $\tdep$ in the {\tt LGR4} series.

\begin{deluxetable*}{lCCCCCCCCCCC}
    \tablecaption{Time-averaged Properties in Simulations\label{tbl:result}}
    \tablehead{
      \colhead{Model} &
      \dcolhead{(t_1,t_2)} &
      \dcolhead{\Sgas} &
      \dcolhead{\Ssfr} &
      \dcolhead{\tdep} &
      \dcolhead{\Sigma_{\rm sp}} &
      \dcolhead{\nH} &
      \dcolhead{H} &
      \dcolhead{\szavg} &
      \dcolhead{\szturb} &
      \dcolhead{\szth} &
      \dcolhead{\szmag}
      \\
      \colhead{} &
      \colhead{$\mathrm{(Myr, Myr)}$} &
      \colhead{$\mathrm{M_{\odot}\,pc^{-2}}$} &
      \colhead{$\mathrm{M_{\odot}\,kpc^{-2}\,yr^{-1}}$} &
      \colhead{$\mathrm{Gyr}$} &
      \colhead{$\mathrm{M_{\odot}\,pc^{-2}}$} &      
      \colhead{$\mathrm{cm^{-3}}$} &
      \colhead{$\mathrm{pc}$} &
      \colhead{$\mathrm{km/s}$} &
      \colhead{$\mathrm{km/s}$} &
      \colhead{$\mathrm{km/s}$} &
      \colhead{$\mathrm{km/s}$}
    }
    \colnumbers
    \startdata
    {\tt S05-Z1.0  } & (  614, 1703) &      4.07 &  1.13\cdot 10^{-4} &      48.9 &     0.242 &     0.135 &       626 &      9.46 &      6.22 &      4.96 &      5.11\\
    {\tt S05-Z0.1  } & (  614, 1830) &      4.76 &  4.74\cdot 10^{-5} &       189 &    0.0616 &     0.167 &       494 &      8.17 &      3.84 &      6.25 &      3.58\\
    \hline
    {\tt R8-Z3.0   } & (  438,  977) &      8.74 &  3.74\cdot 10^{-3} &      2.51 &      3.48 &      1.22 &       219 &      12.4 &      7.64 &      5.09 &      8.32\\
    {\tt R8-Z1.0   } & (  438,  977) &      9.22 &  2.97\cdot 10^{-3} &      3.79 &      2.55 &      1.16 &       235 &      12.8 &      8.38 &      5.92 &      7.61\\
    {\tt R8-Z0.3   } & (  438,  977) &      9.96 &  1.92\cdot 10^{-3} &      5.66 &      1.67 &      1.08 &       203 &      12.5 &      7.17 &      6.48 &      7.89\\
    {\tt R8-Z0.1   } & (  438,  977) &      10.6 &  1.29\cdot 10^{-3} &      14.8 &     0.970 &      1.23 &       188 &      11.9 &      6.36 &      6.60 &      7.60\\
    {\tt R8-Zg0.1Zd0.025} & (  438,  977) &      10.3 &  1.32\cdot 10^{-3} &      12.2 &      1.20 &      1.28 &       200 &      12.1 &      6.89 &      6.55 &      7.47\\
    \hline
    {\tt R8-b10-Z1.0} & (  438,  977) &      8.81 &  2.60\cdot 10^{-3} &      3.77 &      2.97 &      1.14 &       219 &      12.6 &      7.50 &      5.98 &      8.18\\
    {\tt R8-b10-Z0.3} & (  438,  977) &      9.84 &  1.90\cdot 10^{-3} &      6.90 &      1.88 &      1.26 &       203 &      12.2 &      7.17 &      6.39 &      7.52\\
    {\tt R8-b10-Z0.1} & (  438,  977) &      10.4 &  1.29\cdot 10^{-3} &      15.1 &      1.18 &      1.19 &       200 &      12.2 &      6.48 &      6.66 &      7.91\\
    {\tt R8-b10-Zg0.1Zd0.025} & (  438,  977) &      10.3 &  1.29\cdot 10^{-3} &      15.6 &      1.32 &      1.09 &       231 &      13.4 &      7.08 &      6.56 &      9.26\\
    \hline
    {\tt S30-Z1.0  } & (  438,  977) &      16.2 &    0.0114 &      1.53 &      13.1 &      2.62 &       248 &      13.8 &      9.39 &      6.22 &      8.00\\
    {\tt S30-Z0.1  } & (  438,  977) &      22.5 &  7.92\cdot 10^{-3} &      3.47 &      6.67 &      3.03 &       212 &      13.6 &      9.07 &      6.93 &      7.46\\
    \hline
    {\tt LGR4-Z3.0 } & (  204,  444) &      35.7 &    0.0503 &     0.726 &      16.4 &      8.46 &       201 &      13.8 &      9.02 &      4.31 &      9.49\\
    {\tt LGR4-Z1.0 } & (  204,  488) &      36.1 &    0.0343 &      1.08 &      14.6 &      7.38 &       171 &      13.6 &      8.19 &      5.57 &      9.26\\
    {\tt LGR4-Z0.3 } & (  204,  488) &      42.0 &    0.0252 &      1.80 &      8.19 &      6.70 &       175 &      14.0 &      8.58 &      6.26 &      9.11\\
    {\tt LGR4-Z0.1 } & (  204,  488) &      44.6 &    0.0186 &      3.16 &      5.43 &      7.28 &       184 &      13.7 &      7.90 &      6.51 &      9.06\\
    {\tt LGR4-Zg0.1Zd0.025} & (  204,  488) &      42.4 &    0.0202 &      2.21 &      7.14 &      6.45 &       196 &      13.7 &      8.02 &      6.35 &      9.05\\
    \hline
    {\tt LGR4-b10-Z1.0} & (  204,  488) &      33.2 &    0.0278 &      1.29 &      15.1 &      6.25 &       191 &      13.8 &      9.22 &      5.52 &      8.61\\
    {\tt LGR4-b10-Z0.1} & (  204,  488) &      41.4 &    0.0169 &      3.10 &      7.79 &      5.82 &       177 &      13.0 &      7.97 &      6.53 &      8.01\\
    \hline
    {\tt S100-Z1.0r} & (  204,  487) &      54.4 &     0.127 &     0.495 &      42.9 &      14.2 &       276 &      18.4 &      15.7 &      5.77 &      7.83\\
    {\tt S100-Z1.0 } & (  204,  425) &      42.0 &    0.0860 &     0.769 &      50.6 &      10.7 &       321 &      20.9 &      19.0 &      5.65 &      6.64\\
    {\tt S100-Z0.1 } & (  204,  488) &      62.4 &    0.0885 &     0.799 &      34.7 &      12.4 &       215 &      16.3 &      12.4 &      6.95 &      8.01\\
    \hline
    {\tt S150-Om200-Z1.0r} & (  153,  336) &       105 &     0.247 &     0.453 &      49.4 &      36.2 &       113 &      17.1 &      6.50 &      4.87 &      15.1\\
    {\tt S150-Om200-Z1.0} & (  153,  397) &      88.7 &     0.133 &     0.743 &      66.0 &      21.1 &       157 &      20.0 &      6.57 &      4.86 &      18.3\\
    {\tt S150-Om200-Z0.1} & (  153,  412) &       115 &     0.108 &      1.10 &      37.5 &      30.9 &       111 &      17.0 &      5.17 &      6.32 &      14.9\\
    {\tt S150-Om100q0-Z1.0} & (  146,  307) &      69.9 &     0.278 &     0.302 &      74.3 &      26.2 &       277 &      22.2 &      21.3 &      5.74 &      3.02\\
    {\tt S150-Om100q0-Z0.1} & (  146,  307) &      68.7 &     0.261 &     0.319 &      74.5 &      21.4 &       214 &      17.7 &      15.7 &      7.20 &      3.74\\
    \enddata
    \tablecomments{
    Column (1): model name, following convention explained in \autoref{tbl:model} and \autoref{sec:models}. Two models with a tag ``{\tt r}'' (i.e., {\tt S100-Z1.0r} and {\tt S150-Om200-Z1.0r}) are the $Z'=1$ simulations restarted from at the end of the early evolution for the $Z'=0.1$ models. {\tt R8-Z3.0} and {\tt LGR4-Z3.0} are restarted from {\tt R8-Z1.0} and {\tt LGR4-Z1.0} at $t=200$ and 150~Myr, respectively.
    Column (2): time interval over which the mean values are calculated.
    Columns (3)--(5): gas and SFR surface densities and their ratio, as presented in \autoref{sec:sfr}
    Column (6): surface density of star cluster particles formed during the simulation.
    Columns (7)--(12): warm-cold two phase (\twop) gas properties as defined by \autoref{sec:def}.
    }
  \end{deluxetable*}

\begin{deluxetable*}{lCCCCCCCCC}
    \tablecaption{Measured Weight and Midplane Pressures in Simulations\label{tbl:result2}}
    \tablehead{
      \colhead{Model} &
      \dcolhead{\W} &
      \dcolhead{\Ptot} &
      \dcolhead{\Pturb} &
      \dcolhead{\Pth} &
      \dcolhead{\Pimag} &
      \dcolhead{\dPimag} &
      \dcolhead{\oPimag} &
      \dcolhead{\Phot} &
      \dcolhead{\PDE}
    }
    \colnumbers
    \startdata
    {\tt S05-Z1.0  } &     0.125 &     0.141 &    0.0587 &    0.0400 &    0.0422 &    0.0120 &    0.0302 &     0.138 &     0.151\\
    {\tt S05-Z0.1  } &     0.146 &     0.130 &    0.0153 &    0.0883 &    0.0260 &  9.87\cdot 10^{-3} &    0.0162 &     0.160 &     0.172\\
    \hline
    {\tt R8-Z3.0   } &      1.61 &      1.69 &     0.618 &     0.246 &     0.829 &     0.371 &     0.458 &      1.37 &      1.75\\
    {\tt R8-Z1.0   } &      1.80 &      2.06 &     0.835 &     0.439 &     0.786 &     0.322 &     0.463 &      1.57 &      1.91\\
    {\tt R8-Z0.3   } &      1.84 &      2.01 &     0.571 &     0.575 &     0.862 &     0.327 &     0.535 &      1.41 &      2.04\\
    {\tt R8-Z0.1   } &      1.85 &      1.88 &     0.336 &     0.696 &     0.850 &     0.238 &     0.612 &      1.51 &      2.12\\
    {\tt R8-Zg0.1Zd0.025} &      1.84 &      1.85 &     0.322 &     0.653 &     0.876 &     0.271 &     0.605 &      1.68 &      2.07\\
    \hline
    {\tt R8-b10-Z1.0} &      1.67 &      1.77 &     0.558 &     0.402 &     0.807 &     0.333 &     0.474 &      1.35 &      1.79\\
    {\tt R8-b10-Z0.3} &      1.80 &      1.85 &     0.465 &     0.589 &     0.791 &     0.303 &     0.487 &      1.42 &      1.98\\
    {\tt R8-b10-Z0.1} &      1.87 &      1.87 &     0.308 &     0.682 &     0.882 &     0.247 &     0.635 &      1.46 &      2.11\\
    {\tt R8-b10-Zg0.1Zd0.025} &      1.96 &      1.94 &     0.324 &     0.597 &      1.02 &     0.268 &     0.752 &      1.57 &      2.24\\
    \hline
    {\tt S30-Z1.0  } &      4.30 &      4.69 &      1.86 &      1.07 &      1.76 &     0.948 &     0.809 &      3.64 &      3.97\\
    {\tt S30-Z0.1  } &      5.67 &      6.21 &      2.03 &      2.04 &      2.14 &     0.959 &      1.18 &      4.46 &      5.91\\
    \hline
    {\tt LGR4-Z3.0 } &      10.9 &      12.7 &      6.41 &      1.57 &      4.69 &      2.20 &      2.50 &      10.2 &      9.82\\
    {\tt LGR4-Z1.0 } &      10.4 &      11.7 &      5.00 &      2.43 &      4.31 &      1.86 &      2.45 &      10.2 &      9.56\\
    {\tt LGR4-Z0.3 } &      11.8 &      12.7 &      3.88 &      3.18 &      5.65 &      1.83 &      3.82 &      11.9 &      12.1\\
    {\tt LGR4-Z0.1 } &      12.6 &      13.2 &      3.37 &      3.97 &      5.91 &      2.07 &      3.83 &      11.4 &      13.1\\
    {\tt LGR4-Zg0.1Zd0.025} &      11.8 &      13.3 &      5.00 &      3.24 &      5.07 &      1.89 &      3.18 &      10.1 &      12.1\\
    \hline
    {\tt LGR4-b10-Z1.0} &      8.99 &      10.1 &      4.23 &      2.00 &      3.82 &      1.34 &      2.48 &      9.04 &      8.55\\
    {\tt LGR4-b10-Z0.1} &      10.9 &      11.5 &      3.44 &      3.41 &      4.62 &      1.66 &      2.96 &      10.2 &      11.5\\
    \hline
    {\tt S100-Z1.0r} &      27.7 &      32.8 &      19.6 &      5.70 &      7.43 &      3.20 &      4.23 &      24.1 &      21.0\\
    {\tt S100-Z1.0 } &      22.2 &      35.4 &      27.8 &      3.93 &      3.66 &      1.74 &      1.93 &      18.4 &      15.5\\
    {\tt S100-Z0.1 } &      28.2 &      31.6 &      12.7 &      8.53 &      10.4 &      5.24 &      5.13 &      24.9 &      24.1\\
    \hline
    {\tt S150-Om200-Z1.0r} &      70.9 &      72.2 &      15.9 &      8.81 &      47.5 &      14.3 &      33.2 &      69.1 &      56.3\\
    {\tt S150-Om200-Z1.0} &      58.7 &      57.4 &      8.78 &      3.82 &      44.8 &      6.21 &      38.6 &      65.1 &      45.1\\
    {\tt S150-Om200-Z0.1} &      72.7 &      72.0 &      8.66 &      12.4 &      51.0 &      9.55 &      41.4 &      70.1 &      64.7\\
    {\tt S150-Om100q0-Z1.0} &      53.4 &       112 &      98.7 &      10.6 &      3.04 &      2.48 &     0.566 &      34.6 &      33.0\\
    {\tt S150-Om100q0-Z0.1} &      42.0 &      53.1 &      32.0 &      16.2 &      4.86 &      3.79 &      1.08 &      39.5 &      29.1\\
    \enddata
    \tablecomments{Columns (2)--(8): weight and pressure/stress for the warm-cold two phase (\twop) gas as defined in \autoref{sec:def}. 
    Column (9): total pressure of the $\hot$ phase.
    Column (10): weight estimator (\autoref{eq:PDE}).
    All values are in units of $10^4k_B\Punit$, where $k_B$ is the Boltzmann constant.
    }
  \end{deluxetable*}

\begin{figure*}
  \includegraphics[width=0.49\linewidth]{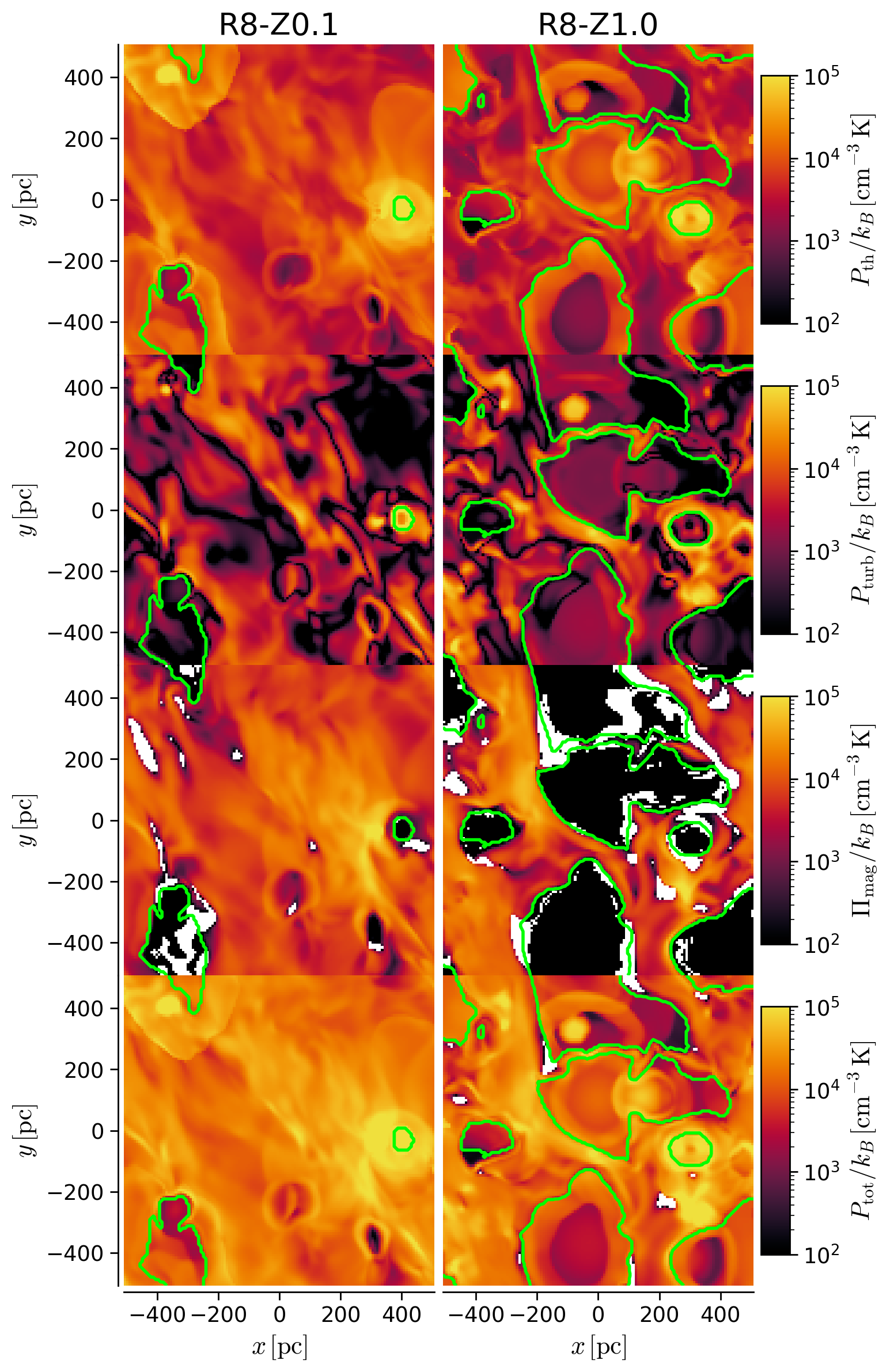}
  \includegraphics[width=0.49\linewidth]
  {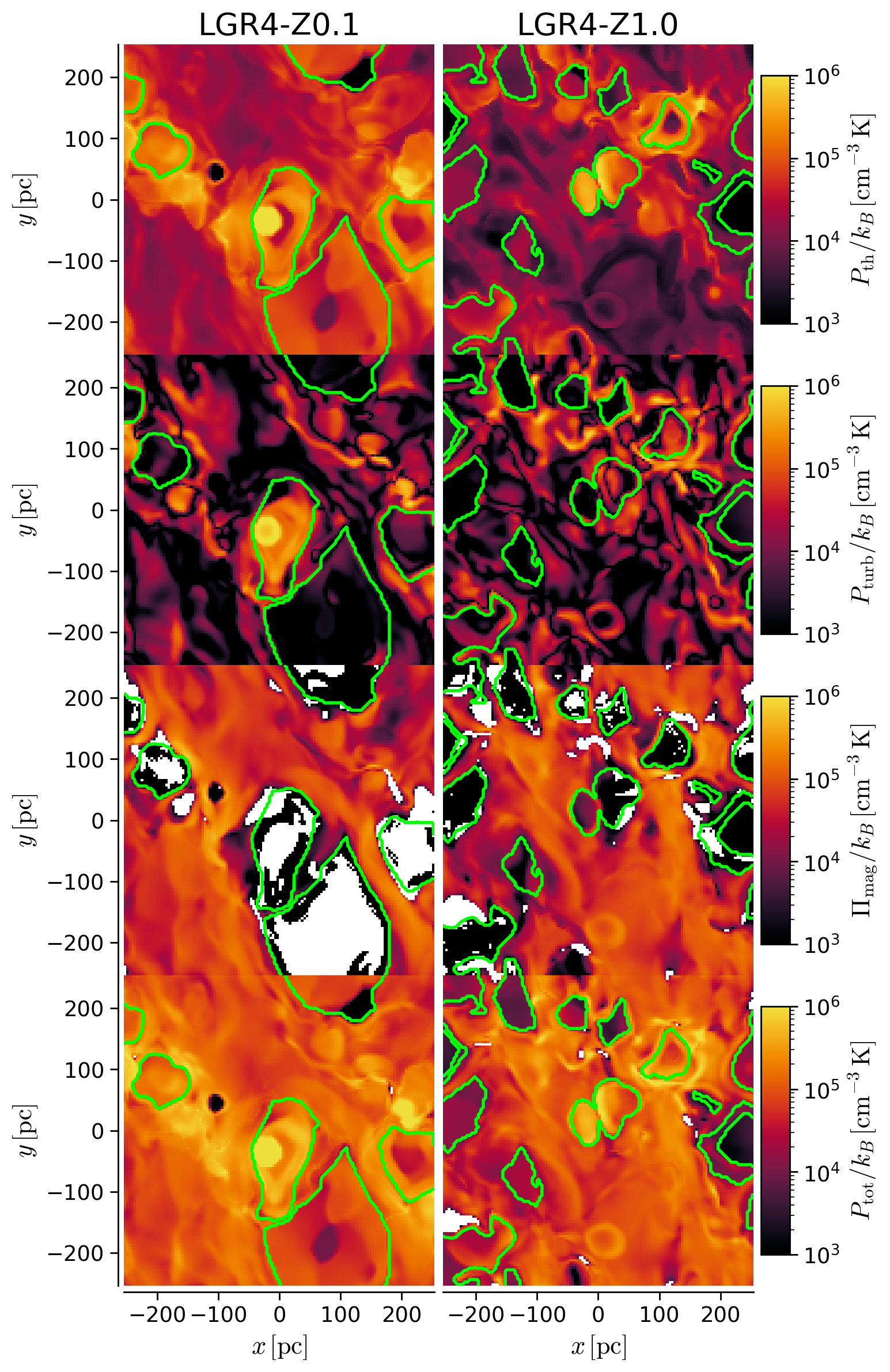}
  \caption{Example pressure/stress component maps for two selected metallicity models ($Z'=0.1$ and 1) for the {\tt R8} (left) and {\tt LGR4} (right) series using the same snapshot time as shown in \autoref{fig:R8-map} and \autoref{fig:LGR4-map}, respectively. The green contours denote the boundary between \WIM{} and \WHIM{}, equivalent to a temperature boundary at $T=3.5\times10^4\Kel$. \label{fig:Pmaps}}
\end{figure*}

\section{PRFM with Varying Metallicities}\label{sec:prfm_lowz}

To understand the metallicity dependence of SFRs seen in \autoref{sec:sfr}, we analyze the simulation results in the context of the PRFM theory of the star-forming ISM.
We begin by demonstrating that in all cases the system is in vertical dynamical equilibrium, meaning that the total midplane pressure provides the support required to balance the total gravitational weight.
We show also that the gas weight is insensitive to changes in metallicity (\autoref{sec:vert_equil}).
This implies that the demand for pressure support remains the same, irrespective of metallicity.
We then show that the thermal pressure (and magnetic stress to a lesser extent) increases at lower metallicity for a given SFR, meaning that low-metallicity systems are more efficient at recovering energy losses via stellar feedback (\autoref{sec:yield}).
We quantify the metallicity dependence of the feedback yield for each pressure/stress component.
Finally, we provide an effective equation of state, the relation between the gas density and total pressure, averaged over the simulation domain (\autoref{sec:eEoS}).
Using the effective equation of state we also compute effective vertical velocity dispersions; these are insensitive to metallicity, which is part of the reason why the weight is metallicity-insensitive.
The PRFM analysis explains the physics underpinning longer gas depletion times measured at low metallicity (\autoref{fig:sfr_history}): a higher yield in the conversion of stellar feedback energy to ISM pressure at low $\Zg$ and $\Zd$ means that energy lost to dissipation in the ISM can be recouped at a lower SFR.

\subsection{Definition of measured quantities}\label{sec:def}

We first define and measure a variety of quantities that describe the vertical momentum conservation (i.e., vertical dynamical equilibrium in a steady state) and characterize the ISM disk.
We construct horizontally averaged vertical profiles.
We use the angle brackets $\abrackets{q} \equiv \int q dx dy/A$ with $A=L_x L_y$ to denote the horizontal average of any quantity $q$.
To separately measure quantities in each phase, we use the Heaviside step function $\Theta({\rm ph})$ that returns 1 for the cell satisfying each phase definition or 0 otherwise (see \autoref{sec:result_map} for the phase definition).
For example, the area filling factor of each phase at a given height $z$ can be defined as
\begin{equation}\label{eq:fvol}
  f_{A,{\rm ph}}(z) = \abrackets{\Theta({\rm ph})}.
\end{equation}

We then define the vertical profiles of ``typical'' density and pressure components, respectively, by
\begin{equation}\label{eq:den}
  \rho_{\rm ph}(z) = \frac{\abrackets{\rho \Theta({\rm ph})}}{f_{A,{\rm ph}}(z)}
  \end{equation}
and
\begin{equation}\label{eq:pressure}
  P_{\rm comp, ph}(z) = \frac{\abrackets{P_{\rm comp}\Theta({\rm ph})}}{f_{A,{\rm ph}}(z)},
\end{equation}
where comp=th, turb, and mag to denote the thermal component $P_{\rm th}$, turbulent component (Reynolds stress) $P_{\rm turb}\equiv \rho v_z^2$, and magnetic component (vertical Maxwell stress) $\Pi_{\rm mag}\equiv B^2/(8\pi) - B_z^2/(4\pi)$ \citep[e.g.,][]{1990ApJ...365..544B}.
In addition to the total Maxwell stress $\Pi_{\rm mag}$, we also measure a decomposition of the stress into the mean $\Pi_{\overline{B}}$ and turbulent $\Pi_{\delta B}$ using $\overline{\bm{B}}\equiv \abrackets{\bm{B}}$ and $\delta{\bm B}\equiv \bm{B} - \overline{\bm{B}}$.
The total pressure is then simply
\begin{equation}\label{eq:Ptot}
  P_{\rm tot, ph} = P_{\rm th, ph} + P_{\rm turb, ph} + \Pi_{\rm mag, ph}.
\end{equation}

Next, we calculate the weight profile $\W(|z|) = (\W_+ + \W_-)/2$ using the integrals from a distance $|z|$ away from the midplane to the top and bottom of the box:
\begin{equation}\label{eq:weight}
  \W_\pm (|z|) = \int_{\pm|z|}^{\pm L_z/2} \abrackets{\rho \frac{d\Phi}{dz}} dz,
\end{equation}
where $\Phi$ is the total gravitational potential, which consists of the gravitational potential of gas, stars (young and old), and dark matter.
We often decompose the weight into two terms; the weight from gas self-gravity ($\Wsg$) and from the external gravity of stars and dark matter ($\Wext$).
Where $\W$ is used without an argument, it denotes the midplane ($z=0$) value.

Additionally, we measure the mass-weighted mean effective vertical velocity dispersion and gas scale height of the warm-cold two-phase gas (\twop=\CNM+\UNM+\WNM) defined by, respectively,
\begin{equation}\label{eq:sigma_avg}
  \szavg \equiv
  \rbrackets{\frac{\int \abrackets{P_{\rm tot}\Theta({\rm 2p})} dz}{\int \abrackets{\rho \Theta({\rm 2p})} dz}}^{1/2}
\end{equation}
and
\begin{equation}\label{eq:H}
  H =
  \rbrackets{\frac{\int \abrackets{\rho \Theta({\rm 2p})} z^2 dz}{\int \abrackets{\rho \Theta({\rm 2p})}dz}}^{1/2}.
\end{equation}
Note that the effective velocity dispersion is a quadratic sum of thermal ($\szth$), turbulent ($\szturb$), and magnetic ($\szmag$) components.
Each of these  may be defined by using \autoref{eq:sigma_avg} but with $P_{\rm th}$, $P_{\rm turb}$, and $\Pi_{\rm mag}$ instead of $\Ptot$.

We summarize quantities measured from our simulations in \autoref{tbl:result} and \autoref{tbl:result2}; values we report are averaged over a period during which the evolution is in a quasi-steady state.
Column (2) of \autoref{tbl:result} lists time ranges over which the time averages are taken, in units of Myr. \edit1{We use a snapshot interval of $\sim1$ Myr for the {\tt S05}, {\tt R8}, and {\tt S30} models and $\sim 0.5$ Myr for the {\tt LGR4}, {\tt S100}, and {\tt S150} models, providing typically  $\sim500$ snapshots for each model ($\sim300$ for {\tt S150} and $1000$ for {\tt S05}).}
Columns (3)--(5) list the mean gas surface density, SFR surface density, and gas depletion time discussed in \autoref{sec:sfr}, now including all models.
Column (6) lists the mean star particle surface density.
Columns (7)--(12) list the midplane hydrogen number density, gas scale height, and effective velocity dispersions of the warm-cold two-phase gas.
Note that the number density of hydrogen at the midplane is $\nHtwo=\rho_{\rm 2p}(0)/(\mu_{\rm H} m_{\rm H})$, where $\mu_{\rm H} = 1.4$ is the mean molecular weight per hydrogen nucleus. \autoref{tbl:result2} presents all midplane pressure and weight measurements that are used in the PRFM analysis in the following sections.

\edit1{
\autoref{fig:Pmaps} shows example midplane pressure/stress components of two selected metallicity models ($Z'=0.1$ and $1$) from the {\tt R8} and {\tt LGR4} series shown in \autoref{fig:R8-map} and \autoref{fig:LGR4-map}. Overall, the spatial variation of pressure/stress (in each component and total) within the two-phase gas is about an order of magnitude. Due to the diverging velocity field centered near the midplane, the hot superbubbles (bounded by the green contours) often have low turbulent stress. The magnetic stress also drops significantly within the hot gas -- it can become negative if the vertical component dominates the magnetic field. The thermal pressure compensates somewhat to make the total pressure more uniform than each pressure/stress component even across the different gas phases. We note that these particular snapshots have $\Pturb$ lower than is typical for the $Z'=1$ models, where $\Pturb\sim 2\Pth$ on average (see \autoref{tbl:result2}).  As we shall show in the subsequent sections, despite the large spatial and temporal variation of pressure \citep[e.g.,][]{2022ApJ...936..137O,2023ApJ...946....3K}, a well-defined mean midplane pressure can be robustly determined, allowing us to investigate the quasi-steady equilibrium state of the pressure/stress components.
}

A commonly adopted analytic weight estimator is
\begin{equation}\label{eq:PDE}
  \PDE \equiv \frac{1}{2}\pi G \Sgas^2
  + \Sgas\szavg (2G\rho_{\rm sd})^{1/2},
\end{equation}
where the midplane stellar+dark matter volume density $\rho_{\rm sd} = \Sstar/(2z_*) + \rho_{\rm dm}$.
In the last column in \autoref{tbl:result2}, we present results for $\PDE$, but take $\Sgas^2 \rightarrow \Sgas(\Sgas+\Sigma_{\rm sp})$ in the first term, where $\Sigma_{\rm sp}$ is the surface density of star cluster particles formed in the simulation, since gas and young star clusters have similar vertical distribution.
The contribution from $\Sigma_{\rm sp}$ is only significant in the {\tt S100} and {\tt S150} models (see \autoref{tbl:result}, column (6)).
This weight estimator has been frequently adopted in observational work; an implicit assumption is that the gas disk's thickness is smaller than that of the stellar disk and dark matter halo.

For succinctness, in \autoref{tbl:result} and {\autoref{tbl:result2}} as well as the rest of the paper, we omit the subscript ``\twop'' for the two-phase gas and the function argument ``(0)'' for the midplane value, i.e., $\Ptottwo(0) \rightarrow \Ptot$, unless they are necessary to clarify the meaning.
The weight integral without a function argument denotes the total weight integrated all the way to the midplane $z=0$ from the top/bottom of the box.

\begin{figure*}
  \includegraphics[width=\linewidth]{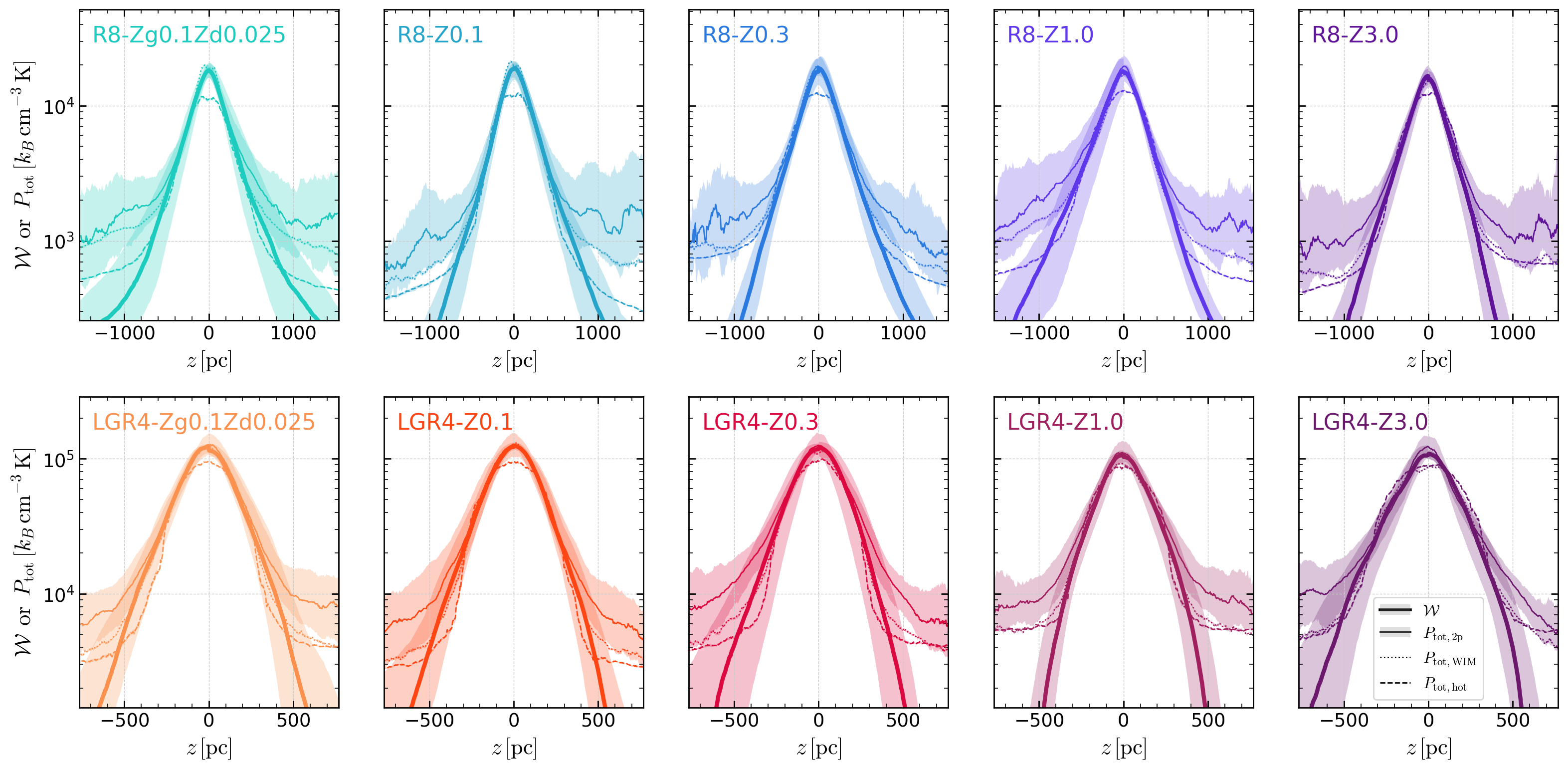}
  \caption{
    Demonstration of vertical dynamical equilibrium using the vertical profiles of the {\tt R8} (top row) and {\tt LGR4} (bottom row) models. Metallicity increases from left to right.
    We plot the vertical profiles of gas weight ($\W$, thick solid), total pressure of the \twop{} ($\Ptottwo$, thin solid), \WIM{} ($P_{\rm tot,WIM}$, dotted), and \hot{} ($\Phot$, dashed) phases.
    The lines and shaded areas show the median and 16th to 84th percentiles for the time range in \autoref{tbl:result}. Note that for visual clarity, we only show $z=\pm L_z/4$ rather than the full extent of $z=\pm L_z/2$.
  } \label{fig:VDE}
\end{figure*}

\begin{figure*}
  \includegraphics[width=\linewidth]{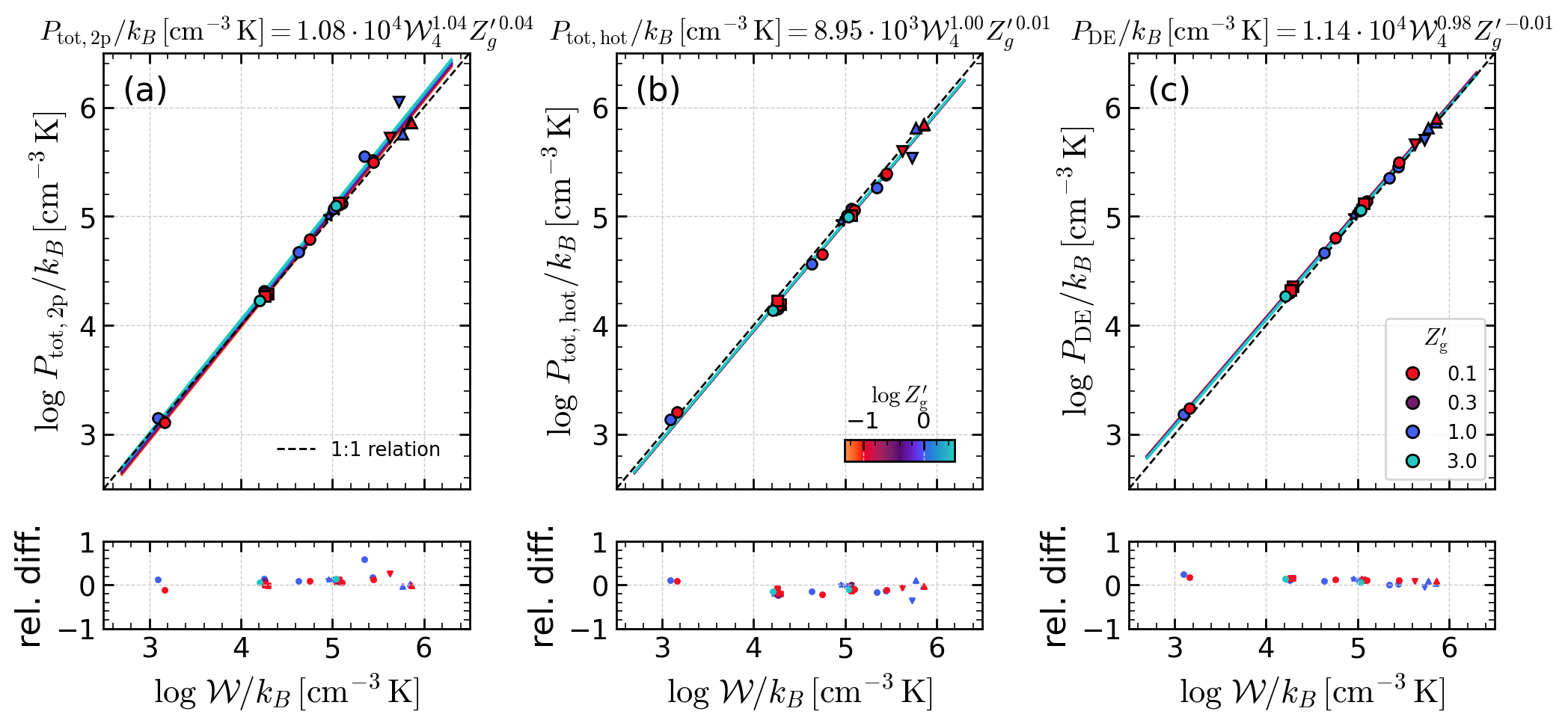}
  \caption{Demonstration of vertical dynamical equilibrium for all models.
  We plot the measured total midplane pressures of the \twop{} and \hot{} phases ((a) $\Ptot$ and (b) $\Phot$, respectively) as well as (c) the weight estimator $\PDE$ presented in \autoref{eq:PDE} as a function of the total weight $\W$.
  The bottom panel gives relative differences between the ordinates and abscissas.
  Each point represents the mean value of each model over the time range in \autoref{tbl:result}.
  The color of points shows the gas metallicity with circles for $\Zg=\Zd$ and squares for $\Zd = 0.025$.
  The star symbols are for the {\tt b10} models (they are hardly visible as they overlap with circles so well).
  The lower and upper triangles are for the {\tt S150-Om100q0} and {\tt S150-Om200} models.
  The colored lines are the fitting results of a bi-variate power-law model.
  The dashed black line is for the one-to-one relation.}
  \label{fig:VDE_fit}
\end{figure*}

\begin{figure*}
  \centering
  \includegraphics[width=\linewidth]{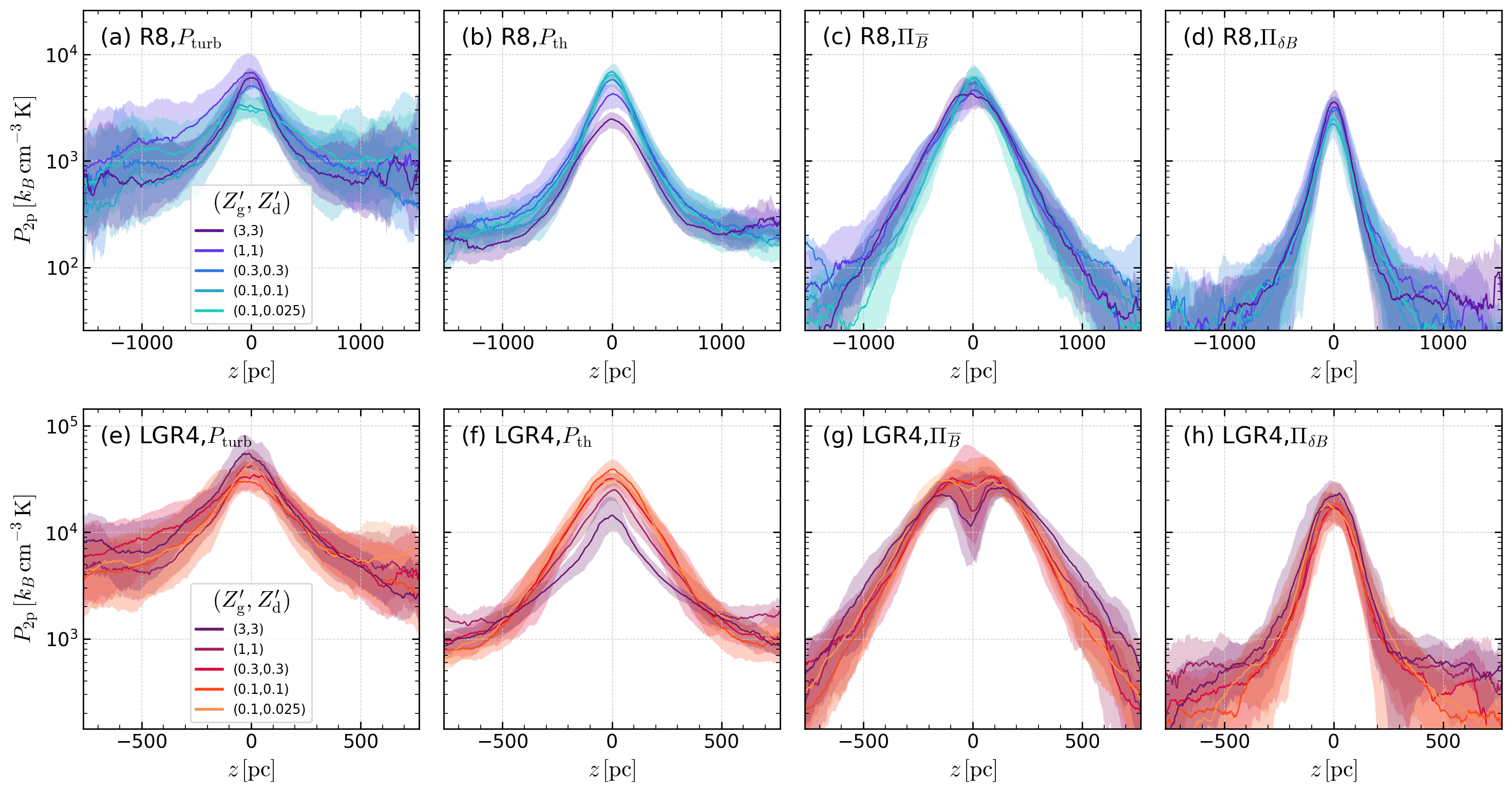}
  \caption{
    Vertical profiles of pressure components in the \twop{} phase as defined by \autoref{eq:pressure}.
    From left to right, we show turbulent, thermal, mean magnetic, and turbulent magnetic components, while different colors denote different metallicity in {\tt R8} (top) and {\tt LGR4} (bottom).
    The lines and shaded areas show the median and 16th to 84th percentiles for the time range in \autoref{tbl:result}.
  } \label{fig:Pcomp}
\end{figure*}

\subsection{Vertical dynamical equilibrium}\label{sec:vert_equil}

To investigate the behavior of SFR at different metallicities using the PRFM theory, we first check the establishment of vertical dynamical equilibrium. We again use the {\tt R8} and {\tt LGR4} model series for in depth investigations. \autoref{fig:VDE} shows the weight profiles along with the total pressure profiles for the \twop, \WIM, and \hot{} phases.
Within one or two gas scale heights, the total pressure profiles of all phases are in good agreement with the weight profiles and with each other.
Only the midplane total pressure of the \hot{} phase is smaller than those of the other phases.
This is because the hot gas near the midplane represents the centers of expanding bubbles whose radial velocity increases outward.
The excess pressure at high-$z$ in the \twop{} phase is dominated by the turbulent component arising from warm fountain flows (\autoref{fig:Pcomp}; see also \citealt{2018ApJ...853..173K,2020ApJ...900...61K}).

Since the external vertical gravity is fixed at a given galactic condition, the total weights (or weight profiles) vary due to the change in total gas surface density, giving rise to a slightly higher weight at lower metallicity (see \autoref{tbl:result}).
Modulo this effect, the shape of total pressure and density profiles is more or less similar within each model series irrespective of metallicity, characterized by similar scale height and velocity dispersions.

To give a more quantitative view,
\autoref{fig:VDE_fit}(a) plots the measured total midplane pressure of the \twop{} phase as a function of the total weight for all models.
This pressure and the weight are in excellent agreement, validating
vertical dynamical equilibrium.\footnote{Only two models ({\tt S150-Om100q0-Z1.0} and {\tt S100-Z1.0}) show significant deviation from equilibrium, with $\Ptot>\W$ by a factor of $\sim 2.1$ and $1.6$, respectively.
A deeper inspection of these models reveals that they are still in vertical dynamical equilibrium when we account for all phases and the \hot{} phase occupies the majority of the volume for all $z$.
In this case, the warm/cold \twop{} phase is no longer volume-filling and compressed into clumpy blobs, making the \twop{} phase not a representative phase for vertical dynamical equilibrium.}
For the same galactic condition, the models with different metallicities (denoted by different colors) are clustered at similar weights with a tendency to have slightly higher weights at lower metallicities.
Again, this is because, at lower metallicity, gas surface density decreases more slowly due to lower SFR surface density and longer gas depletion time.

\autoref{fig:VDE_fit}(b) shows that total pressure equilibrium is satisfied between the \twop{} and \hot{} phases. \autoref{fig:VDE_fit}(c) shows that the analytic weight estimator in \autoref{eq:PDE} using the measured effective velocity dispersion (from the mass-weighed average along the vertical direction) provides a good estimate of the true weight and hence the total pressure.
These conclusions hold irrespective of metallicity.
While above each panel we report the results of a bi-variate fit of pressure to $\W$ and $\Zg$, the purpose of this is simply to demonstrate (1) how accurately vertical equilibrium is satisfied, and (2) how negligible the measured dependence on metallicity is.

Taken together, and using the definition of weight in \autoref{eq:weight}, the results of \autoref{fig:VDE_fit}  show that for all gas phases,
\begin{equation}\label{eq:Pandg}
\Ptot \approx {\W} = \frac{1}{2}\Sgas \langle g_z \rangle
\end{equation}
where
\begin{equation}\label{eq:gz}
\langle g_z\rangle \approx \pi G \Sgas + 2\szavg (2G\rho_{\rm sd})^{1/2}
\end{equation}
is an estimator of the mean vertical gravity within the gas layer.
We caution, though, that the weight estimator in \autoref{eq:PDE} and gravity estimator in \autoref{eq:gz} are derived assuming that the gas disk is thinner than the stellar disk and dark matter halo.
This assumption, and hence application of \autoref{eq:PDE} is acceptable for the current simulation suite and most nearby normal star-forming galaxies \citep[e.g.,][]{2020ApJ...892..148S,2021MNRAS.503.3643B}.
However, more extreme, gas-rich systems \citep[e.g.,][]{2021ApJ...909...12G}, may have more vertically extended gas disks where \autoref{eq:PDE} overestimates the weight.
For this reason, in the future, we recommend adopting the generalized formulae for weight and vertical gravity presented in S. Hassan et al (2024, submitted), which may be applied to gas disks that are either thinner or thicker than the corresponding stellar disks.

\autoref{fig:Pcomp} shows the vertical profiles of pressure components in the \twop{} phase.
Unlike the total pressure and weight profiles, these show noticeable metallicity dependence.
In particular, the thermal pressure increases as gas metallicity decreases, while there is a corresponding reduction in turbulent Reynolds stress.
The Maxwell stresses have somewhat less consistent behavior; for {\tt R8} the mean Maxwell stress is higher and the turbulent Maxwell stress is lower at low metallicity, while for {\tt LGR4} these trends are less clear.

We conclude that consistent with expectations, the disk's vertical structure satisfies the quasi-steady-state vertical momentum conservation law such that the total pressure support matches the weight, which is insensitive to metallicity.
However, the detailed ISM physics governed by MHD and thermodynamics with radiation and mechanical feedback is responsible for setting the individual pressure components.
To fulfill the required total pressure support, the pressure components are constrained to add up to the same value regardless of metallicity, but the relative importance of different components varies in interesting ways.
From now on, we shall quantify the variation of the relative importance of each pressure component as a function of weight and metallicity.

\begin{figure*}
  \gridline{\fig{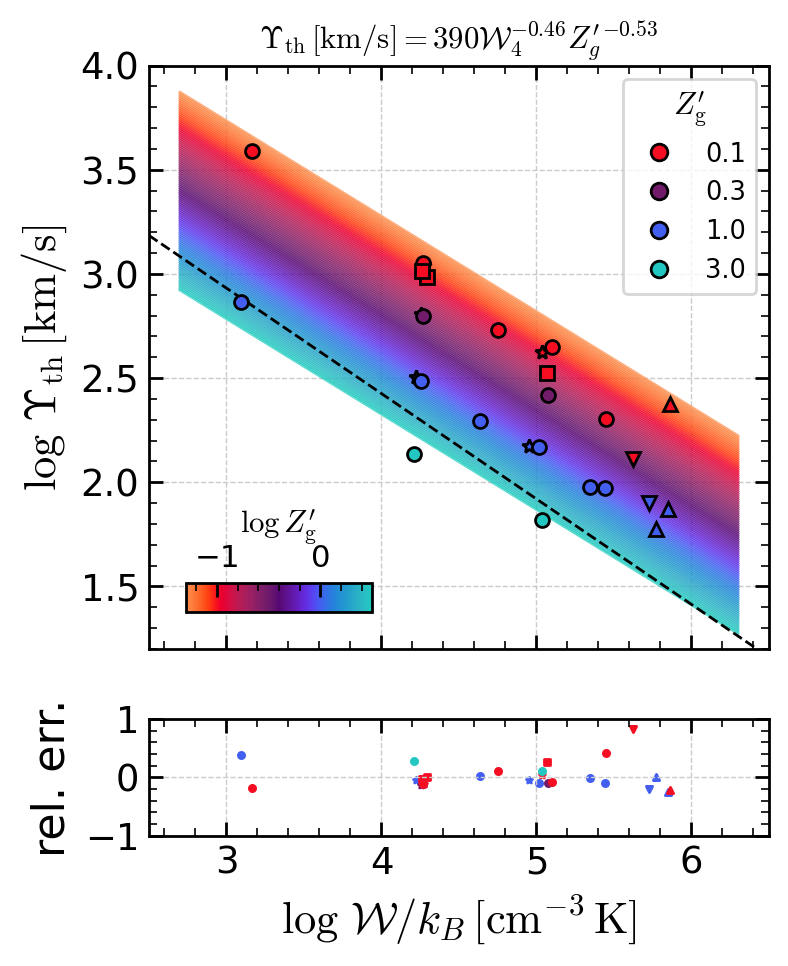}{0.45\textwidth}{(a) Thermal Feedback Yield}
  \fig{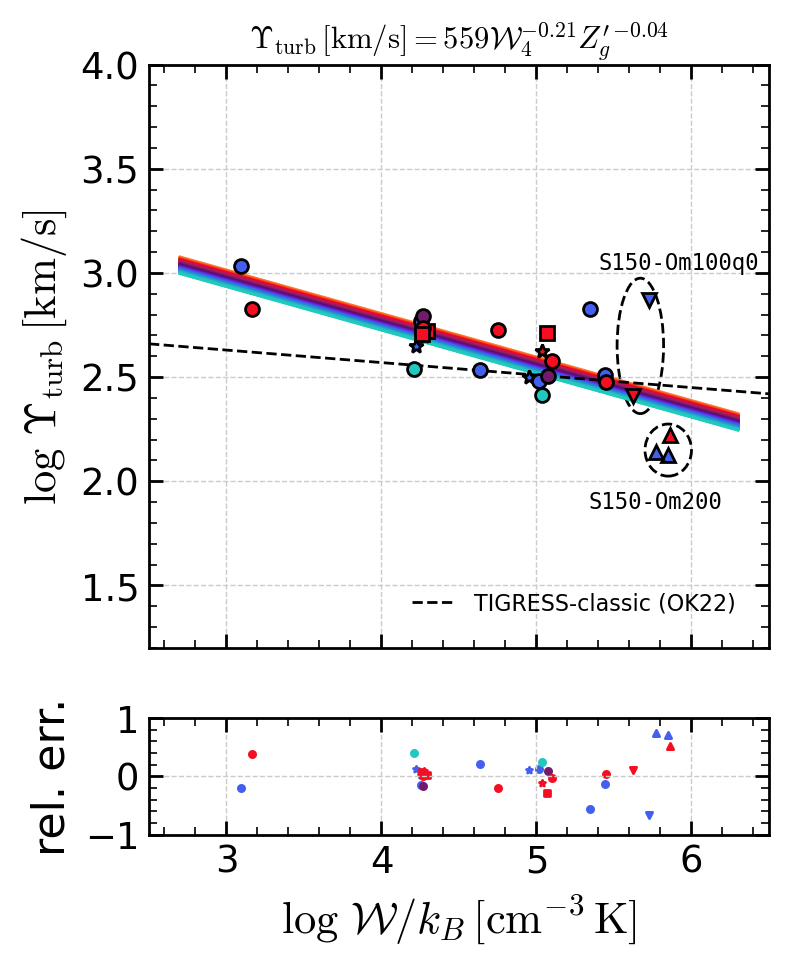}{0.45\textwidth}{(b) Turbulent Feedback Yield}}
  \gridline{\fig{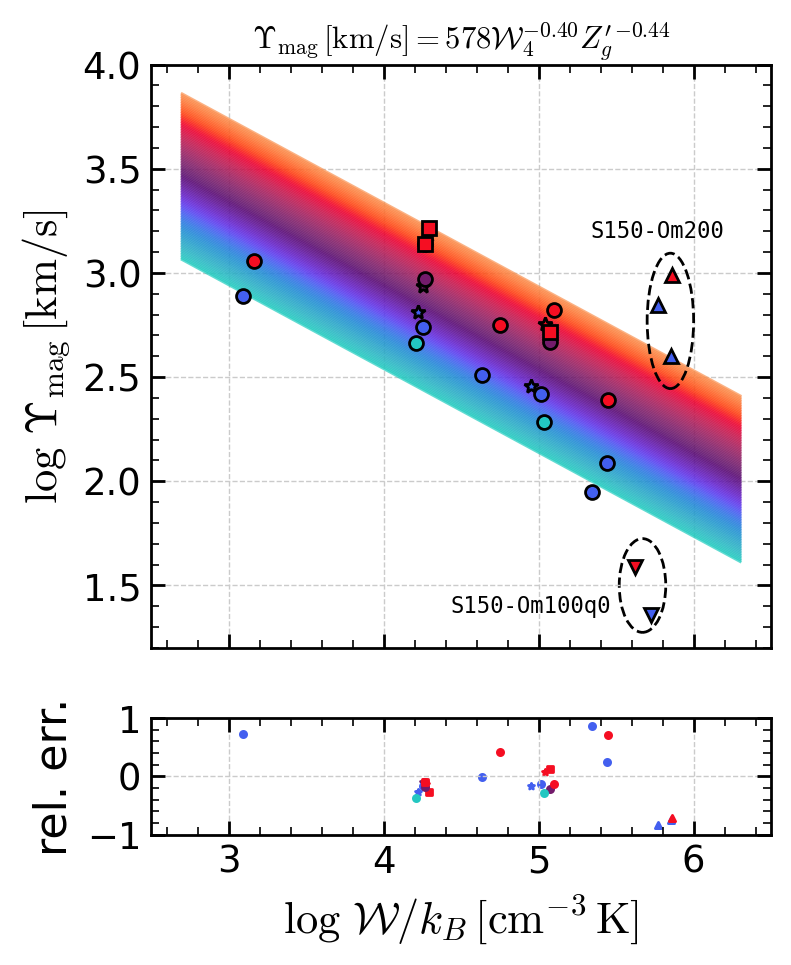}{0.45\textwidth}{(c) Magnetic Feedback Yield}
  \fig{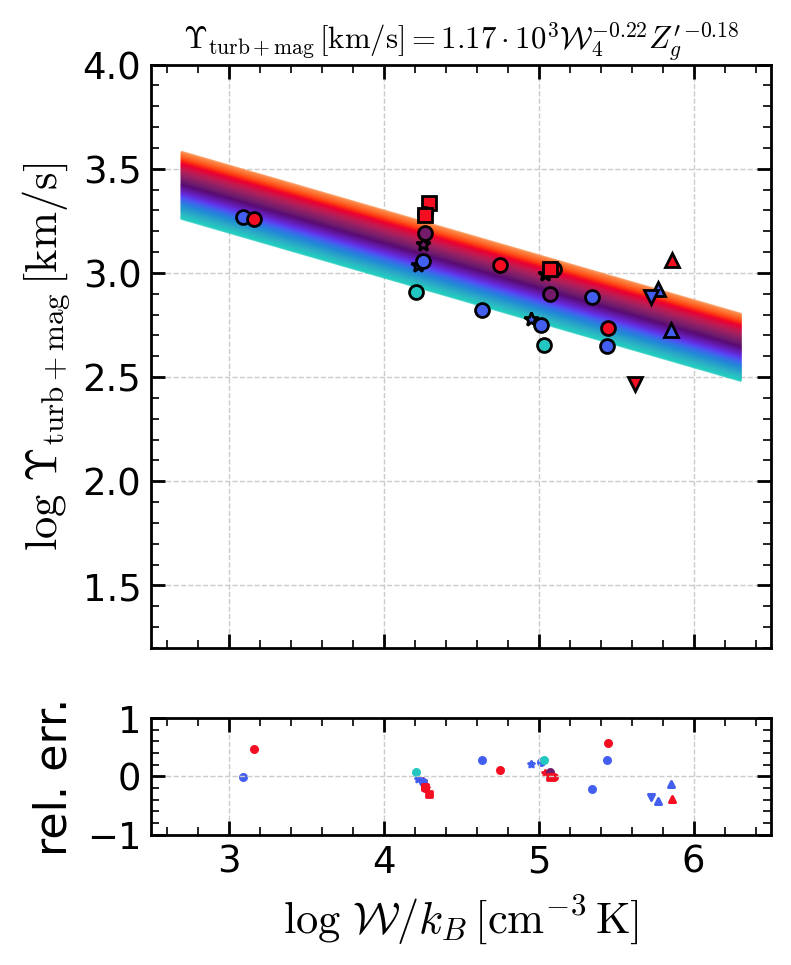}{0.45\textwidth}{(d) Non-Thermal Feedback Yield}}
  \caption{Metallicity and weight dependence of each feedback yield component.
  Symbols denote the mean value of feedback yield and weight over the time range in \autoref{tbl:result}.
  The color of points shows the gas metallicity with circles for $\Zg=\Zd$ and squares for $\Zd = 0.025$.
  The star symbols are for the {\tt b10} models. The lower and upper triangles are for the {\tt S150-Om100q0} and {\tt S150-Om200} models.
  The fitting results are presented as the equation in the panel title and as colored lines.
  The relative errors of the fitting results (i.e., model/simulation $- 1$) are presented at the bottom of each panel.
  In panels (a) and (b), the fits to the TIGRESS-classic suite (with $Z'=1$) presented in \citet{2022ApJ...936..137O} are shown as the black dashed line.}
  \label{fig:yield}
\end{figure*}

\begin{figure*}
  \includegraphics[width=0.48\linewidth]{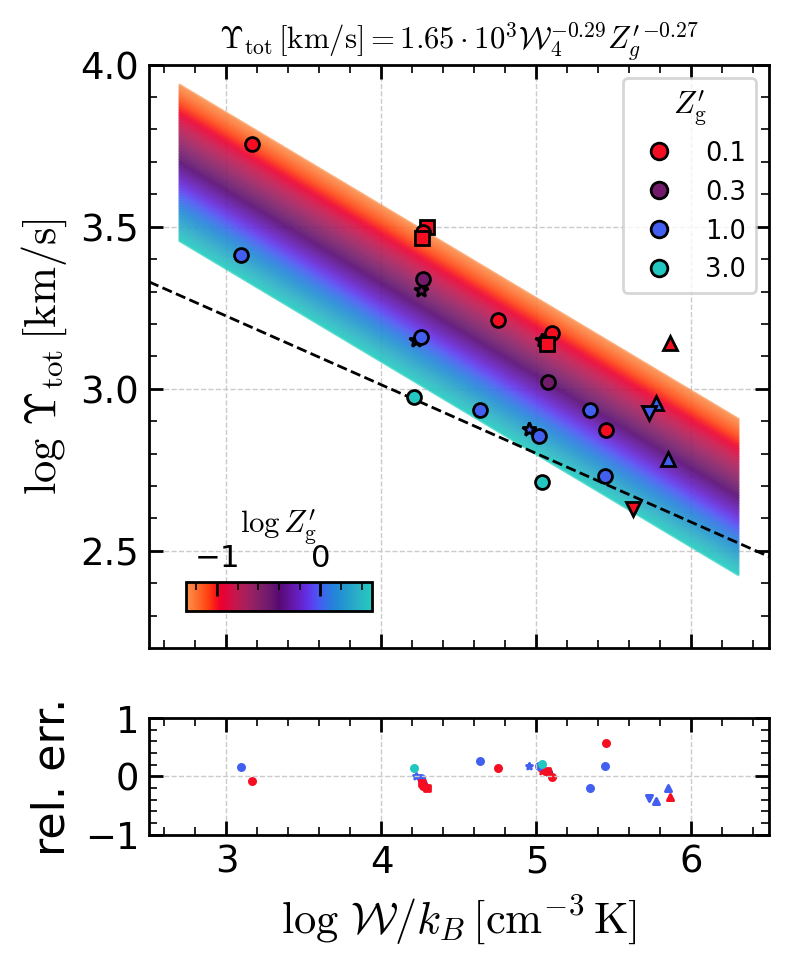}
  \includegraphics[width=0.48\linewidth]{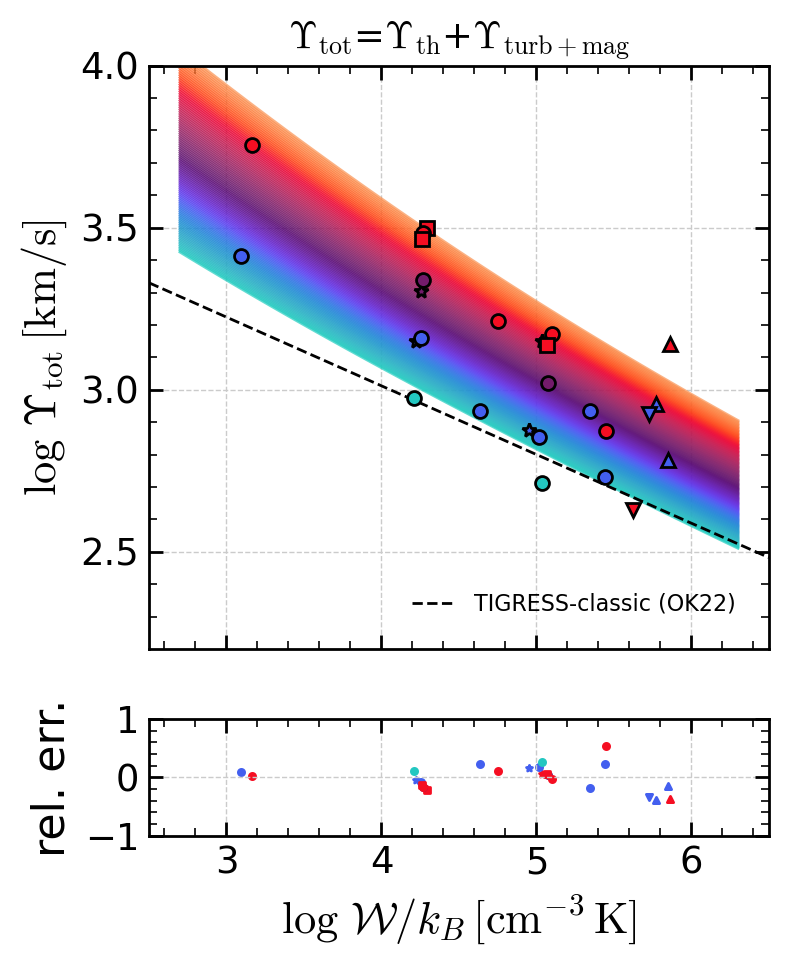}
  \caption{Two fitting results for the total feedback yield. {\bf Left}: total feedback yield model obtained by direct fitting for measured $\Ytot$, i.e., \autoref{eq:Ytot}. {\bf Right}: total feedback yield model obtained by the sum of two bi-variate power fitting results in \autoref{fig:yield}(a) and \autoref{fig:yield}(d), i.e., \autoref{eq:Yth} plus \autoref{eq:Ynonth}.
  The fit to the TIGRESS-classic suite presented in \citet{2022ApJ...936..137O} is shown as the black dashed line.
  Symbols and lines have the same meaning as in \autoref{fig:yield}.}
  \label{fig:yield_tot}
\end{figure*}

\subsection{Feedback yields}\label{sec:yield}

In the PRFM theory, we postulate that the ISM weight is balanced by pressure support that responds to star formation feedback, which can vary as needed to fulfill its role.
The connection between pressure and SFR surface density can be characterized in terms of the feedback yield parameter \citep[see][and references therein, and   \citealt{2015ApJ...815...67K} for an in-depth focus on magnetic effects]{2022ApJ...936..137O}.
The total feedback yield, and feedback yields of individual stress components, are defined by
\begin{equation}\label{eq:yield}
  \Upsilon_{\rm tot} \equiv \frac{\Ptot}{\Ssfr};\ \
  \Upsilon_{\rm comp} \equiv \frac{P_{\rm comp}}{\Ssfr}
\end{equation}
using the pressure/stress at the midplane reported in \autoref{tbl:result2}.
The feedback yield has units of velocity.

\autoref{fig:yield} shows each feedback yield measured from all models as a function of the weight and metallicity.
We fit the resulting feedback yield using a bi-variate power-law model for the weight and gas metallicity.
The fitting is carried out by the orthogonal distance regression method implemented in the {\tt scipy} package.
Since the goal here is to provide the calibration of the feedback yield employing the \emph{equilibrium} assumption made in the PRFM theory,\footnote{It is of great interest to develop a theoretical model including the effects of mutually correlated time evolution of pressure and SFR surface density.
However, such a model requires a deeper understanding of the time scales of energy gain and loss as well as dynamical evolution under changing gravity and pressure, which we defer to future work.} we simply use the mean values measured over long-term evolution covering \edit1{$4-6$ star formation/feedback and outflow/inflow} cycles and weigh them equally.
We do not attempt to reduce the bias arising from the uneven sampling of the parameter space given that the underlying true parameter distribution is not well understood.
We remark that, if the true physical relationship follows this theoretical power-law, then this regression should converge to the true parameters regardless of parameter priors.
The bi-variate fitting results are presented at the top of the following figures.

The fitting results are
\begin{equation}\label{eq:Yth}
  \Yth = 390\kms {\W}_4^{-0.46}\Zg^{-0.53},
\end{equation}
\begin{equation}\label{eq:Yturb}
  \Yturb = 561\kms {\W}_4^{-0.21}\Zg^{-0.04},
\end{equation}
\begin{equation}\label{eq:Ymag}
  \Ymag = 578\kms {\W}_4^{-0.40}\Zg^{-0.44},
\end{equation}
where $\W_4 \equiv \W/(10^4k_{\rm B}\Punit)$.
In the high pressure regime, the different galactic rotation speeds lead to diverging magnetic and turbulent feedback yields (upper vs. lower triangles).
There is a rough correspondence between low(high) $\Ymag$ and high(low) $\Yturb$, implying an exchange between turbulent and magnetic energy.
It is thus reasonable to combine turbulent and magnetic stresses and measure the associated feedback yield as in \autoref{fig:yield}(d), which gives
\begin{equation}\label{eq:Ynonth}
  \Ynonth = 1.17\times10^3\kms {\W}_4^{-0.22}\Zg^{-0.18}.
\end{equation}

The decreasing trend of $\Yth$ with increasing $\W=\Ptot$ and $\Zg$ is the cleanest relation seen in \autoref{fig:yield}.
As we will show below, this behavior is mainly driven by the increased attenuation of FUV radiation in environments with higher density (corresponding to higher weight and pressure) and higher dust abundance.
A decreasing trend of $\Yth$ with increasing pressure and weight at solar metallicity was previously reported in \citet{2022ApJ...936..137O} based on the TIGRESS-classic suite at $Z'=1$ (shown as a dashed line in \autoref{fig:yield}(a)).
In the TIGRESS-classic framework, however, the global radiation attenuation effect is \emph{imposed} by a theoretical model for the radiation field based on a simple plane-parallel approximation (see Equation 11 in \citealt{2022ApJ...936..137O}).
With explicit UV radiation transfer using ART as implemented in TIGRESS-NCR, in \citet{2023ApJ...946....3K} we confirmed for {\tt R8} and {\tt LGR4} (at solar metallicity) consistency in the scaling of $\Yth$ with $\W$ with that of the TIGRESS-classic suite, with a normalization about a factor of 1.5 higher. \autoref{fig:yield}(a) reaffirms the evidence that $\Yth$ decreases with $\W$, but now over a wider range of parameter space and at different metallicities.
This is an indirect confirmation of the effective attenuation model used in the TIGRESS-classic framework \citep{2020ApJ...900...61K,2022ApJ...936..137O}; a direct comparison of the resulting FUV radiation field with different approximate solutions will be presented in N. Linzer et al. (submitted).

To understand the metallicity dependence of the thermal feedback yield, multiple factors must be considered simultaneously.
The simplest estimation of the thermal pressure is from the assumption of the balance between metal cooling \edit1{( $\nH^2 \Lambda_{\rm metal}$, especially \ion{C}{2} and \ion{O}{1}) and grain PE heating ($\nH\Gamma_{\rm PE}$) in the \CNM{},} and pressure equilibrium between the \CNM{} and \WNM{}.
The assumption of thermal equilibrium gives rise to $n_{\rm eq} = \Gamma_{\rm PE}/\Lambda_{\rm metal}$, and hence one can write
\begin{equation}\label{eq:Pth_Z}
  P_{\rm th,eq}/k_B = n_{\rm eq} T =  \frac{\Gamma_{\rm PE}T}{\Lambda_{\rm metal}(T, x_{\rm e})} \propto \frac{\epsilon_{\rm PE} J_{\rm FUV} \Zd}{\Zg}
\end{equation}
where in the final proportionality we focus just on the factors that depend strongly on metallicity.
Dividing \autoref{eq:Pth_Z} by $\Sigma_{\rm SFR}$, we obtain
\begin{equation}\label{eq:Yth_Z}
  \Yth \propto  \frac{\Zd}{\Zg}\epsilon_{\rm PE}f_\tau.
\end{equation}
Here, $f_\tau \equiv 4\pi J_{\rm FUV}/\Sigma_{\rm FUV}$ is an attenuation factor relating the FUV radiation field to the injection rate of FUV per unit area in the disk, $\Sigma_{\rm FUV}$, which is linearly proportional to $\Ssfr$ averaged over an appropriate time bin (e.g., $t_{\rm bin} \sim 10 \Myr$ for FUV).

Although the metallicity effects cancel out as long as $\Zg$ varies linearly with $\Zd$ to the lowest order, the higher order effects associated with radiation transfer ($f_\tau$) and the thermal and ionization state of gas ($\epsilon_{\rm PE}$) still lead to a metallicity dependence of the thermal feedback yield in the \CNM{}.
On the one hand, as clearly seen in \autoref{fig:R8-map} and \autoref{fig:LGR4-map}, the FUV radiation field (third row; $\chi_{\rm FUV}\equiv J_{\rm FUV}/J_{\rm FUV, Draine}$) is much higher at low dust abundance (with similar $\Ssfr$) because attenuation is vastly reduced, making $f_\tau$ significantly higher.
On the other hand, $\epsilon_{\rm PE}$ depends (inversely) on the grain charging parameter $\propto \chi_{\rm FUV} T^{1/2}/n_e$ \citep{1994ApJ...427..822B,2001ApJS..134..263W}, with $\epsilon_{\rm PE}$ larger in colder, higher density gas, and sensitive to the ionization fraction.
In the neutral ISM, the main source of free electrons is from H ionization by low-energy CRs (with electrons from C$^+$ becoming important at $\Zg>1$), so the grain charging parameter is relatively insensitive to metallicity for a given FUV field and CR ionization rate (see Fig. 6 of \citealt{2023ApJS..264...10K}).
In {\tt R8} and {\tt LGR4}, we find that the reduction from $\Zg=1$ to 0.1 gives rise to a factor of 2.6 and 3 increase in $f_\tau$ and 0.9 and 0.6 decrease in $\epsilon_{\rm PE}$, respectively.
This explains the larger thermal feedback yield at lower metallicity.

Finally, there is no difference seen in the model with reduced dust abundance $\Zd=0.025$ at $\Zg=0.1$ (square symbols).
We find that this is because, at this low metallicity, the PE heating is comparable to or weaker than the CR heating which has no dependence on dust abundance.
At the same time, metallicity-dependent cooling is no longer dominant.
Rather, the \ion{H}{1} Ly$\alpha$ cooling in the \WNM{} becomes dominant as the \CNM{} fraction decreases with decreasing $\Zg$, removing the metallicity dependence below $\Zg<0.1$.

The turbulent feedback yield shown in \autoref{fig:yield}(b) varies less sensitively with weight and metallicity.
This is qualitatively consistent with the expectation that the total momentum injected by each SN is not a sensitive function of gas density and metallicity \citep[e.g.,][]{1998ApJ...500...95T,2015ApJ...802...99K,2017ApJ...834...25K,2023ApJS..264...10K,2015MNRAS.450..504M,2020ApJ...896...66K,2020MNRAS.495.1035S,2022ApJS..262....9O}.
In a high-pressure environment, the behavior becomes somewhat irregular and more sensitive to other environmental parameters (e.g., galactic rotation speed).
The apparent behavior depending on galactic rotation (upper and lower triangles at high pressure/weight) is related to magnetic fields.

In the models with high rotation/shear ({\tt S150-Om200}; upper triangles), the mean magnetic field gets much stronger than in the low shear models ({\tt S150-Om100q0}; lower triangles), dominating the overall support against the total weight ($\Ymag \sim 10^3\kms$ with $\Yturb$ and $\Yth\sim 10^2\kms$).
The gas depletion time is longer in {\tt S150-Om200}, with significantly lower turbulent velocity dispersion $\szturb\sim 5$--$7\kms$ as opposed to $15$--$20\kms$ in {\tt S150-Om100q0}.
In the {\tt S100} and {\tt S150-Om100q0} models where SN feedback is very strong and blows away a lot of mass together with magnetic fields, the build-up of the mean magnetic field is hindered significantly.
These models behave like hydrodynamics models without much contribution from magnetic feedback yield ($\Ymag\lesssim 100\kms$), which is compensated by large turbulent feedback yield.

The total feedback yield is the sum of all components.
In \autoref{fig:yield_tot}, we show the total feedback yield measurements with (a) a fitting result to the total feedback yield directly and (b) a two-component model using thermal and non-thermal feedback yields (i.e., the sum of \autoref{eq:Yth} and \autoref{eq:Ynonth}).
The direct fitting result is
\begin{equation}\label{eq:Ytot}
  \Ytot = 1.65\times10^3\kms {\W}_4^{-0.29}\Zg^{-0.27}.
\end{equation}
There is only a small improvement in the mean of the residual from (b) compared to (a).
This implies that these two are virtually equivalent within the parameter space explored in this paper.
But, the two-component model would behave better in extreme conditions by respectively capturing the dominance of thermal and non-thermal pressure supports in low and high pressure regimes.
Compared to the reference line showing the result from the TIGRESS-classic suite \citep{2022ApJ...936..137O}, TIGRESS-NCR gives quite similar values near $\W/k_B =10^5 \pcc \Kel$ for $\Zg=1$ but has a steeper slope, with feedback yields larger by $\sim 0.2$ dex at low pressure.
Additionally, $\Ytot$ is larger by $\sim 0.3$ dex at $\Zg=0.1$ compared to $\Zg=1$.

We attribute the enhancement of total feedback yield to the effect of ionizing UV, which was missing in TIGRESS-classic.
Further investigation of the role of individual feedback channels will be addressed in a subsequent paper.

\begin{figure}
    \centering
    \includegraphics[width=\linewidth]{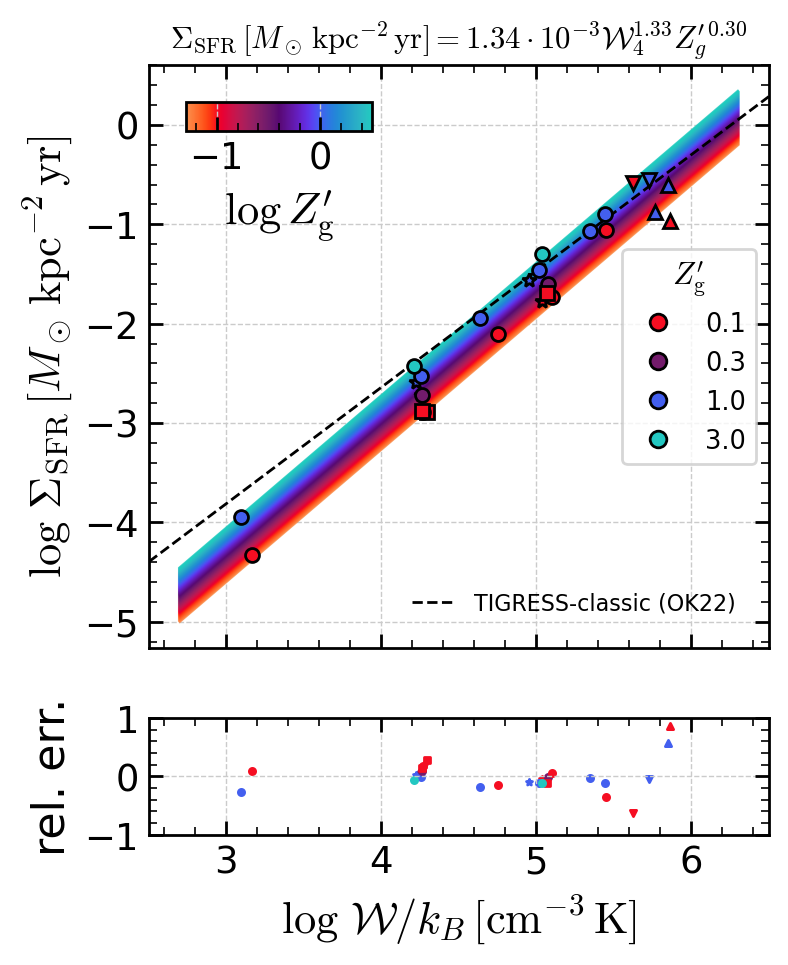}
    \caption{
    Metallicity and weight dependence of SFR surface density for our full simulation suite.
    The fit to the TIGRESS-classic suite presented in \citet{2022ApJ...936..137O} is shown as the black dashed line.
    Symbols and lines have the same meaning as in \autoref{fig:yield}.
    The observable manifestation of lower feedback efficiency under higher pressure (or density) and higher metallicity conditions is the slightly superlinear dependence of $\Ssfr$ on $\W$, and the moderate positive dependence of $\Ssfr$ on $Z$, quantified in the fit shown.
    }
    \label{fig:sfr_tdep_W}
\end{figure}

Finally, \autoref{fig:sfr_tdep_W} shows the relationship between $\Ssfr$, $\W$, and $\Zg$ for our full simulation suite, together with a bi-variate fit.\footnote{The fit is slightly different from what would be obtained from $\Ssfr = \W/\Ytot$ using \autoref{eq:Ytot} because the fitting results are not precise at the level of the significant figure reported here.}
The superlinear dependence on $\W$ and positive dependence on $\Zg$ express the fact that all forms of feedback become less efficient in regions where the pressure (and density) and the metallicity are higher.
Since, however, the dependencies on $\W$ and $\Zg$ in \autoref{eq:Ytot} are comparable, while the variation in $\W$ is larger than the variation in $\Zg$ if one considers the range of conditions under which most of the stars in the universe have formed, the primary environmental factor affecting star formation is $\W$.

\begin{figure*}
  \centering
  \includegraphics[width=0.48\linewidth]{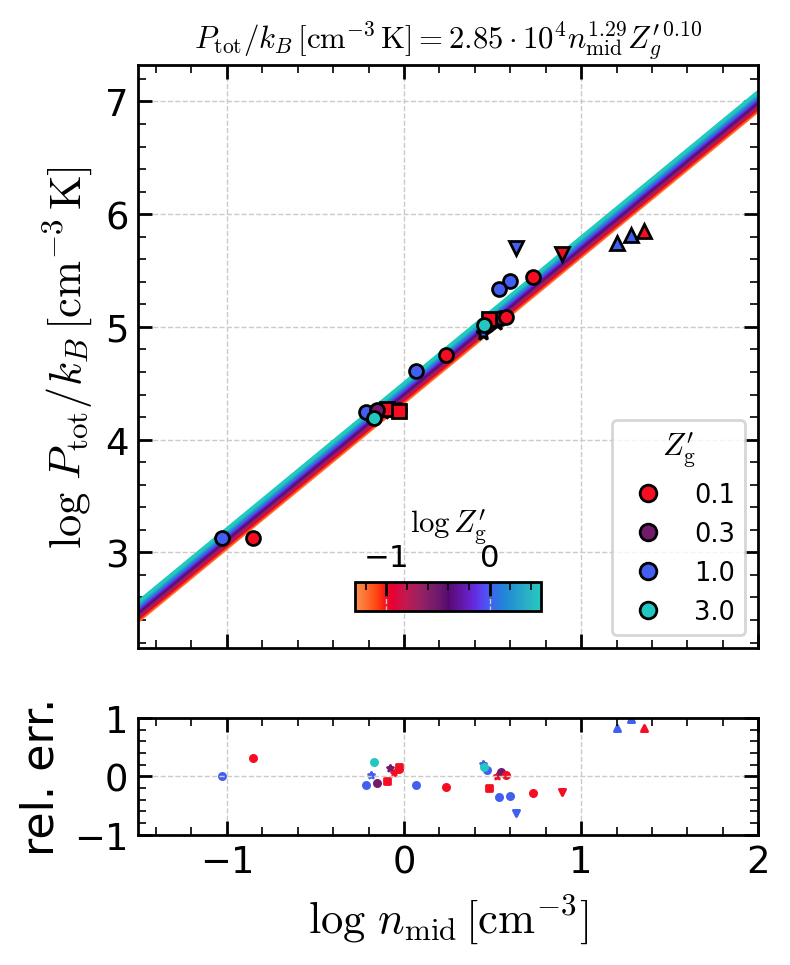}
  \includegraphics[width=0.48\linewidth]{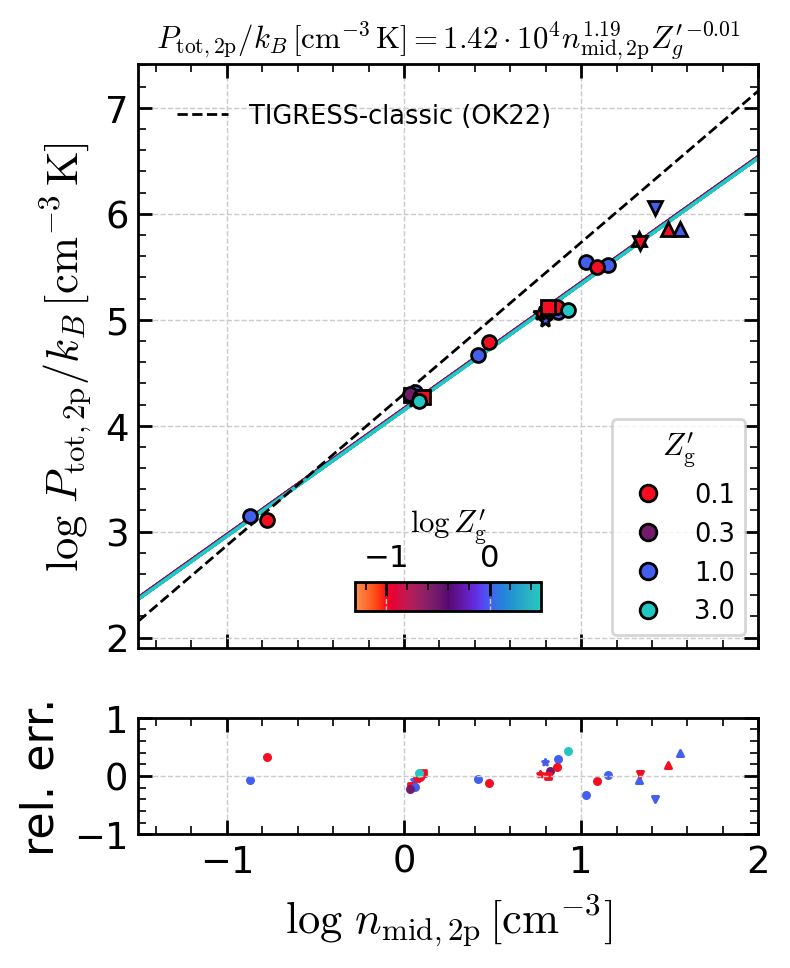}
  \caption{
    Calibration of the effective equation of state;
    i.e., the relation between total pressure and gas number density averaged over all gas (\emph{left}) and two phase gas (\emph{right}).
    A slightly shallower relation $\Ptot\propto\nmidplane^{1.29}$ with a weak metallicity dependence is found in the NCR suite compared to that of the TIGRESS-classic ($\Ptot\propto\nmidplane^{1.43}$, black dashed in right panel; \citealt{2022ApJ...936..137O}).
    Symbols and lines have the same meaning as in \autoref{fig:yield}.
  }\label{fig:eEoS}
\end{figure*}

\begin{figure*}
    \centering
    \includegraphics[width=0.48\linewidth]{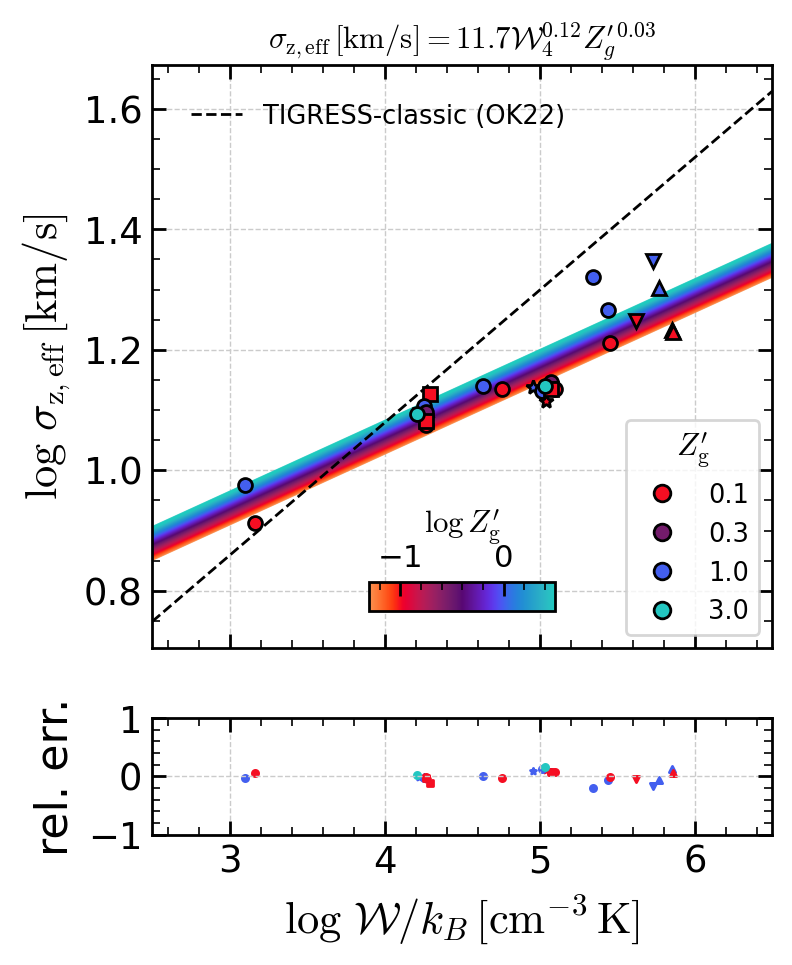}
    \includegraphics[width=0.48\linewidth]{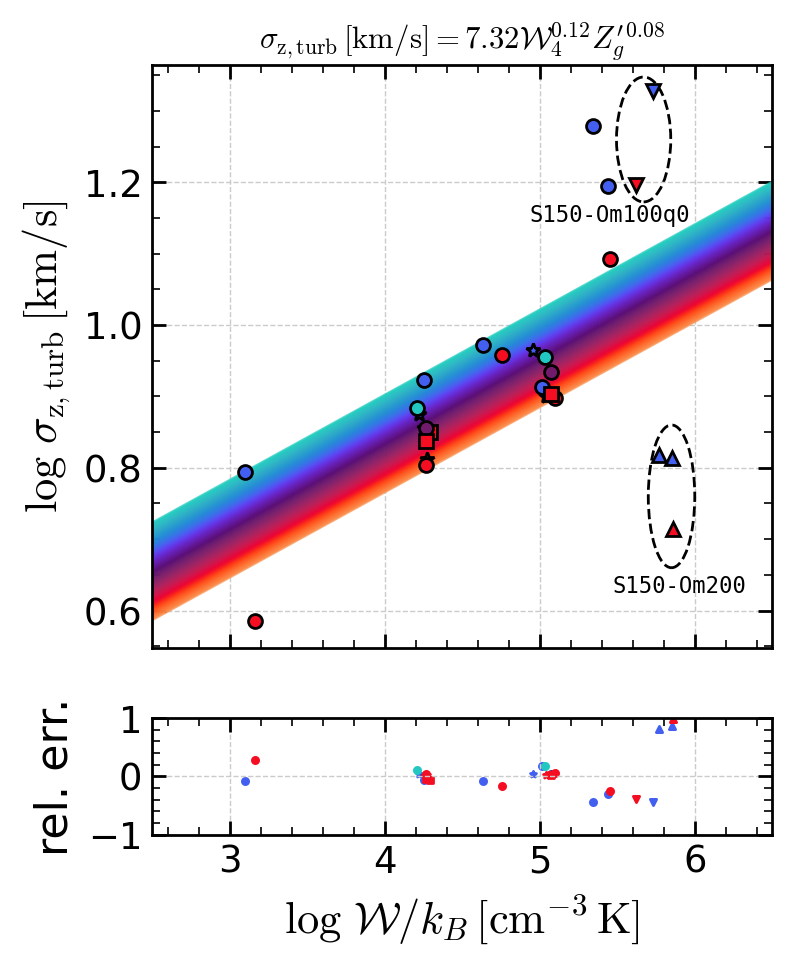}
    \caption{Mass-weighted mean vertical velocity dispersions of the \twop\ phase as defined by \autoref{eq:sigma_avg} for the total (left) and turbulent (right) pressure.
    On the left, the fit to the TIGRESS-classic suite presented in \citet{2022ApJ...936..137O} is shown as the black dashed line.
    Symbols and lines have the same meaning as in \autoref{fig:yield}.
    }
    \label{fig:veld}
\end{figure*}

\subsection{Effective equation of state}\label{sec:eEoS}

\autoref{fig:eEoS} presents the mean relation between total midplane pressure and gas density of all gas (left) and just the \twop{} gas (right).
The former relation represents an effective equation of state in a volume-averaged sense, encapsulating the system-level behavior of the multiphase star-forming ISM gas that develops as a consequence of the thermodynamic and MHD responses to intermittent energy injection by feedback.
We find that the effective equation of state is not sensitive to changes in metallicity.
The best fit bi-variate power law relation gives
\begin{equation}\label{eq:eos}
  \Ptot/k_B = 2.85\times10^4\Punit
  \rbrackets{\frac{\nmidplane}{\pcc}}^{1.29} \Zg^{-0.10}.
\end{equation}
From the TIGRESS-classic simulations, we previously found a slightly steeper (1.43) slope when considering just the \twop{} gas; this comparison is shown in the right panel.

It should be borne in mind that the above effective equation of state represents a numerical average over large space and time scales, which physically corresponds to multiple cycles of star formation and feedback.
Within a given local cycle, a smaller-scale region in a disk will experience first an increase in density and pressure (from compression driven by turbulence and gravity), and subsequently, an increase in specific energy  (due to feedback) that leads to a drop in density.
The effective equation of state represents an average over the complex dynamic and thermodynamic cycles that take place within the ISM under different galactic environmental conditions.
By relating the quasi-equilibrium midplane density and pressure values, the effective equation of state differs from an adiabatic pressure-density relation (applicable over time for a given fluid element) or polytropic pressure-density relation (applicable throughout space for a given equilibrium structure).
The effective equation of state is more analogous to a polytropic relation, in that it can be used to compute the equilibrium thickness of the disk given its surface density, similar to the mass-radius relation for stellar polytropes.

The right panel of \autoref{fig:eEoS}
can be translated into an effective midplane vertical velocity dispersion of the \twop{} gas,
\begin{eqnarray}\label{eq:sigma_eff_mid}
  \szmid &=& \rbrackets{\frac{\Ptottwo}{\rho_{\rm 2p}}}^{1/2} \nonumber\\
  &=&  8.9\kms \rbrackets{\frac{P_{\rm tot,2p}/k_B}{10^4\Punit}}^{0.08}\Zg^{-0.005}.
\end{eqnarray}
The midplane effective velocity dispersion can be compared with the mass-weighted mean considering the warm and cold gas over the whole vertical domain, shown in the left panel of \autoref{fig:veld}, which has a power-law fit
\begin{equation}\label{eq:sigma_eff_avg}
  \szeff = 11.7 \kms\W_4^{0.12}\Zg^{0.03}.
\end{equation}
The mass-weighted mean is $\sim 30\%$ higher than the midplane value.
By comparison, the effective mass-weighted velocity dispersion reported from TIGRESS-classic simulations in \citet{2022ApJ...936..137O} is comparable to \autoref{eq:sigma_eff_avg} for Solar neighborhood conditions, but increases faster in high-pressure environments ($\propto {\W}^{0.22}$).
The lower effective velocity dispersion for TIGRESS-NCR in the high-weight regime is mainly because SN-driven turbulence ends up being reduced when ionizing radiation (early feedback) is included.

The right panel of \autoref{fig:veld} shows the turbulent component of the mass-weighted mean velocity dispersion.
Similar to the total, the turbulent component also scales weakly with weight and metallicity.
In the high-pressure regime, there is a wide range of turbulent velocity dispersions, similar to the divergence among models of the turbulent feedback yield due to greatly enhanced magnetic support in rapidly rotating models.

\subsection{Application to subgrid modeling of the SFR}
\label{sec:subgrid}

The results presented in this section can be incorporated into a new subgrid star formation prescription based on the PRFM theory, for implementation in large-scale galaxy formation models where the ISM, star formation, and feedback are unresolved.
For this application,
S. Hassan et al (2024, submitted) presents detailed procedures for computing the required quantities from simulation variables, including the calculation of the equilibrium weight $\W$ considering the contribution from gaseous and stellar disks as well as a dark matter halo.
While the simple weight estimate in \autoref{eq:PDE} is applicable for normal star-forming galaxies where the gas disk is thinner than the stellar disk, the more general weight estimator in S. Hassan et al (2024, submitted) is applicable for any thickness ratio between gaseous and stellar disks.

Once the weight $\W$ is determined, the total feedback yield can be used to obtain the SFR.
Within the parameter space explored in this paper, either the single fit to the total feedback yield (\autoref{eq:Ytot}) or the sum of fits for thermal (\autoref{eq:Yth}) and turbulent+magnetic (\autoref{eq:Ynonth}) components can be used to obtain $\Ytot$.
While not recommended, if extrapolation beyond the parameter space is unavoidable, using the latter ($\Ytot=\Yth + \Ynonth$) is safer as the expected behavior would be captured in low and high pressure regimes where thermal and turbulent+magnetic support respectively dominates.

For a disk galaxy, we can express $\Ssfr$ in terms of star formation efficiency per dynamical time, $\epsilon_{\rm dyn}$:
\begin{equation}\label{eq:sfr_edyn}
    \Ssfr \equiv \epsilon_{\rm dyn} \frac{\Sgas}{t_{\rm dyn}}.
\end{equation}
In the above,
\begin{equation}\label{eq:tdyn}
    t_{\rm dyn}\equiv\frac{2H}{\szeff}=\frac{2\szeff}{\langle g_z\rangle}
\end{equation}
is the vertical dynamical time scale, where we have used $\Sgas = 2\rho_{\rm mid}H$ and $\Ptot = \rho_{\rm mid}\szeff^2= \Sgas \langle g_z\rangle/2$.
\autoref{eq:sfr_edyn} is completely equivalent to \autoref{eq:tdep}.
Since these expressions relate to gas mass and SFR, they are equally applicable for any mass element that averages spatially over the multiphase ISM and temporally over several cycles of star formation and feedback.
We could thus write, for a large-scale, unresolved mass element $m_\mathrm{gas}$, a mean SFR
\begin{equation}\label{eq:SFR_masselement}
    \dot{m}_\star= \frac{m_\mathrm{gas}}{\tdep} = \epsilon_{\rm dyn} \frac{m_\mathrm{gas}}{t_{\rm dyn}} .
\end{equation}

Using \autoref{eq:Pandg} and \autoref{eq:yield}, the gas depletion time can be expressed in terms of vertical gravity and feedback yield, or alternatively using \autoref{eq:tdyn} in terms of the dynamical time:
\begin{equation}\label{eq:tdepUps}
    \tdep = \frac{2 \Ytot}{\langle g_z\rangle}=\frac{\Ytot}{\szeff} \tdyn.
\end{equation}
\autoref{sec:vert_equil} shows that the vertical gravity is nearly independent of metallicity, and this can also be seen explicitly from \autoref{eq:gz} and \autoref{eq:sigma_eff_avg}.
Because $\langle g_z\rangle$ is essentially independent of metallicity, the first equality in \autoref{eq:tdepUps} implies that metallicity is expected to affect the timescale for converting gas to stars only through the feedback yield, $\Ytot$.
The quantities $\langle g_z\rangle$ and $\tdyn$ can be easily estimated from properties of galaxies (in simulations or observations) that are measured on $\sim \kpc$ scales;  \autoref{eq:gz} gives $\langle g_z\rangle$ for the case where the gas disk is thinner than the stellar disk, and S. Hassan et al (2024, submitted) present more general algebraic formulae.

From our fits for $\Ytot$ in \autoref{eq:Ytot} and $\szeff$ in \autoref{eq:sigma_eff_avg}, we can obtain an expression for the dependence of star formation efficiency on $\W$ and $\Zg$:
\begin{equation}\label{eq:edyn}
    \epsilon_{\rm dyn} \equiv \frac{\tdyn}{\tdep}=\frac{\szeff}{\Ytot}= 0.0071\W_4^{0.41}\Zg^{0.30}.
\end{equation}
We note that in this equation, an alternative to using \autoref{eq:Ytot} is to use  $\Ytot = \Yth + \Ynonth$ with the calibrations given in \autoref{eq:Yth} and \autoref{eq:Ynonth}.
From the TIGRESS-classic results in \citet{2022ApJ...936..137O}, we obtain a similar scaling with $\W$ and slightly larger normalization,  $\epsilon_{\rm dyn}=0.012\W_4^{0.43}$.

In systems where gas gravity dominates (e.g., self-gravitating clouds), a similar expression to \autoref{eq:SFR_masselement} is often used to characterize star formation in terms of an efficiency per gas free fall time \citep[e.g.,][]{2007ApJ...654..304K,2014prpl.conf...77P,2022ApJ...929L..18E}, with $ \epsilon_{\rm dyn}\rightarrow \epsilon_{\rm ff}$ and $t_{\rm dyn} \rightarrow t_{\rm ff} =[3\pi/(32 G\bar{\rho}_{\rm gas})]^{1/2}$.
The free-fall time-based star formation recipe (often called a Schmidt-type star formation recipe; \citealt{1959ApJ...129..243S}) has been widely used in galaxy formation simulations where the multiphase ISM is not resolved \citep[e.g.,][for reviews]{2015ARA&A..53...51S,2017ARA&A..55...59N}.
However, adopting a prescription based solely on the gaseous free-fall time on galactic scales, where gravity from the stellar disk (and sometimes dark matter) is of equal or greater importance, is not well justified.

The advantages of our new subgrid model approach compared to currently-used Schmidt-type recipes are thus twofold.
(1) The dynamical time estimate (as opposed to gas free-fall time) takes into account the gravitational contributions from both gas and external stellar and dark matter gravity.
(2) The efficiency per dynamical time is taken to depend on the galactic environment, via the ISM weight and metallicity, rather than being a fixed constant.
In addition, a major advance of this paper \citep[along with][]{2022ApJ...936..137O} is that the calibrations of $\Ytot$ and $\szeff$ are based on holistic simulations of the ISM with star formation and feedback, with accurate treatments of radiative transfer and photochemistry.
This allows us to obtain a physics-based (as opposed to empirical) star formation prescription.
Finally, the choice to calibrate  $\epsilon_{\rm dyn}$ as a function of $\W$ and $\Zg$ makes it possible to apply this subgrid prescription even in large-box galaxy formation simulations where the vertical structure of the disk is unresolved.
In this situation, $\bar\rho_{\rm gas}$ would be lower than it \edit1{would be at higher resolution,} making $t_{\rm ff}$ larger than it should be.
\edit1{The underestimation of density at low resolution is due to the limit on density set by the smoothing scale and mass resolution, which may be lower than the true density should be for a given surface density and gravitational potential.}
However, it is still possible to obtain a robust estimate of $\W$ as well as $\tdyn$ from available simulation variables (see S. Hassan et al 2024, submitted).

\section{Discussion}\label{sec:discussion}

\subsection{When does thermal pressure become important?}

One of the main results of this paper is the metallicity dependence of SFRs (\autoref{fig:sfr_history} and \autoref{fig:sfr_tdep_W}).
We attribute the reduced SFRs at low metallicities to the enhancement of the thermal feedback yield under conditions of reduced FUV attenuation (\autoref{sec:yield}).
Because radiation is much less attenuated in low surface density conditions, $\Yth$  also increases rapidly with decreasing weight (which depends non-linearly on surface density).
The bi-variate fit in \autoref{eq:Yth} is $\Yth \propto \W^{-0.46}\Zg^{-0.53}$.
This means that at lower metallicities and gas surface densities, we expect more efficient thermal pressure regulation by UV radiative feedback.
The turbulent feedback yield $\Yturb$ depends on the specific momentum injected by supernovae (or mechanical feedback in general), which decreases as cooling increases.
Since supernova remnants cool when the shock velocity is slightly higher under conditions of higher ambient density and metallicity \citep{2015ApJ...802...99K,2023ApJS..264...10K}, the turbulent yield follows similar trends to the thermal yield, but with shallower dependencies; combining turbulent with magnetic yields we find $\Ynonth\propto  \W^{-0.22}\Zg^{-0.18}$.
Because of the different yield scalings, thermal pressure would begin to dominate over nonthermal pressure roughly when ${\W}\Zg$ falls below $100 \Punit$; these combined low pressure and low metallicity conditions may be found in dwarfs or the outer parts of spirals.

For the total yield, \autoref{eq:Ytot} gives a dependence on weight and metallicity $\Ytot\propto \W^{-0.29}\Zg^{-0.27}$, which compromises between the steeper thermal and shallower nonthermal dependencies.
From \autoref{eq:tdepUps}, the increase in total feedback yield at low metallicity would imply a corresponding increase in the gas depletion time $\tdep \propto \Ytot$, for a given galactic gravitational field.
\autoref{fig:sfr_tdep_W} shows our results for the dependence of $\Ytot$ on $\W$ and $Z$ for the full simulation suite.
Overall, the enhancement in total feedback yield at $Z'=0.1$ compared to $Z' = 1$ is about a factor 2--2.5.
Given the relatively small predicted variation in $\Ssfr$ or $\tdep$ for varying abundances, and potentially large uncertainties in observational measurements, quantitative confirmation of these predictions may require large ensemble averages of regions sharing similar conditions.

While comprehensive observational comparisons are needed, there is already some observational evidence aligned with the predictions of our metallicity-dependent models.
\citet{2017ApJ...835..201H} have measured thermal pressure in the CNM for nearby galaxies using \emph{Herchel} [\ion{C}{2}] 158$\mu$m observations combined with \ion{H}{1} and CO data.
The KINGFISH sample they analyzed provides a large dataset for $>500$ atomic-dominated regions with typical sizes of $\sim 1\kpc^2$, showing a statistically significant correlation between thermal pressure and $\Ssfr$ (see their Figure 5).
Using Oxygen abundance measurements from \citet{2010ApJS..190..233M}, \citet{2017ApJ...835..201H} reported a systematic trend of increasing thermal pressure at lower metallicities for a given $\Ssfr$.
This is consistent with higher thermal feedback yield at lower metallicities at a given weight predicted in our results (\autoref{eq:Yth}).

Our results have important implications for galaxy formation modeling, where it is difficult to follow the thermal structure of the ISM via explicit treatments of cooling and heating processes.
Given the large dynamic range that must be covered in global or cosmological galaxy simulations, it has been challenging to properly capture the thermal state of the ISM.
Since turbulent pressure dominates over thermal pressure in the Milky Way and other nearby galaxies, an implicit assumption has often been that exact recovery of the thermal pressure is not required for accurately predicting the regulation of star formation, with more emphasis instead placed on the momentum injection from SNe that drives turbulent pressure \citep[e.g.,][]{2023MNRAS.519.3154H}.
However, because thermal pressure is expected to exceed turbulent pressure in low-metallicity, low-pressure environments, care must be taken to ensure that simulations of these environments include the necessary physical ingredients for realistic heating and cooling.

A subtlety that is often overlooked is that PE heating depends on the electron abundance through the grain charging parameter.
It is often assumed in galactic and cosmological simulations that the only source of ionization is UV radiation (either attenuated metagalactic or locally-produced UV).
In the neutral ISM, FUV is able to ionize weakly-bound electrons to make C$^+$, but the primary source of ionization is believed to be H-ionization by low-energy CRs.
When CR ionization is not included, the electron fraction is unphysically low for the neutral ISM, leading to  PE heating rates that are too low \citep[see][Section 8.2]{2023ApJS..264...10K}.
This results in a much higher CNM fraction and lower thermal pressure than is realistic.
In some simulations, mechanical feedback is boosted by multiplying the terminal radial momentum per SN by a large factor (typically by a factor of 5) \citep{2013ApJ...770...25A,2018ApJ...861..107L,2017ApJ...845..133S}.
It is possible that this boosting is required to make up for missing thermal support (as well as magnetic support in purely hydrodynamical simulations).
While this enhancement of turbulent pressure may produce reasonable SFRs, a concern is that there may be other unintended consequences (e.g., for outflows).

\subsection{When do SFRs depend on metallicity?}

In the set of simulations presented here, with $\Zg\ge 0.1$, the dominant heating source in the \twop{} phase is from the PE effect, proportional to dust abundance.
At sufficiently low dust abundance, however, the PE heating will fall below CR heating as the dominant term \citep[see e.g. Fig. 6 in both of][]{2023ApJS..264...10K,2019ApJ...881..160B}, and at sufficiently low dust and metal abundance, the primary cooling in both warm and hot diffuse gas will be from hydrogen  \citep[see e.g. Fig. 3, 4, 13  and Fig. 1, 3 in ][respectively]{2023ApJS..264...10K,2019ApJ...881..160B}.
Thus, sensitivity to metallicity (for both thermal pressure and SN momentum injection) must eventually drop when abundances are low enough.
However, the exact point when this occurs depends on the details of the cooling and heating processes, some of which are still uncertain.

Two major pieces of modeling uncertainty affect quantitative predictions for the dominant heating mechanism.
First, the dust model adopted in the TIGRESS-NCR framework does not include any metallicity dependence.
In reality, metallicity-dependent changes in dust properties would be important for both radiation attenuation and modeling PE heating, which is dominated by small grains and PAHs.
There is empirical evidence of metallicity dependence of the PAH abundance, \edit1{dust-to-metal} ratio, and grain size distribution \citep[e.g.,][]{2007ApJ...663..866D,2012ApJ...744...20S,2014A&A...563A..31R,2019A&A...623A...5D,2022ApJ...928...90R,2023ApJ...944L..11C}.
All of these can cause metallicity dependence of attenuation of UV radiation and the PE heating efficiency \citep{1994ApJ...427..822B,2001ApJS..134..263W}.
It has been proposed, for example, that the reduction of the PAH abundance at low metallicity may not be gradual but a step-like function at $Z'\lesssim 0.2$ \citep[][but see \citealt{2020ApJ...889..150A}]{2007ApJ...663..866D}.
In that case, the overall metallicity dependence we find at $Z'>0.1$ may still hold, while CR heating would begin to dominate earlier at $Z'\sim0.2$.

Second, our current treatment of CR ionization rate is still provisional even though it is motivated by empirical relations and first-order physical arguments.
The CR transport problem in a multiphase, highly dynamic ISM is not fully understood.
Recent advances have been made by using two-moment evolution equations for CR transport in order to capture advection, diffusion, and (at high scattering rate) streaming limited by the Alfv\'en speed \citep[e.g.][]{2018ApJ...854....5J,2020MNRAS.491..993G,2022MNRAS.516.3470H}.
In most studies, the CR scattering rates are imposed as a function of energy, but some implementations \citep{2021ApJ...922...11A,2022MNRAS.517.5413H} have considered the self-confinement paradigm \citep[e.g.][]{2013PhPl...20e5501Z} that is believed to hold at $E\lesssim 100$GeV, in which scattering rates are set by a balance between streaming-driven wave excitation and damping.
Provided that the full multiphase structure and dynamics of the ISM are sufficiently resolved by the underlying MHD model (including a hot, high-velocity component that rapidly advects CRs out of the disk, a primarily-neutral component where CRs are highly diffusive, and a warm ionized component where transport is limited by Alfv\`enic streaming), realistic properties of GeV CRs are obtained \citep{2024ApJ...964...99A}.
An extension of this work to allow for multiple CR energy groups, coupled with the TIGRESS-NCR framework, will be able to shed light on the environmental dependence of both ISM heating from MeV CRs and ISM dynamics driven by GeV CRs.

In addition to uncertainties in the ISM physics, there are also uncertainties related to the initial mass function and population synthesis model, especially at low metallicities and high redshifts \citep[e.g.,][]{2013ARA&A..51..393C,2022ARA&A..60..455E}.
In this paper, we simply keep both stellar models unchanged from the standard choice we made for solar neighborhood conditions, i.e., STARBURST99 \citep{1999ApJS..123....3L,2014ApJS..212...14L} coupled with a Kroupa IMF \citep{2001MNRAS.322..231K} and the Geneva evolutionary tracks for non-rotating stars.
More realistically, at low metallicities UV radiation may be stronger for the same population of stars \citep[e.g.,][]{2021ApJ...908..241G}, shifting up the normalization for thermal feedback yield related to the ratio of FUV radiation to SFR, $\Sigma_{\rm FUV}/\Sigma_{\rm SFR}$.
Stellar rotation and binary evolution can also change the UV photon production rate as well as SN rate \citep[e.g.,][]{2012ApJ...751...67L,2022ARA&A..60..455E}.
There is also evidence that the IMF may have been more bottom-heavy in environments that produce very high stellar densities, based on observations of massive elliptical galaxies \citep[e.g.][]{2012ApJ...760...71C}.
Further numerical experiments that vary the model ingredients using TIGRESS-NCR or similar frameworks will be critical to understanding parameter sensitivity in observational diagnostics and underlying physical properties, such as SFRs and ISM properties, with the ultimate goal of constraining the uncertain model parameters.

\subsection{Comparison with 
previous numerical work}

\edit1{The most similar numerical simulations to those analyzed here were presented in \citet{2021ApJ...920...44H}. They} ran a set of simulations representing a kpc patch of solar neighborhood, \edit1{similar to our {\tt R8} series}, using the Lagrangian code {\tt GIZMO} \citep{2015MNRAS.450...53H}, and varying metallicities over the range $Z'=0.1-3$.
\edit1{Their simulations include time-dependent hydrogen chemistry, as well as resolved supernova feedback similar to TIGRESS-NCR. Taking advantage of high resolution in the dense gas obtained by the Lagrangian approach, they focus on the effect of metallicity in the atomic-to-molecular transition and the distribution of carbon-bearing species that are key observables (\ion{C}{2}, \ion{C}{1}, and CO).} 

\edit1{The long-term mean SFR  from the solar metallicity simulation of \citet{2021ApJ...920...44H} is similar to that in our $Z'=1$ TIGRESS-NCR model, which is also in agreement with observations and with our previous TIGRESS simulations with solar neighborhood conditions \citep{2017ApJ...846..133K}}.
\edit1{Although SFRs were not the main focus of their work, it is worth noting that unlike us,} \citet{2021ApJ...920...44H} did not find systematic metallicity dependence of mean SFRs.
Since they focused on chemical properties, it is difficult to pinpoint the reason.
While the gas surface density and gravitational potential were similar to our {\tt R8} models and their heating and cooling prescriptions were similar, 
\edit1{a potential key difference is in the treatment of the UV radiation field.}
Instead \edit1{of the direct UV radiation transfer approach of this paper}, at any time they first compute a uniform background UV radiation field by scaling proportional to $\Ssfr$, using the observed solar neighborhood radiation field and $\Ssfr$ to normalize.
They then apply local dust shielding based on column density, calculated using a tree method with a fixed shielding length of 100 pc.
Without direct radiative transfer on large scales, this approach did not capture the overall enhancement in the ratio of the FUV radiation field to the SFR \edit1{at low metallicity}.
In our simulations, the reduced large-scale attenuation of UV at low metallicity is what leads to higher thermal feedback yield and ultimately enables a lower SFR.

The difference in outcomes 
\edit1{suggests that
accounting for large-scale} UV radiation transfer 
\edit1{may be important} 
when aiming to quantitatively capture self-regulation of SFRs, especially for the regime where thermal pressure plays a major role.
If an expensive radiation transfer calculation is not feasible, an effective model calibrated based on more direct radiation transfer results like ours is an alternative approach.
As pointed out in \citet{2023ApJ...946....3K}, the simple plane-parallel approximation proposed in \citet[][see also \citealt{2022ApJ...936..137O}]{2010ApJ...721..975O} provides a reasonable model for large-scale attenuation of FUV radiation \citep[see also][]{2020ApJ...903...62B}.
More extensive analysis of radiation fields obtained from ray-tracing in the TIGRESS-NCR suite, and comparison with other approximate models, will be presented in N. Linzer et al. (submitted).

Although their mean SFRs did not depend on metallicity, \citet{2021ApJ...920...44H} observed an increase of burstiness in SFRs as metallicity decreases. We find a similar qualitative trend for both {\tt R8} and {\tt LGR4} (see \autoref{fig:sfr_history}).
The reason for the increase in SFR burstiness in our simulations is twofold: (1) faster quenching of star formation (due to heating of the cold phase) over a larger region at lower $Z$ as the radiation is less attenuated, and (2) slower recovery of cold gas due to longer cooling times from reduced metal cooling.

\subsection{Caveats and future perspectives}\label{sec:future}

We find that the mean metallicity increases by a factor of 1.5-3 over the course of our simulations.
However, we ignore this metal enrichment as we do not follow dust evolution explicitly, and we also do not include other aspects of global evolution (such as changing gas surface density from accretion).
\citet{2023ApJ...952..140H} included a model of dust formation and destruction in their simulations and found that dust growth is fast and reaches high dust abundance quickly in dense, star-forming gas.
We cover a range of dust abundance at low metallicity, but we did not consider spatially varying dust abundance.
In the future, it would be interesting to couple our TIGRESS-NCR framework to dust evolution models and investigate the \edit1{effect of locally varying metallicity, dust abundance, and dust-to-metal ratios.}

Finally, it is worth mentioning that the regime of high weight/pressure ($\W/k_B \gtrsim 10^{5.5} \pcc \Kel$) is still subject to some uncertainty, as evident in the results presented here as well as other on-going studies.
In particular, the saturation level of magnetic fields and turbulence properties in the high weight/pressure regime may have some sensitivity to simulation box size as it sets the largest scale of collapse and hence clustering of SNe.
In addition, \edit1{at low redshift,} high weight/pressure regimes are usually found within distinct galactic structures such as spiral arms \citep{2020ApJ...898...35K} and nuclear rings fed by bars \citep{2023ApJ...946..114M}.
To reach firm conclusions regarding the quantitative results for turbulent and magnetic feedback yields in the high weight/pressure regime and the effective equation of state, further investigation is warranted.
Recent efforts in coupling various photochemistry treatments with radiation transfer (usually based on the two-moment method with the M1 closure; e.g., \citealt{2020MNRAS.499.5732K,2020MNRAS.496.5160L,2022arXiv221104626K}) in cosmological and global isolated galaxy modeling will help shed light on exploring more extreme conditions.
Nevertheless, box size and global geometry would not alter the robustness of our conclusions regarding the qualitative behavior of feedback yields with metallicity and weight, or the insensitivity of vertical dynamical equilibrium and effective equation of state to metallicity.

\section{Summary}\label{sec:summary}

In this paper, we present the first results from a new suite of local galactic patch simulations with varying galactic conditions (weight and metallicity) using the TIGRESS-NCR framework \citep{2023ApJ...946....3K}.
The new simulation suite includes a total of 28 models covering a wide range of gas and stellar surface densities $\Sgas=5-150\Surf$ and $\Sstar=1-50\Surf$.
We vary gas metallicity and dust abundance up to 3 times solar values, metals down to 0.1 times solar, and dust down to 0.025 times solar.
After running for at least one orbit time to pass through an early transient stage, we run all models for more than 2 orbit times, corresponding to evolution periods ranging from 150 Myr to 1.5 Gyr (shorter duration for higher gas and SFR surface density conditions).
The emergent midplane total pressure from our suite covers a range $\Ptot/k_B=10^3-10^6\Punit$, while the range of SFR surface density is $\Ssfr=10^{-4}-0.5\sfrunit$.

The TIGRESS-NCR framework represents a significant advance from the original TIGRESS framework \citep[][referred to as ``TIGRESS-classic'']{2017ApJ...846..133K}.
The key improvements over TIGRESS-classic are explicit UV radiation transfer using ART \citep{2017ApJ...851...93K} from star cluster particles for both non-ionizing and ionizing UV, and a photochemistry module which enables realistic ISM cooling and heating over a range of metal and dust abundances \citep{2023ApJS..264...10K}.
TIGRESS-classic adopted much simpler temperature-dependent cooling and spatially constant (but SFR-dependent) heating, and was only applicable for solar metallicity.
Other major dynamical and feedback processes -- including MHD, galactic sheared rotation, and self-gravity, as well as star cluster formation in gravitational collapse, and resolved supernova feedback -- are the same in TIGRESS-classic and TIGRESS-NCR.

The primary focus of this paper is to investigate the metallicity dependence of SFRs in the context of the PRFM star formation theory \citep[e.g.,][]{2022ApJ...936..137O}.
A key motivation is to calibrate subgrid star formation models that may be implemented in cosmological galaxy formation simulations to follow evolution over a range of redshifts.

Our main findings are as follows:
\begin{itemize}
  \item When other conditions are fixed, SFRs drop at lower metallicities (\autoref{sec:sfr}). From $Z'=1$ to $Z'=0.1$, we find that the mean $\Ssfr$ is reduced by a factor of 2-3.
  In terms of gas depletion time $\tdep=\Sgas/\Ssfr$, we find a comparable or slightly larger enhancement in mean $\tdep$ at lower metallicities.
  This is because reduced $\Ssfr$ in the early evolution results in slightly higher $\Sgas$ in the later evolution.
  At $\Zg=0.1$, a reduction in dust abundance from $\Zd=0.1$ to $\Zd=0.025$ does not cause a further reduction in $\Ssfr$.

  \item In all simulations, vertical dynamical equilibrium is satisfied (\autoref{sec:vert_equil}).
  This means that at the disk midplane, the total pressure $\Ptot$ matches the weight per unit area $\W$; vertical profiles of pressure and weight also match each other within $|z|\lesssim 500 \pc$.
  The simple dynamical equilibrium estimator (\autoref{eq:PDE}) is in good agreement with the actual weight irrespective of metallicity.
  Since the weight is determined mainly by $\Sgas$ and $\Sstar$  and is insensitive to metallicity, the total midplane pressure is also insensitive to metallicity.
  However, the relative contributions to $\Ptot$ of the thermal pressure ($\Pth$), the vertical component of Reynolds stress ($\Pturb = \rho v_z^2$), and the vertical component of Maxwell stress ($\Pimag = B^2/8\pi - B_z^2/4\pi$)  vary with metallicity.
  The increase of $\Pth/\Ptot$ at low metallicity (by more than a factor of two from $\Zg=3$ to $\Zg=0.1$ in the {\tt R8} series) implies a corresponding reduction in the other fractional pressure contributions.
  The \hot{}, \WIM, and \twop{} phases are in total pressure equilibrium within the scale height of the \twop{} phase.

  \item We provide a bi-variate power-law fit for each feedback yield component and the total feedback yield, as functions of weight and metallicity (\autoref{sec:yield}).
  We calculate the mean values of the stresses and SFR surface density over an extended interval during the later evolution covering multiple star-formation and feedback cycles.
  We then measure the ratios of each stress component to the SFR surface density, $\Ssfr$.
  These ratios are termed  \emph{feedback yields} in the PRFM theory, denoted $\Ytot = \Ptot/\Ssfr$, $\Yturb = \Pturb/\Ssfr$, $\Yth = \Pth/\Ssfr$, and $\Ymag = \Pimag/\Ssfr$.
  We confirm, consistent with \citet{2022ApJ...936..137O}, a decreasing trend in all feedback yields as the weight increases.
  We find a clear increasing trend in $\Yth$  with decreasing metallicity down to $\Zg=0.1$.
  There is also a strong increase in $\Yth$ with decreasing weight because environments with low pressure ($\Ptot \approx {\W}$) have low density such that radiation is only weakly attenuated and efficiently heats the gas.
  There is almost no metallicity dependence in $\Yturb$.
  While the individual behavior of turbulent and magnetic feedback yields becomes irregular at high weights ($\W/k_B>10^{5.5}\Punit$), the combined non-thermal (turbulent+magnetic) feedback yield $\Ynonth$ is better described by a power-law with overall reduced scatter.
  $\Ynonth$ depends less sensitively on both weight and metallicity, compared to $\Yth$.

  \item We provide a new calibration for the effective equation of state for star-forming multiphase gas (\autoref{sec:eEoS}).
  We obtain a shallower exponent than TIGRESS-classic (1.29 vs 1.43 from \citet{2022ApJ...936..137O}) for the total pressure and density relation with no metallicity dependence.
  We also provide a fit for the effective vertical velocity dispersion of the warm-cold \twop{} medium, which increases weakly at higher pressure but is insensitive to metallicity.

  \item We describe how our calibrations of feedback yield and effective equation of state can be used in a new subgrid star formation model based on the PRFM theory (\autoref{sec:subgrid}). This model expresses the SFR in terms of the dynamical time $\tdyn$ and an efficiency factor $\epsilon_{\rm dyn}$ (\autoref{eq:SFR_masselement}).
  We provide a fit for $\epsilon_{\rm dyn}$ based on our calibrations for $\Ytot$ and $\szeff$ in terms of weight $\W$ and metallicity (\autoref{eq:edyn}).
  The advantages of this new model are that it is grounded in explicit simulations of the ISM with accurate physics treatments and widely ranging galactic environments and that it takes into account the gravity of stars and dark matter as well as gas.
  By using weight rather than density as the input parameter, our SFR model is designed to be robustly applicable in large-box galaxy formation simulations in which gas disk scale heights are unresolved (and hence mean density is resolution dependent).
\end{itemize}

The main focus of this work was on studying the variation of SFRs under a wide range of galactic conditions, covering environments observed in both massive galaxies and dwarfs, and in the low- and high-redshift universe.
Here we employed the ``feedback yield'' to quantitatively characterize the equilibrium multiphase ISM pressure response to energy inputs from recently formed stars.
An equally interesting numerical problem, and equally important for developing new cosmological subgrid models, is to quantify how galactic multiphase outflows are produced due to star formation feedback.
Following \citet{2020ApJ...900...61K,2020ApJ...903L..34K}, galactic outflows can be characterized in terms of overall loading factors as well as distributions of outflow speed and sound speed.
The present simulation suite, and additional TIGRESS-NCR simulations with augmented physics and parameter coverage, will enable extensions of previous wind-driving analyses to cover a range of metallicities.

\software{{\tt Athena} \citep{2008ApJS..178..137S,2009NewA...14..139S},
{\tt astropy} \citep{astropy:2013,astropy:2018,astropy:2022},
{\tt scipy} \citep{2020SciPy-NMeth},
{\tt numpy} \citep{vanderWalt2011},
{\tt IPython} \citep{Perez2007},
{\tt matplotlib} \citep{Hunter:2007},
{\tt xarray} \citep{hoyer2017xarray},
{\tt pandas} \citep{mckinney-proc-scipy-2010},
{\tt CMasher} \citep{CMasher},
{\tt adstex} (\url{https://github.com/yymao/adstex})
}

\acknowledgments
{We acknowledge the anonymous referee for a useful report. We are grateful to Amiel Sternberg for constructive comments on the manuscript.}
This work was supported by grant 10013948 from the Simons Foundation to Princeton University, sponsoring the Learning the Universe collaboration.
The work of C.-G.K. was partly supported by NASA ATP grant No. 80NSSC22K0717. J.-G.K. acknowledges support from the EACOA Fellowship awarded by the East Asia Core Observatories Association. GLB acknowledges support from the NSF (AST-2108470 and AST-2307419, ACCESS), a NASA TCAN award, and the Simons Foundation. SH acknowledges support for Program number HST-HF2-51507 provided by NASA through a grant from the Space Telescope Science Institute, which is operated by the Association of Universities for Research in Astronomy, incorporated, under NASA contract NAS5-26555. The Center for Computational Astrophysics at the Flatiron Institute is supported by the Simons Foundation.
Resources supporting this work were provided in part by the NASA High-End Computing (HEC) Program through the NASA Advanced Supercomputing (NAS) Division at Ames Research Center and in part by the Princeton Institute for Computational Science and Engineering (PICSciE) and the Office of Information Technology's High Performance Computing Center.
This research has made use of NASA's Astrophysics Data System.

\bibliography{ref,software}

\begin{thebibliography}{}
\expandafter\ifx\csname natexlab\endcsname\relax\def\natexlab#1{#1}\fi
\providecommand{\url}[1]{\href{#1}{#1}}
\providecommand{\dodoi}[1]{doi:~\href{http://doi.org/#1}{\nolinkurl{#1}}}
\providecommand{\doeprint}[1]{\href{http://ascl.net/#1}{\nolinkurl{http://ascl.net/#1}}}
\providecommand{\doarXiv}[1]{\href{https://arxiv.org/abs/#1}{\nolinkurl{https://arxiv.org/abs/#1}}}

\bibitem[{{Agertz} {et~al.}(2013){Agertz}, {Kravtsov}, {Leitner}, \&
  {Gnedin}}]{2013ApJ...770...25A}
{Agertz}, O., {Kravtsov}, A.~V., {Leitner}, S.~N., \& {Gnedin}, N.~Y. 2013,
  \apj, 770, 25, \dodoi{10.1088/0004-637X/770/1/25}

\bibitem[{{Aniano} {et~al.}(2020){Aniano}, {Draine}, {Hunt}, {Sandstrom},
  {Calzetti}, {Kennicutt}, {Dale}, {Galametz}, {Gordon}, {Leroy}, {Smith},
  {Roussel}, {Sauvage}, {Walter}, {Armus}, {Bolatto}, {Boquien}, {Crocker}, {De
  Looze}, {Donovan Meyer}, {Helou}, {Hinz}, {Johnson}, {Koda}, {Miller},
  {Montiel}, {Murphy}, {Rela{\~n}o}, {Rix}, {Schinnerer}, {Skibba}, {Wolfire},
  \& {Engelbracht}}]{2020ApJ...889..150A}
{Aniano}, G., {Draine}, B.~T., {Hunt}, L.~K., {et~al.} 2020, \apj, 889, 150,
  \dodoi{10.3847/1538-4357/ab5fdb}

\bibitem[{{Armillotta} {et~al.}(2021){Armillotta}, {Ostriker}, \&
  {Jiang}}]{2021ApJ...922...11A}
{Armillotta}, L., {Ostriker}, E.~C., \& {Jiang}, Y.-F. 2021, \apj, 922, 11,
  \dodoi{10.3847/1538-4357/ac1db2}

\bibitem[{{Armillotta} {et~al.}(2024){Armillotta}, {Ostriker}, {Kim}, \&
  {Jiang}}]{2024ApJ...964...99A}
{Armillotta}, L., {Ostriker}, E.~C., {Kim}, C.-G., \& {Jiang}, Y.-F. 2024,
  \apj, 964, 99, \dodoi{10.3847/1538-4357/ad1e5c}

\bibitem[{{Asplund} {et~al.}(2009){Asplund}, {Grevesse}, {Sauval}, \&
  {Scott}}]{2009ARA&A..47..481A}
{Asplund}, M., {Grevesse}, N., {Sauval}, A.~J., \& {Scott}, P. 2009, \araa, 47,
  481, \dodoi{10.1146/annurev.astro.46.060407.145222}

\bibitem[{{Astropy Collaboration} {et~al.}(2013){Astropy Collaboration},
  {Robitaille}, {Tollerud}, {Greenfield}, {Droettboom}, {Bray}, {Aldcroft},
  {Davis}, {Ginsburg}, {Price-Whelan}, {Kerzendorf}, {Conley}, {Crighton},
  {Barbary}, {Muna}, {Ferguson}, {Grollier}, {Parikh}, {Nair}, {Unther},
  {Deil}, {Woillez}, {Conseil}, {Kramer}, {Turner}, {Singer}, {Fox}, {Weaver},
  {Zabalza}, {Edwards}, {Azalee Bostroem}, {Burke}, {Casey}, {Crawford},
  {Dencheva}, {Ely}, {Jenness}, {Labrie}, {Lim}, {Pierfederici}, {Pontzen},
  {Ptak}, {Refsdal}, {Servillat}, \& {Streicher}}]{astropy:2013}
{Astropy Collaboration}, {Robitaille}, T.~P., {Tollerud}, E.~J., {et~al.} 2013,
  \aap, 558, A33, \dodoi{10.1051/0004-6361/201322068}

\bibitem[{{Astropy Collaboration} {et~al.}(2018){Astropy Collaboration},
  {Price-Whelan}, {Sip{\H{o}}cz}, {G{\"u}nther}, {Lim}, {Crawford}, {Conseil},
  {Shupe}, {Craig}, {Dencheva}, {Ginsburg}, {Vand erPlas}, {Bradley},
  {P{\'e}rez-Su{\'a}rez}, {de Val-Borro}, {Aldcroft}, {Cruz}, {Robitaille},
  {Tollerud}, {Ardelean}, {Babej}, {Bach}, {Bachetti}, {Bakanov}, {Bamford},
  {Barentsen}, {Barmby}, {Baumbach}, {Berry}, {Biscani}, {Boquien}, {Bostroem},
  {Bouma}, {Brammer}, {Bray}, {Breytenbach}, {Buddelmeijer}, {Burke},
  {Calderone}, {Cano Rodr{\'\i}guez}, {Cara}, {Cardoso}, {Cheedella}, {Copin},
  {Corrales}, {Crichton}, {D'Avella}, {Deil}, {Depagne}, {Dietrich}, {Donath},
  {Droettboom}, {Earl}, {Erben}, {Fabbro}, {Ferreira}, {Finethy}, {Fox},
  {Garrison}, {Gibbons}, {Goldstein}, {Gommers}, {Greco}, {Greenfield},
  {Groener}, {Grollier}, {Hagen}, {Hirst}, {Homeier}, {Horton}, {Hosseinzadeh},
  {Hu}, {Hunkeler}, {Ivezi{\'c}}, {Jain}, {Jenness}, {Kanarek}, {Kendrew},
  {Kern}, {Kerzendorf}, {Khvalko}, {King}, {Kirkby}, {Kulkarni}, {Kumar},
  {Lee}, {Lenz}, {Littlefair}, {Ma}, {Macleod}, {Mastropietro}, {McCully},
  {Montagnac}, {Morris}, {Mueller}, {Mumford}, {Muna}, {Murphy}, {Nelson},
  {Nguyen}, {Ninan}, {N{\"o}the}, {Ogaz}, {Oh}, {Parejko}, {Parley}, {Pascual},
  {Patil}, {Patil}, {Plunkett}, {Prochaska}, {Rastogi}, {Reddy Janga},
  {Sabater}, {Sakurikar}, {Seifert}, {Sherbert}, {Sherwood-Taylor}, {Shih},
  {Sick}, {Silbiger}, {Singanamalla}, {Singer}, {Sladen}, {Sooley},
  {Sornarajah}, {Streicher}, {Teuben}, {Thomas}, {Tremblay}, {Turner},
  {Terr{\'o}n}, {van Kerkwijk}, {de la Vega}, {Watkins}, {Weaver}, {Whitmore},
  {Woillez}, {Zabalza}, \& {Astropy Contributors}}]{astropy:2018}
{Astropy Collaboration}, {Price-Whelan}, A.~M., {Sip{\H{o}}cz}, B.~M., {et~al.}
  2018, \aj, 156, 123, \dodoi{10.3847/1538-3881/aabc4f}

\bibitem[{{Astropy Collaboration} {et~al.}(2022){Astropy Collaboration},
  {Price-Whelan}, {Lim}, {Earl}, {Starkman}, {Bradley}, {Shupe}, {Patil},
  {Corrales}, {Brasseur}, {N{"o}the}, {Donath}, {Tollerud}, {Morris},
  {Ginsburg}, {Vaher}, {Weaver}, {Tocknell}, {Jamieson}, {van Kerkwijk},
  {Robitaille}, {Merry}, {Bachetti}, {G{"u}nther}, {Aldcroft},
  {Alvarado-Montes}, {Archibald}, {B{'o}di}, {Bapat}, {Barentsen}, {Baz{'a}n},
  {Biswas}, {Boquien}, {Burke}, {Cara}, {Cara}, {Conroy}, {Conseil}, {Craig},
  {Cross}, {Cruz}, {D'Eugenio}, {Dencheva}, {Devillepoix}, {Dietrich},
  {Eigenbrot}, {Erben}, {Ferreira}, {Foreman-Mackey}, {Fox}, {Freij}, {Garg},
  {Geda}, {Glattly}, {Gondhalekar}, {Gordon}, {Grant}, {Greenfield}, {Groener},
  {Guest}, {Gurovich}, {Handberg}, {Hart}, {Hatfield-Dodds}, {Homeier},
  {Hosseinzadeh}, {Jenness}, {Jones}, {Joseph}, {Kalmbach}, {Karamehmetoglu},
  {Ka{l}uszy{'n}ski}, {Kelley}, {Kern}, {Kerzendorf}, {Koch}, {Kulumani},
  {Lee}, {Ly}, {Ma}, {MacBride}, {Maljaars}, {Muna}, {Murphy}, {Norman},
  {O'Steen}, {Oman}, {Pacifici}, {Pascual}, {Pascual-Granado}, {Patil},
  {Perren}, {Pickering}, {Rastogi}, {Roulston}, {Ryan}, {Rykoff}, {Sabater},
  {Sakurikar}, {Salgado}, {Sanghi}, {Saunders}, {Savchenko}, {Schwardt},
  {Seifert-Eckert}, {Shih}, {Jain}, {Shukla}, {Sick}, {Simpson},
  {Singanamalla}, {Singer}, {Singhal}, {Sinha}, {Sip{H{o}}cz}, {Spitler},
  {Stansby}, {Streicher}, {{{S}}umak}, {Swinbank}, {Taranu}, {Tewary},
  {Tremblay}, {Val-Borro}, {Van Kooten}, {Vasovi{'c}}, {Verma}, {de Miranda
  Cardoso}, {Williams}, {Wilson}, {Winkel}, {Wood-Vasey}, {Xue}, {Yoachim},
  {Zhang}, {Zonca}, \& {Astropy Project Contributors}}]{astropy:2022}
{Astropy Collaboration}, {Price-Whelan}, A.~M., {Lim}, P.~L., {et~al.} 2022,
  apj, 935, 167, \dodoi{10.3847/1538-4357/ac7c74}

\bibitem[{{Bakes} \& {Tielens}(1994)}]{1994ApJ...427..822B}
{Bakes}, E.~L.~O., \& {Tielens}, A.~G.~G.~M. 1994, \apj, 427, 822,
  \dodoi{10.1086/174188}

\bibitem[{{Barrera-Ballesteros} {et~al.}(2021){Barrera-Ballesteros},
  {S{\'a}nchez}, {Heckman}, {Wong}, {Bolatto}, {Ostriker}, {Rosolowsky},
  {Carigi}, {Vogel}, {Levy}, {Colombo}, {Luo}, \& {Cao}}]{2021MNRAS.503.3643B}
{Barrera-Ballesteros}, J.~K., {S{\'a}nchez}, S.~F., {Heckman}, T., {et~al.}
  2021, \mnras, 503, 3643, \dodoi{10.1093/mnras/stab755}

\bibitem[{{Bialy}(2020)}]{2020ApJ...903...62B}
{Bialy}, S. 2020, \apj, 903, 62, \dodoi{10.3847/1538-4357/abb804}

\bibitem[{{Bialy} \& {Sternberg}(2019)}]{2019ApJ...881..160B}
{Bialy}, S., \& {Sternberg}, A. 2019, \apj, 881, 160,
  \dodoi{10.3847/1538-4357/ab2fd1}

\bibitem[{{Bieri} {et~al.}(2023){Bieri}, {Naab}, {Geen}, {Coles}, {Pakmor}, \&
  {Walch}}]{2023MNRAS.523.6336B}
{Bieri}, R., {Naab}, T., {Geen}, S., {et~al.} 2023, \mnras, 523, 6336,
  \dodoi{10.1093/mnras/stad1710}

\bibitem[{{Boulares} \& {Cox}(1990)}]{1990ApJ...365..544B}
{Boulares}, A., \& {Cox}, D.~P. 1990, \apj, 365, 544, \dodoi{10.1086/169509}

\bibitem[{{Buck} {et~al.}(2020){Buck}, {Obreja}, {Macci{\`o}}, {Minchev},
  {Dutton}, \& {Ostriker}}]{2020MNRAS.491.3461B}
{Buck}, T., {Obreja}, A., {Macci{\`o}}, A.~V., {et~al.} 2020, \mnras, 491,
  3461, \dodoi{10.1093/mnras/stz3241}

\bibitem[{{Chastenet} {et~al.}(2023){Chastenet}, {Sutter}, {Sandstrom},
  {Belfiore}, {Egorov}, {Larson}, {Leroy}, {Liu}, {Rosolowsky}, {Thilker},
  {Watkins}, {Williams}, {Barnes}, {Bigiel}, {Boquien}, {Chevance}, {Chiang},
  {Dale}, {Kruijssen}, {Emsellem}, {Grasha}, {Groves}, {Hassani}, {Hughes},
  {Kreckel}, {Meidt}, {Rickards Vaught}, {Sardone}, \&
  {Schinnerer}}]{2023ApJ...944L..11C}
{Chastenet}, J., {Sutter}, J., {Sandstrom}, K., {et~al.} 2023, \apjl, 944, L11,
  \dodoi{10.3847/2041-8213/acadd7}

\bibitem[{{Chevance} {et~al.}(2023){Chevance}, {Krumholz}, {McLeod},
  {Ostriker}, {Rosolowsky}, \& {Sternberg}}]{2023ASPC..534....1C}
{Chevance}, M., {Krumholz}, M.~R., {McLeod}, A.~F., {et~al.} 2023, in
  Astronomical Society of the Pacific Conference Series, Vol. 534, Protostars
  and Planets VII, ed. S.~{Inutsuka}, Y.~{Aikawa}, T.~{Muto}, K.~{Tomida}, \&
  M.~{Tamura}, 1

\bibitem[{{Colling} {et~al.}(2018){Colling}, {Hennebelle}, {Geen}, {Iffrig}, \&
  {Bournaud}}]{2018A&A...620A..21C}
{Colling}, C., {Hennebelle}, P., {Geen}, S., {Iffrig}, O., \& {Bournaud}, F.
  2018, \aap, 620, A21, \dodoi{10.1051/0004-6361/201833161}

\bibitem[{{Conroy}(2013)}]{2013ARA&A..51..393C}
{Conroy}, C. 2013, \araa, 51, 393, \dodoi{10.1146/annurev-astro-082812-141017}

\bibitem[{{Conroy} \& {van Dokkum}(2012)}]{2012ApJ...760...71C}
{Conroy}, C., \& {van Dokkum}, P.~G. 2012, \apj, 760, 71,
  \dodoi{10.1088/0004-637X/760/1/71}

\bibitem[{{Cox}(1972)}]{1972ApJ...178..159C}
{Cox}, D.~P. 1972, \apj, 178, 159, \dodoi{10.1086/151775}

\bibitem[{{De Vis} {et~al.}(2019){De Vis}, {Jones}, {Viaene}, {Casasola},
  {Clark}, {Baes}, {Bianchi}, {Cassara}, {Davies}, {De Looze}, {Galametz},
  {Galliano}, {Lianou}, {Madden}, {Manilla-Robles}, {Mosenkov}, {Nersesian},
  {Roychowdhury}, {Xilouris}, \& {Ysard}}]{2019A&A...623A...5D}
{De Vis}, P., {Jones}, A., {Viaene}, S., {et~al.} 2019, \aap, 623, A5,
  \dodoi{10.1051/0004-6361/201834444}

\bibitem[{{Deng} {et~al.}(2024){Deng}, {Li}, {Kannan}, {Smith}, {Vogelsberger},
  \& {Bryan}}]{2024MNRAS.527..478D}
{Deng}, Y., {Li}, H., {Kannan}, R., {et~al.} 2024, \mnras, 527, 478,
  \dodoi{10.1093/mnras/stad3202}

\bibitem[{{Draine}(1978)}]{1978ApJS...36..595D}
{Draine}, B.~T. 1978, \apjs, 36, 595, \dodoi{10.1086/190513}

\bibitem[{{Draine}(2011)}]{2011piim.book.....D}
---. 2011, {Physics of the Interstellar and Intergalactic Medium} (Princeton
  University Press)

\bibitem[{{Draine} {et~al.}(2007){Draine}, {Dale}, {Bendo}, {Gordon}, {Smith},
  {Armus}, {Engelbracht}, {Helou}, {Kennicutt}, {Li}, {Roussel}, {Walter},
  {Calzetti}, {Moustakas}, {Murphy}, {Rieke}, {Bot}, {Hollenbach}, {Sheth}, \&
  {Teplitz}}]{2007ApJ...663..866D}
{Draine}, B.~T., {Dale}, D.~A., {Bendo}, G., {et~al.} 2007, \apj, 663, 866,
  \dodoi{10.1086/518306}

\bibitem[{{Dubois} {et~al.}(2021){Dubois}, {Beckmann}, {Bournaud}, {Choi},
  {Devriendt}, {Jackson}, {Kaviraj}, {Kimm}, {Kraljic}, {Laigle}, {Martin},
  {Park}, {Peirani}, {Pichon}, {Volonteri}, \& {Yi}}]{2021A&A...651A.109D}
{Dubois}, Y., {Beckmann}, R., {Bournaud}, F., {et~al.} 2021, \aap, 651, A109,
  \dodoi{10.1051/0004-6361/202039429}

\bibitem[{{El-Badry} {et~al.}(2019){El-Badry}, {Ostriker}, {Kim}, {Quataert},
  \& {Weisz}}]{2019MNRAS.490.1961E}
{El-Badry}, K., {Ostriker}, E.~C., {Kim}, C.-G., {Quataert}, E., \& {Weisz},
  D.~R. 2019, \mnras, 490, 1961, \dodoi{10.1093/mnras/stz2773}

\bibitem[{{Eldridge} \& {Stanway}(2022)}]{2022ARA&A..60..455E}
{Eldridge}, J.~J., \& {Stanway}, E.~R. 2022, \araa, 60, 455,
  \dodoi{10.1146/annurev-astro-052920-100646}

\bibitem[{{Evans} {et~al.}(2022){Evans}, {Kim}, \&
  {Ostriker}}]{2022ApJ...929L..18E}
{Evans}, N.~J., {Kim}, J.-G., \& {Ostriker}, E.~C. 2022, \apjl, 929, L18,
  \dodoi{10.3847/2041-8213/ac6427}

\bibitem[{{Feldmann} {et~al.}(2023){Feldmann}, {Quataert},
  {Faucher-Gigu{\`e}re}, {Hopkins}, {{\c{C}}atmabacak}, {Kere{\v{s}}},
  {Bassini}, {Bernardini}, {Bullock}, {Cenci}, {Gensior}, {Liang}, {Moreno}, \&
  {Wetzel}}]{2023MNRAS.522.3831F}
{Feldmann}, R., {Quataert}, E., {Faucher-Gigu{\`e}re}, C.-A., {et~al.} 2023,
  \mnras, 522, 3831, \dodoi{10.1093/mnras/stad1205}

\bibitem[{{Ferland} {et~al.}(2017){Ferland}, {Chatzikos}, {Guzm{\'a}n},
  {Lykins}, {van Hoof}, {Williams}, {Abel}, {Badnell}, {Keenan}, {Porter}, \&
  {Stancil}}]{2017RMxAA..53..385F}
{Ferland}, G.~J., {Chatzikos}, M., {Guzm{\'a}n}, F., {et~al.} 2017, \rmxaa, 53,
  385, \dodoi{10.48550/arXiv.1705.10877}

\bibitem[{{Field} {et~al.}(1969){Field}, {Goldsmith}, \&
  {Habing}}]{1969ApJ...155L.149F}
{Field}, G.~B., {Goldsmith}, D.~W., \& {Habing}, H.~J. 1969, \apjl, 155, L149,
  \dodoi{10.1086/180324}

\bibitem[{{Fielding} {et~al.}(2018){Fielding}, {Quataert}, \&
  {Martizzi}}]{2018MNRAS.481.3325F}
{Fielding}, D., {Quataert}, E., \& {Martizzi}, D. 2018, \mnras, 481, 3325,
  \dodoi{10.1093/mnras/sty2466}

\bibitem[{{Gatto} {et~al.}(2017){Gatto}, {Walch}, {Naab}, {Girichidis},
  {W{\"u}nsch}, {Glover}, {Klessen}, {Clark}, {Peters}, {Derigs}, {Baczynski},
  \& {Puls}}]{2017MNRAS.466.1903G}
{Gatto}, A., {Walch}, S., {Naab}, T., {et~al.} 2017, \mnras, 466, 1903,
  \dodoi{10.1093/mnras/stw3209}

\bibitem[{{Geen} {et~al.}(2021){Geen}, {Bieri}, {Rosdahl}, \& {de
  Koter}}]{2021MNRAS.501.1352G}
{Geen}, S., {Bieri}, R., {Rosdahl}, J., \& {de Koter}, A. 2021, \mnras, 501,
  1352, \dodoi{10.1093/mnras/staa3705}

\bibitem[{{Geen} {et~al.}(2016){Geen}, {Hennebelle}, {Tremblin}, \&
  {Rosdahl}}]{2016MNRAS.463.3129G}
{Geen}, S., {Hennebelle}, P., {Tremblin}, P., \& {Rosdahl}, J. 2016, \mnras,
  463, 3129, \dodoi{10.1093/mnras/stw2235}

\bibitem[{{Geen} {et~al.}(2020){Geen}, {Pellegrini}, {Bieri}, \&
  {Klessen}}]{2020MNRAS.492..915G}
{Geen}, S., {Pellegrini}, E., {Bieri}, R., \& {Klessen}, R. 2020, \mnras, 492,
  915, \dodoi{10.1093/mnras/stz3491}

\bibitem[{{Gentry} {et~al.}(2019){Gentry}, {Krumholz}, {Madau}, \&
  {Lupi}}]{2019MNRAS.483.3647G}
{Gentry}, E.~S., {Krumholz}, M.~R., {Madau}, P., \& {Lupi}, A. 2019, \mnras,
  483, 3647, \dodoi{10.1093/mnras/sty3319}

\bibitem[{{Girard} {et~al.}(2021){Girard}, {Fisher}, {Bolatto}, {Abraham},
  {Bassett}, {Glazebrook}, {Herrera-Camus}, {Jim{\'e}nez}, {Lenki{\'c}}, \&
  {Obreschkow}}]{2021ApJ...909...12G}
{Girard}, M., {Fisher}, D.~B., {Bolatto}, A.~D., {et~al.} 2021, \apj, 909, 12,
  \dodoi{10.3847/1538-4357/abd5b9}

\bibitem[{{Girichidis} {et~al.}(2018){Girichidis}, {Naab}, {Hanasz}, \&
  {Walch}}]{2018MNRAS.479.3042G}
{Girichidis}, P., {Naab}, T., {Hanasz}, M., \& {Walch}, S. 2018, \mnras, 479,
  3042, \dodoi{10.1093/mnras/sty1653}

\bibitem[{{Girichidis} {et~al.}(2020){Girichidis}, {Pfrommer}, {Hanasz}, \&
  {Naab}}]{2020MNRAS.491..993G}
{Girichidis}, P., {Pfrommer}, C., {Hanasz}, M., \& {Naab}, T. 2020, \mnras,
  491, 993, \dodoi{10.1093/mnras/stz2961}

\bibitem[{{Girichidis} {et~al.}(2016){Girichidis}, {Walch}, {Naab}, {Gatto},
  {W{\"u}nsch}, {Glover}, {Klessen}, {Clark}, {Peters}, {Derigs}, \&
  {Baczynski}}]{2016MNRAS.456.3432G}
{Girichidis}, P., {Walch}, S., {Naab}, T., {et~al.} 2016, \mnras, 456, 3432,
  \dodoi{10.1093/mnras/stv2742}

\bibitem[{{Gnat} \& {Ferland}(2012)}]{2012ApJS..199...20G}
{Gnat}, O., \& {Ferland}, G.~J. 2012, \apjs, 199, 20,
  \dodoi{10.1088/0067-0049/199/1/20}

\bibitem[{{Gnat} \& {Sternberg}(2007)}]{2007ApJS..168..213G}
{Gnat}, O., \& {Sternberg}, A. 2007, \apjs, 168, 213, \dodoi{10.1086/509786}

\bibitem[{{Grasha} {et~al.}(2021){Grasha}, {Roy}, {Sutherland}, \&
  {Kewley}}]{2021ApJ...908..241G}
{Grasha}, K., {Roy}, A., {Sutherland}, R.~S., \& {Kewley}, L.~J. 2021, \apj,
  908, 241, \dodoi{10.3847/1538-4357/abd6bf}

\bibitem[{{Haffner} {et~al.}(2009){Haffner}, {Dettmar}, {Beckman}, {Wood},
  {Slavin}, {Giammanco}, {Madsen}, {Zurita}, \&
  {Reynolds}}]{2009RvMP...81..969H}
{Haffner}, L.~M., {Dettmar}, R.~J., {Beckman}, J.~E., {et~al.} 2009, Reviews of
  Modern Physics, 81, 969, \dodoi{10.1103/RevModPhys.81.969}

\bibitem[{{Haid} {et~al.}(2018){Haid}, {Walch}, {Seifried}, {W{\"u}nsch},
  {Dinnbier}, \& {Naab}}]{2018MNRAS.478.4799H}
{Haid}, S., {Walch}, S., {Seifried}, D., {et~al.} 2018, \mnras, 478, 4799,
  \dodoi{10.1093/mnras/sty1315}

\bibitem[{{Herrera-Camus} {et~al.}(2017){Herrera-Camus}, {Bolatto}, {Wolfire},
  {Ostriker}, {Draine}, {Leroy}, {Sandstrom}, {Hunt}, {Kennicutt}, {Calzetti},
  {Smith}, {Croxall}, {Galametz}, {de Looze}, {Dale}, {Crocker}, \&
  {Groves}}]{2017ApJ...835..201H}
{Herrera-Camus}, R., {Bolatto}, A., {Wolfire}, M., {et~al.} 2017, \apj, 835,
  201, \dodoi{10.3847/1538-4357/835/2/201}

\bibitem[{{Hollenbach} \& {McKee}(1979)}]{1979ApJS...41..555H}
{Hollenbach}, D., \& {McKee}, C.~F. 1979, \apjs, 41, 555,
  \dodoi{10.1086/190631}

\bibitem[{{Hopkins}(2015)}]{2015MNRAS.450...53H}
{Hopkins}, P.~F. 2015, \mnras, 450, 53, \dodoi{10.1093/mnras/stv195}

\bibitem[{{Hopkins} {et~al.}(2022{\natexlab{a}}){Hopkins}, {Butsky},
  {Panopoulou}, {Ji}, {Quataert}, {Faucher-Gigu{\`e}re}, \&
  {Kere{\v{s}}}}]{2022MNRAS.516.3470H}
{Hopkins}, P.~F., {Butsky}, I.~S., {Panopoulou}, G.~V., {et~al.}
  2022{\natexlab{a}}, \mnras, 516, 3470, \dodoi{10.1093/mnras/stac1791}

\bibitem[{{Hopkins} {et~al.}(2014){Hopkins}, {Kere{\v{s}}}, {O{\~n}orbe},
  {Faucher-Gigu{\`e}re}, {Quataert}, {Murray}, \&
  {Bullock}}]{2014MNRAS.445..581H}
{Hopkins}, P.~F., {Kere{\v{s}}}, D., {O{\~n}orbe}, J., {et~al.} 2014, \mnras,
  445, 581, \dodoi{10.1093/mnras/stu1738}

\bibitem[{{Hopkins} {et~al.}(2022{\natexlab{b}}){Hopkins}, {Squire}, {Butsky},
  \& {Ji}}]{2022MNRAS.517.5413H}
{Hopkins}, P.~F., {Squire}, J., {Butsky}, I.~S., \& {Ji}, S.
  2022{\natexlab{b}}, \mnras, 517, 5413, \dodoi{10.1093/mnras/stac2909}

\bibitem[{{Hopkins} {et~al.}(2018){Hopkins}, {Wetzel}, {Kere{\v{s}}},
  {Faucher-Gigu{\`e}re}, {Quataert}, {Boylan-Kolchin}, {Murray}, {Hayward},
  {Garrison-Kimmel}, {Hummels}, {Feldmann}, {Torrey}, {Ma},
  {Angl{\'e}s-Alc{\'a}zar}, {Su}, {Orr}, {Schmitz}, {Escala}, {Sanderson},
  {Grudi{\'c}}, {Hafen}, {Kim}, {Fitts}, {Bullock}, {Wheeler}, {Chan},
  {Elbert}, \& {Narayanan}}]{2018MNRAS.480..800H}
{Hopkins}, P.~F., {Wetzel}, A., {Kere{\v{s}}}, D., {et~al.} 2018, \mnras, 480,
  800, \dodoi{10.1093/mnras/sty1690}

\bibitem[{{Hopkins} {et~al.}(2023){Hopkins}, {Wetzel}, {Wheeler}, {Sanderson},
  {Grudi{\'c}}, {Sameie}, {Boylan-Kolchin}, {Orr}, {Ma}, {Faucher-Gigu{\`e}re},
  {Kere{\v{s}}}, {Quataert}, {Su}, {Moreno}, {Feldmann}, {Bullock}, {Loebman},
  {Angl{\'e}s-Alc{\'a}zar}, {Stern}, {Necib}, {Choban}, \&
  {Hayward}}]{2023MNRAS.519.3154H}
{Hopkins}, P.~F., {Wetzel}, A., {Wheeler}, C., {et~al.} 2023, \mnras, 519,
  3154, \dodoi{10.1093/mnras/stac3489}

\bibitem[{Hoyer \& Hamman(2017)}]{hoyer2017xarray}
Hoyer, S., \& Hamman, J. 2017, Journal of Open Research Software, 5, 10,
  \dodoi{10.5334/jors.148}

\bibitem[{{Hu} {et~al.}(2017){Hu}, {Naab}, {Glover}, {Walch}, \&
  {Clark}}]{2017MNRAS.471.2151H}
{Hu}, C.-Y., {Naab}, T., {Glover}, S. C.~O., {Walch}, S., \& {Clark}, P.~C.
  2017, \mnras, 471, 2151, \dodoi{10.1093/mnras/stx1773}

\bibitem[{{Hu} {et~al.}(2021){Hu}, {Sternberg}, \& {van
  Dishoeck}}]{2021ApJ...920...44H}
{Hu}, C.-Y., {Sternberg}, A., \& {van Dishoeck}, E.~F. 2021, \apj, 920, 44,
  \dodoi{10.3847/1538-4357/ac0dbd}

\bibitem[{{Hu} {et~al.}(2023{\natexlab{a}}){Hu}, {Sternberg}, \& {van
  Dishoeck}}]{2023ApJ...952..140H}
---. 2023{\natexlab{a}}, \apj, 952, 140, \dodoi{10.3847/1538-4357/acdcfa}

\bibitem[{{Hu} {et~al.}(2019){Hu}, {Zhukovska}, {Somerville}, \&
  {Naab}}]{2019MNRAS.487.3252H}
{Hu}, C.-Y., {Zhukovska}, S., {Somerville}, R.~S., \& {Naab}, T. 2019, \mnras,
  487, 3252, \dodoi{10.1093/mnras/stz1481}

\bibitem[{{Hu} {et~al.}(2023{\natexlab{b}}){Hu}, {Smith}, {Teyssier}, {Bryan},
  {Verbeke}, {Emerick}, {Somerville}, {Burkhart}, {Li}, {Forbes}, \&
  {Starkenburg}}]{2023ApJ...950..132H}
{Hu}, C.-Y., {Smith}, M.~C., {Teyssier}, R., {et~al.} 2023{\natexlab{b}}, \apj,
  950, 132, \dodoi{10.3847/1538-4357/accf9e}

\bibitem[{{Hunter}(2007)}]{Hunter:2007}
{Hunter}, J.~D. 2007, Computing in Science and Engineering, 9, 90,
  \dodoi{10.1109/MCSE.2007.55}

\bibitem[{{Iffrig} \& {Hennebelle}(2015)}]{2015A&A...576A..95I}
{Iffrig}, O., \& {Hennebelle}, P. 2015, \aap, 576, A95,
  \dodoi{10.1051/0004-6361/201424556}

\bibitem[{{Indriolo} {et~al.}(2015){Indriolo}, {Neufeld}, {Gerin}, {Schilke},
  {Benz}, {Winkel}, {Menten}, {Chambers}, {Black}, {Bruderer}, {Falgarone},
  {Godard}, {Goicoechea}, {Gupta}, {Lis}, {Ossenkopf}, {Persson},
  {Sonnentrucker}, {van der Tak}, {van Dishoeck}, {Wolfire}, \&
  {Wyrowski}}]{2015ApJ...800...40I}
{Indriolo}, N., {Neufeld}, D.~A., {Gerin}, M., {et~al.} 2015, \apj, 800, 40,
  \dodoi{10.1088/0004-637X/800/1/40}

\bibitem[{{Jeffreson} {et~al.}(2021){Jeffreson}, {Krumholz}, {Fujimoto},
  {Armillotta}, {Keller}, {Chevance}, \& {Kruijssen}}]{2021MNRAS.505.3470J}
{Jeffreson}, S. M.~R., {Krumholz}, M.~R., {Fujimoto}, Y., {et~al.} 2021,
  \mnras, 505, 3470, \dodoi{10.1093/mnras/stab1536}

\bibitem[{{Jeffreson} {et~al.}(2022){Jeffreson}, {Sun}, \&
  {Wilson}}]{2022MNRAS.515.1663J}
{Jeffreson}, S. M.~R., {Sun}, J., \& {Wilson}, C.~D. 2022, \mnras, 515, 1663,
  \dodoi{10.1093/mnras/stac1874}

\bibitem[{{Jiang} \& {Oh}(2018)}]{2018ApJ...854....5J}
{Jiang}, Y.-F., \& {Oh}, S.~P. 2018, \apj, 854, 5,
  \dodoi{10.3847/1538-4357/aaa6ce}

\bibitem[{{Joung} \& {Mac Low}(2006)}]{2006ApJ...653.1266J}
{Joung}, M.~K.~R., \& {Mac Low}, M.-M. 2006, \apj, 653, 1266,
  \dodoi{10.1086/508795}

\bibitem[{{Joung} {et~al.}(2009){Joung}, {Mac Low}, \&
  {Bryan}}]{2009ApJ...704..137J}
{Joung}, M.~R., {Mac Low}, M.-M., \& {Bryan}, G.~L. 2009, \apj, 704, 137,
  \dodoi{10.1088/0004-637X/704/1/137}

\bibitem[{{Kannan} {et~al.}(2020{\natexlab{a}}){Kannan}, {Marinacci},
  {Simpson}, {Glover}, \& {Hernquist}}]{2020MNRAS.491.2088K}
{Kannan}, R., {Marinacci}, F., {Simpson}, C.~M., {Glover}, S. C.~O., \&
  {Hernquist}, L. 2020{\natexlab{a}}, \mnras, 491, 2088,
  \dodoi{10.1093/mnras/stz3078}

\bibitem[{{Kannan} {et~al.}(2020{\natexlab{b}}){Kannan}, {Marinacci},
  {Vogelsberger}, {Sales}, {Torrey}, {Springel}, \&
  {Hernquist}}]{2020MNRAS.499.5732K}
{Kannan}, R., {Marinacci}, F., {Vogelsberger}, M., {et~al.} 2020{\natexlab{b}},
  \mnras, 499, 5732, \dodoi{10.1093/mnras/staa3249}

\bibitem[{{Karpov} {et~al.}(2020){Karpov}, {Martizzi}, {Macias},
  {Ramirez-Ruiz}, {Kolborg}, \& {Naiman}}]{2020ApJ...896...66K}
{Karpov}, P.~I., {Martizzi}, D., {Macias}, P., {et~al.} 2020, \apj, 896, 66,
  \dodoi{10.3847/1538-4357/ab8f23}

\bibitem[{{Katz}(2022)}]{2022MNRAS.512..348K}
{Katz}, H. 2022, \mnras, 512, 348, \dodoi{10.1093/mnras/stac423}

\bibitem[{{Katz} {et~al.}(2022){Katz}, {Liu}, {Kimm}, {Rey}, {Andersson},
  {Cameron}, {Rodriguez-Montero}, {Agertz}, {Devriendt}, \&
  {Slyz}}]{2022arXiv221104626K}
{Katz}, H., {Liu}, S., {Kimm}, T., {et~al.} 2022, arXiv e-prints,
  arXiv:2211.04626.
\newblock \doarXiv{2211.04626}

\bibitem[{{Kennicutt} \& {Evans}(2012)}]{2012ARA&A..50..531K}
{Kennicutt}, R.~C., \& {Evans}, N.~J. 2012, \araa, 50, 531,
  \dodoi{10.1146/annurev-astro-081811-125610}

\bibitem[{{Kim} {et~al.}(2023{\natexlab{a}}){Kim}, {Kim}, {Gong}, \&
  {Ostriker}}]{2023ApJ...946....3K}
{Kim}, C.-G., {Kim}, J.-G., {Gong}, M., \& {Ostriker}, E.~C.
  2023{\natexlab{a}}, \apj, 946, 3, \dodoi{10.3847/1538-4357/acbd3a}

\bibitem[{{Kim} {et~al.}(2011){Kim}, {Kim}, \&
  {Ostriker}}]{2011ApJ...743...25K}
{Kim}, C.-G., {Kim}, W.-T., \& {Ostriker}, E.~C. 2011, \apj, 743, 25,
  \dodoi{10.1088/0004-637X/743/1/25}

\bibitem[{{Kim} \& {Ostriker}(2015{\natexlab{a}})}]{2015ApJ...802...99K}
{Kim}, C.-G., \& {Ostriker}, E.~C. 2015{\natexlab{a}}, \apj, 802, 99,
  \dodoi{10.1088/0004-637X/802/2/99}

\bibitem[{{Kim} \& {Ostriker}(2015{\natexlab{b}})}]{2015ApJ...815...67K}
---. 2015{\natexlab{b}}, \apj, 815, 67, \dodoi{10.1088/0004-637X/815/1/67}

\bibitem[{{Kim} \& {Ostriker}(2017)}]{2017ApJ...846..133K}
---. 2017, \apj, 846, 133, \dodoi{10.3847/1538-4357/aa8599}

\bibitem[{{Kim} \& {Ostriker}(2018)}]{2018ApJ...853..173K}
---. 2018, \apj, 853, 173, \dodoi{10.3847/1538-4357/aaa5ff}

\bibitem[{{Kim} {et~al.}(2017{\natexlab{a}}){Kim}, {Ostriker}, \&
  {Raileanu}}]{2017ApJ...834...25K}
{Kim}, C.-G., {Ostriker}, E.~C., \& {Raileanu}, R. 2017{\natexlab{a}}, \apj,
  834, 25, \dodoi{10.3847/1538-4357/834/1/25}

\bibitem[{{Kim} {et~al.}(2020{\natexlab{a}}){Kim}, {Ostriker}, {Somerville},
  {Bryan}, {Fielding}, {Forbes}, {Hayward}, {Hernquist}, \&
  {Pandya}}]{2020ApJ...900...61K}
{Kim}, C.-G., {Ostriker}, E.~C., {Somerville}, R.~S., {et~al.}
  2020{\natexlab{a}}, \apj, 900, 61, \dodoi{10.3847/1538-4357/aba962}

\bibitem[{{Kim} {et~al.}(2020{\natexlab{b}}){Kim}, {Ostriker}, {Fielding},
  {Smith}, {Bryan}, {Somerville}, {Forbes}, {Genel}, \&
  {Hernquist}}]{2020ApJ...903L..34K}
{Kim}, C.-G., {Ostriker}, E.~C., {Fielding}, D.~B., {et~al.}
  2020{\natexlab{b}}, \apjl, 903, L34, \dodoi{10.3847/2041-8213/abc252}

\bibitem[{{Kim} {et~al.}(2023{\natexlab{b}}){Kim}, {Gong}, {Kim}, \&
  {Ostriker}}]{2023ApJS..264...10K}
{Kim}, J.-G., {Gong}, M., {Kim}, C.-G., \& {Ostriker}, E.~C.
  2023{\natexlab{b}}, \apjs, 264, 10, \dodoi{10.3847/1538-4365/ac9b1d}

\bibitem[{{Kim} {et~al.}(2018){Kim}, {Kim}, \&
  {Ostriker}}]{2018ApJ...859...68K}
{Kim}, J.-G., {Kim}, W.-T., \& {Ostriker}, E.~C. 2018, \apj, 859, 68,
  \dodoi{10.3847/1538-4357/aabe27}

\bibitem[{{Kim} {et~al.}(2017{\natexlab{b}}){Kim}, {Kim}, {Ostriker}, \&
  {Skinner}}]{2017ApJ...851...93K}
{Kim}, J.-G., {Kim}, W.-T., {Ostriker}, E.~C., \& {Skinner}, M.~A.
  2017{\natexlab{b}}, \apj, 851, 93, \dodoi{10.3847/1538-4357/aa9b80}

\bibitem[{{Kim} {et~al.}(2021){Kim}, {Ostriker}, \&
  {Filippova}}]{2021ApJ...911..128K}
{Kim}, J.-G., {Ostriker}, E.~C., \& {Filippova}, N. 2021, \apj, 911, 128,
  \dodoi{10.3847/1538-4357/abe934}

\bibitem[{{Kim} {et~al.}(2020{\natexlab{c}}){Kim}, {Kim}, \&
  {Ostriker}}]{2020ApJ...898...35K}
{Kim}, W.-T., {Kim}, C.-G., \& {Ostriker}, E.~C. 2020{\natexlab{c}}, \apj, 898,
  35, \dodoi{10.3847/1538-4357/ab9b87}

\bibitem[{{Kimm} \& {Cen}(2014)}]{2014ApJ...788..121K}
{Kimm}, T., \& {Cen}, R. 2014, \apj, 788, 121,
  \dodoi{10.1088/0004-637X/788/2/121}

\bibitem[{{Kroupa}(2001)}]{2001MNRAS.322..231K}
{Kroupa}, P. 2001, \mnras, 322, 231, \dodoi{10.1046/j.1365-8711.2001.04022.x}

\bibitem[{{Krumholz} {et~al.}(2019){Krumholz}, {McKee}, \&
  {Bland-Hawthorn}}]{2019ARA&A..57..227K}
{Krumholz}, M.~R., {McKee}, C.~F., \& {Bland-Hawthorn}, J. 2019, \araa, 57,
  227, \dodoi{10.1146/annurev-astro-091918-104430}

\bibitem[{{Krumholz} \& {Tan}(2007)}]{2007ApJ...654..304K}
{Krumholz}, M.~R., \& {Tan}, J.~C. 2007, \apj, 654, 304, \dodoi{10.1086/509101}

\bibitem[{{Kuijken} \& {Gilmore}(1989)}]{1989MNRAS.239..571K}
{Kuijken}, K., \& {Gilmore}, G. 1989, \mnras, 239, 571,
  \dodoi{10.1093/mnras/239.2.571}

\bibitem[{{Lancaster} {et~al.}(2021{\natexlab{a}}){Lancaster}, {Ostriker},
  {Kim}, \& {Kim}}]{2021ApJ...914...90L}
{Lancaster}, L., {Ostriker}, E.~C., {Kim}, J.-G., \& {Kim}, C.-G.
  2021{\natexlab{a}}, \apj, 914, 90, \dodoi{10.3847/1538-4357/abf8ac}

\bibitem[{{Lancaster} {et~al.}(2021{\natexlab{b}}){Lancaster}, {Ostriker},
  {Kim}, \& {Kim}}]{2021ApJ...922L...3L}
---. 2021{\natexlab{b}}, \apjl, 922, L3, \dodoi{10.3847/2041-8213/ac3333}

\bibitem[{{Lee} {et~al.}(2022){Lee}, {Kimm}, {Blaizot}, {Katz}, {Lee}, {Sheen},
  {Devriendt}, \& {Slyz}}]{2022ApJ...928..144L}
{Lee}, J., {Kimm}, T., {Blaizot}, J., {et~al.} 2022, \apj, 928, 144,
  \dodoi{10.3847/1538-4357/ac5595}

\bibitem[{{Leitherer} {et~al.}(2014){Leitherer}, {Ekstr{\"o}m}, {Meynet},
  {Schaerer}, {Agienko}, \& {Levesque}}]{2014ApJS..212...14L}
{Leitherer}, C., {Ekstr{\"o}m}, S., {Meynet}, G., {et~al.} 2014, \apjs, 212,
  14, \dodoi{10.1088/0067-0049/212/1/14}

\bibitem[{{Leitherer} {et~al.}(1999){Leitherer}, {Schaerer}, {Goldader},
  {Delgado}, {Robert}, {Kune}, {de Mello}, {Devost}, \&
  {Heckman}}]{1999ApJS..123....3L}
{Leitherer}, C., {Schaerer}, D., {Goldader}, J.~D., {et~al.} 1999, \apjs, 123,
  3, \dodoi{10.1086/313233}

\bibitem[{{Levesque} {et~al.}(2012){Levesque}, {Leitherer}, {Ekstrom},
  {Meynet}, \& {Schaerer}}]{2012ApJ...751...67L}
{Levesque}, E.~M., {Leitherer}, C., {Ekstrom}, S., {Meynet}, G., \& {Schaerer},
  D. 2012, \apj, 751, 67, \dodoi{10.1088/0004-637X/751/1/67}

\bibitem[{{Li} {et~al.}(2024){Li}, {Li}, {Cui}, {Marinacci}, {Sales},
  {Vogelsberger}, \& {Torrey}}]{2024MNRAS.tmp..833L}
{Li}, C., {Li}, H., {Cui}, W., {et~al.} 2024, \mnras, 529, 4073,
  \dodoi{10.1093/mnras/stae797}

\bibitem[{{Li} {et~al.}(2018){Li}, {Gnedin}, \& {Gnedin}}]{2018ApJ...861..107L}
{Li}, H., {Gnedin}, O.~Y., \& {Gnedin}, N.~Y. 2018, \apj, 861, 107,
  \dodoi{10.3847/1538-4357/aac9b8}

\bibitem[{{Li} {et~al.}(2020){Li}, {Vogelsberger}, {Marinacci}, {Sales}, \&
  {Torrey}}]{2020MNRAS.499.5862L}
{Li}, H., {Vogelsberger}, M., {Marinacci}, F., {Sales}, L.~V., \& {Torrey}, P.
  2020, \mnras, 499, 5862, \dodoi{10.1093/mnras/staa3122}

\bibitem[{{Lupi} {et~al.}(2020){Lupi}, {Pallottini}, {Ferrara}, {Bovino},
  {Carniani}, \& {Vallini}}]{2020MNRAS.496.5160L}
{Lupi}, A., {Pallottini}, A., {Ferrara}, A., {et~al.} 2020, \mnras, 496, 5160,
  \dodoi{10.1093/mnras/staa1842}

\bibitem[{{Martizzi} {et~al.}(2015){Martizzi}, {Faucher-Gigu{\`e}re}, \&
  {Quataert}}]{2015MNRAS.450..504M}
{Martizzi}, D., {Faucher-Gigu{\`e}re}, C.-A., \& {Quataert}, E. 2015, \mnras,
  450, 504, \dodoi{10.1093/mnras/stv562}

\bibitem[{{Martizzi} {et~al.}(2016){Martizzi}, {Fielding},
  {Faucher-Gigu{\`e}re}, \& {Quataert}}]{2016MNRAS.459.2311M}
{Martizzi}, D., {Fielding}, D., {Faucher-Gigu{\`e}re}, C.-A., \& {Quataert}, E.
  2016, \mnras, 459, 2311, \dodoi{10.1093/mnras/stw745}

\bibitem[{{McKee} \& {Ostriker}(1977)}]{1977ApJ...218..148M}
{McKee}, C.~F., \& {Ostriker}, J.~P. 1977, \apj, 218, 148,
  \dodoi{10.1086/155667}

\bibitem[{{M}c{K}inney(2010)}]{mckinney-proc-scipy-2010}
{M}c{K}inney, W. 2010, in {P}roceedings of the 9th {P}ython in {S}cience
  {C}onference, ed. {S}t\'efan van~der {W}alt \& {J}arrod {M}illman (Austin,
  TX: SciPy), 56

\bibitem[{{Moon} {et~al.}(2021){Moon}, {Kim}, {Kim}, \&
  {Ostriker}}]{2021ApJ...914....9M}
{Moon}, S., {Kim}, W.-T., {Kim}, C.-G., \& {Ostriker}, E.~C. 2021, \apj, 914,
  9, \dodoi{10.3847/1538-4357/abfa93}

\bibitem[{{Moon} {et~al.}(2023){Moon}, {Kim}, {Kim}, \&
  {Ostriker}}]{2023ApJ...946..114M}
---. 2023, \apj, 946, 114, \dodoi{10.3847/1538-4357/acc250}

\bibitem[{{Moustakas} {et~al.}(2010){Moustakas}, {Kennicutt}, {Tremonti},
  {Dale}, {Smith}, \& {Calzetti}}]{2010ApJS..190..233M}
{Moustakas}, J., {Kennicutt}, Robert~C., J., {Tremonti}, C.~A., {et~al.} 2010,
  \apjs, 190, 233, \dodoi{10.1088/0067-0049/190/2/233}

\bibitem[{{Naab} \& {Ostriker}(2017)}]{2017ARA&A..55...59N}
{Naab}, T., \& {Ostriker}, J.~P. 2017, \araa, 55, 59,
  \dodoi{10.1146/annurev-astro-081913-040019}

\bibitem[{{Nelson} {et~al.}(2019){Nelson}, {Pillepich}, {Springel}, {Pakmor},
  {Weinberger}, {Genel}, {Torrey}, {Vogelsberger}, {Marinacci}, \&
  {Hernquist}}]{2019MNRAS.490.3234N}
{Nelson}, D., {Pillepich}, A., {Springel}, V., {et~al.} 2019, \mnras, 490,
  3234, \dodoi{10.1093/mnras/stz2306}

\bibitem[{{Neufeld} \& {Wolfire}(2017)}]{2017ApJ...845..163N}
{Neufeld}, D.~A., \& {Wolfire}, M.~G. 2017, \apj, 845, 163,
  \dodoi{10.3847/1538-4357/aa6d68}

\bibitem[{{Oku} {et~al.}(2022){Oku}, {Tomida}, {Nagamine}, {Shimizu}, \&
  {Cen}}]{2022ApJS..262....9O}
{Oku}, Y., {Tomida}, K., {Nagamine}, K., {Shimizu}, I., \& {Cen}, R. 2022,
  \apjs, 262, 9, \dodoi{10.3847/1538-4365/ac77ff}

\bibitem[{{Ostriker} \& {Kim}(2022)}]{2022ApJ...936..137O}
{Ostriker}, E.~C., \& {Kim}, C.-G. 2022, \apj, 936, 137,
  \dodoi{10.3847/1538-4357/ac7de2}

\bibitem[{{Ostriker} {et~al.}(2010){Ostriker}, {McKee}, \&
  {Leroy}}]{2010ApJ...721..975O}
{Ostriker}, E.~C., {McKee}, C.~F., \& {Leroy}, A.~K. 2010, \apj, 721, 975,
  \dodoi{10.1088/0004-637X/721/2/975}

\bibitem[{{Ostriker} \& {Shetty}(2011)}]{2011ApJ...731...41O}
{Ostriker}, E.~C., \& {Shetty}, R. 2011, \apj, 731, 41,
  \dodoi{10.1088/0004-637X/731/1/41}

\bibitem[{{Ostriker} \& {McKee}(1988)}]{1988RvMP...60....1O}
{Ostriker}, J.~P., \& {McKee}, C.~F. 1988, Reviews of Modern Physics, 60, 1,
  \dodoi{10.1103/RevModPhys.60.1}

\bibitem[{{Padoan} {et~al.}(2014){Padoan}, {Federrath}, {Chabrier}, {Evans},
  {Johnstone}, {J{\o}rgensen}, {McKee}, \& {Nordlund}}]{2014prpl.conf...77P}
{Padoan}, P., {Federrath}, C., {Chabrier}, G., {et~al.} 2014, in Protostars and
  Planets VI, ed. H.~{Beuther}, R.~S. {Klessen}, C.~P. {Dullemond}, \&
  T.~{Henning}, 77--100

\bibitem[{{Perez} \& {Granger}(2007)}]{Perez2007}
{Perez}, F., \& {Granger}, B.~E. 2007, Computing in Science and Engineering, 9,
  21, \dodoi{10.1109/MCSE.2007.53}

\bibitem[{{Pillepich} {et~al.}(2019){Pillepich}, {Nelson}, {Springel},
  {Pakmor}, {Torrey}, {Weinberger}, {Vogelsberger}, {Marinacci}, {Genel}, {van
  der Wel}, \& {Hernquist}}]{2019MNRAS.490.3196P}
{Pillepich}, A., {Nelson}, D., {Springel}, V., {et~al.} 2019, \mnras, 490,
  3196, \dodoi{10.1093/mnras/stz2338}

\bibitem[{{Rathjen} {et~al.}(2023){Rathjen}, {Naab}, {Walch}, {Seifried},
  {Girichidis}, \& {W{\"u}nsch}}]{2023MNRAS.522.1843R}
{Rathjen}, T.-E., {Naab}, T., {Walch}, S., {et~al.} 2023, \mnras, 522, 1843,
  \dodoi{10.1093/mnras/stad1104}

\bibitem[{{Rathjen} {et~al.}(2021){Rathjen}, {Naab}, {Girichidis}, {Walch},
  {W{\"u}nsch}, {Dinnbier}, {Seifried}, {Klessen}, \&
  {Glover}}]{2021MNRAS.504.1039R}
{Rathjen}, T.-E., {Naab}, T., {Girichidis}, P., {et~al.} 2021, \mnras, 504,
  1039, \dodoi{10.1093/mnras/stab900}

\bibitem[{{R{\'e}my-Ruyer} {et~al.}(2014){R{\'e}my-Ruyer}, {Madden},
  {Galliano}, {Galametz}, {Takeuchi}, {Asano}, {Zhukovska}, {Lebouteiller},
  {Cormier}, {Jones}, {Bocchio}, {Baes}, {Bendo}, {Boquien}, {Boselli},
  {DeLooze}, {Doublier-Pritchard}, {Hughes}, {Karczewski}, \&
  {Spinoglio}}]{2014A&A...563A..31R}
{R{\'e}my-Ruyer}, A., {Madden}, S.~C., {Galliano}, F., {et~al.} 2014, \aap,
  563, A31, \dodoi{10.1051/0004-6361/201322803}

\bibitem[{{Roman-Duval} {et~al.}(2022){Roman-Duval}, {Jenkins}, {Tchernyshyov},
  {Clark}, {De Cia}, {Gordon}, {Hamanowicz}, {Lebouteiller}, {Rafelski},
  {Sandstrom}, {Werk}, \& {Yanchulova Merica-Jones}}]{2022ApJ...928...90R}
{Roman-Duval}, J., {Jenkins}, E.~B., {Tchernyshyov}, K., {et~al.} 2022, \apj,
  928, 90, \dodoi{10.3847/1538-4357/ac5248}

\bibitem[{{Sandstrom} {et~al.}(2012){Sandstrom}, {Bolatto}, {Bot}, {Draine},
  {Ingalls}, {Israel}, {Jackson}, {Leroy}, {Li}, {Rubio}, {Simon}, {Smith},
  {Stanimirovi{\'c}}, {Tielens}, \& {van Loon}}]{2012ApJ...744...20S}
{Sandstrom}, K.~M., {Bolatto}, A.~D., {Bot}, C., {et~al.} 2012, \apj, 744, 20,
  \dodoi{10.1088/0004-637X/744/1/20}

\bibitem[{{Schaye} \& {Dalla Vecchia}(2008)}]{2008MNRAS.383.1210S}
{Schaye}, J., \& {Dalla Vecchia}, C. 2008, \mnras, 383, 1210,
  \dodoi{10.1111/j.1365-2966.2007.12639.x}

\bibitem[{{Schinnerer} {et~al.}(2019){Schinnerer}, {Leroy}, {Blanc},
  {Emsellem}, {Hughes}, {Rosolowsky}, {Schruba}, {Bigiel}, {Escala}, {Groves},
  {Kreckel}, {Kruijssen}, {Lee}, {Meidt}, {Pety}, {Sanchez-Blazquez},
  {Sandstrom}, {Usero}, {Barnes}, {Belfiore}, {Be{\v{s}}li{\'c}}, {Chandar},
  {Chatzigiannakis}, {Chevance}, {Congiu}, {Dale}, {Faesi}, {Gallagher},
  {Garcia-Rodriguez}, {Glover}, {Grasha}, {Henshaw}, {Herrera}, {Ho}, {Hygate},
  {Jimenez-Donaire}, {Kessler}, {Kim}, {Klessen}, {Koch}, {Lang}, {Larson}, {Le
  Reste}, {Liu}, {McElroy}, {Nofech}, {Ostriker}, {Pessa Gutierrez},
  {Puschnig}, {Querejeta}, {Razza}, {Saito}, {Santoro}, {Stuber}, {Sun},
  {Thilker}, {Turner}, {Ubeda}, {Utreras}, {Utomo}, {van Dyk}, {Ward}, \&
  {Whitmore}}]{2019Msngr.177...36S}
{Schinnerer}, E., {Leroy}, A., {Blanc}, G., {et~al.} 2019, The Messenger, 177,
  36, \dodoi{10.18727/0722-6691/5151}

\bibitem[{{Schmidt}(1959)}]{1959ApJ...129..243S}
{Schmidt}, M. 1959, \apj, 129, 243, \dodoi{10.1086/146614}

\bibitem[{{Schneider} \& {Mao}(2024)}]{2024arXiv240212474S}
{Schneider}, E.~E., \& {Mao}, S.~A. 2024, \apj, 966, 37,
  \dodoi{10.3847/1538-4357/ad2e8a}

\bibitem[{{Schneider} {et~al.}(2020){Schneider}, {Ostriker}, {Robertson}, \&
  {Thompson}}]{2020ApJ...895...43S}
{Schneider}, E.~E., {Ostriker}, E.~C., {Robertson}, B.~E., \& {Thompson}, T.~A.
  2020, \apj, 895, 43, \dodoi{10.3847/1538-4357/ab8ae8}

\bibitem[{{Semenov} {et~al.}(2017){Semenov}, {Kravtsov}, \&
  {Gnedin}}]{2017ApJ...845..133S}
{Semenov}, V.~A., {Kravtsov}, A.~V., \& {Gnedin}, N.~Y. 2017, \apj, 845, 133,
  \dodoi{10.3847/1538-4357/aa8096}

\bibitem[{{Smith} {et~al.}(2021){Smith}, {Bryan}, {Somerville}, {Hu},
  {Teyssier}, {Burkhart}, \& {Hernquist}}]{2021MNRAS.506.3882S}
{Smith}, M.~C., {Bryan}, G.~L., {Somerville}, R.~S., {et~al.} 2021, \mnras,
  506, 3882, \dodoi{10.1093/mnras/stab1896}

\bibitem[{{Smith} {et~al.}(2024){Smith}, {Fielding}, {Bryan}, {Kim},
  {Ostriker}, {Somerville}, {Stern}, {Su}, {Weinberger}, {Hu}, {Forbes},
  {Hernquist}, {Burkhart}, \& {Li}}]{2024MNRAS.527.1216S}
{Smith}, M.~C., {Fielding}, D.~B., {Bryan}, G.~L., {et~al.} 2024, \mnras, 527,
  1216, \dodoi{10.1093/mnras/stad3168}

\bibitem[{{Smith} {et~al.}(2020){Smith}, {Tre{\ss}}, {Sormani}, {Glover},
  {Klessen}, {Clark}, {Izquierdo}, {Duarte-Cabral}, \&
  {Zucker}}]{2020MNRAS.492.1594S}
{Smith}, R.~J., {Tre{\ss}}, R.~G., {Sormani}, M.~C., {et~al.} 2020, \mnras,
  492, 1594, \dodoi{10.1093/mnras/stz3328}

\bibitem[{{Somerville} \& {Dav{\'e}}(2015)}]{2015ARA&A..53...51S}
{Somerville}, R.~S., \& {Dav{\'e}}, R. 2015, \araa, 53, 51,
  \dodoi{10.1146/annurev-astro-082812-140951}

\bibitem[{{Springel} \& {Hernquist}(2003)}]{2003MNRAS.339..289S}
{Springel}, V., \& {Hernquist}, L. 2003, \mnras, 339, 289,
  \dodoi{10.1046/j.1365-8711.2003.06206.x}

\bibitem[{{Steinwandel} {et~al.}(2023){Steinwandel}, {Bryan}, {Somerville},
  {Hayward}, \& {Burkhart}}]{2023MNRAS.526.1408S}
{Steinwandel}, U.~P., {Bryan}, G.~L., {Somerville}, R.~S., {Hayward}, C.~C., \&
  {Burkhart}, B. 2023, \mnras, 526, 1408, \dodoi{10.1093/mnras/stad2744}

\bibitem[{{Steinwandel} \& {Goldberg}(2023)}]{2023arXiv231011495S}
{Steinwandel}, U.~P., \& {Goldberg}, J.~A. 2023, arXiv e-prints,
  arXiv:2310.11495, \dodoi{10.48550/arXiv.2310.11495}

\bibitem[{{Steinwandel} {et~al.}(2024){Steinwandel}, {Kim}, {Bryan},
  {Ostriker}, {Somerville}, \& {Fielding}}]{2024ApJ...960..100S}
{Steinwandel}, U.~P., {Kim}, C.-G., {Bryan}, G.~L., {et~al.} 2024, \apj, 960,
  100, \dodoi{10.3847/1538-4357/ad09e1}

\bibitem[{{Steinwandel} {et~al.}(2020){Steinwandel}, {Moster}, {Naab}, {Hu}, \&
  {Walch}}]{2020MNRAS.495.1035S}
{Steinwandel}, U.~P., {Moster}, B.~P., {Naab}, T., {Hu}, C.-Y., \& {Walch}, S.
  2020, \mnras, 495, 1035, \dodoi{10.1093/mnras/staa821}

\bibitem[{{Stone} \& {Gardiner}(2009)}]{2009NewA...14..139S}
{Stone}, J.~M., \& {Gardiner}, T. 2009, \na, 14, 139,
  \dodoi{10.1016/j.newast.2008.06.003}

\bibitem[{{Stone} \& {Gardiner}(2010)}]{2010ApJS..189..142S}
{Stone}, J.~M., \& {Gardiner}, T.~A. 2010, \apjs, 189, 142,
  \dodoi{10.1088/0067-0049/189/1/142}

\bibitem[{{Stone} {et~al.}(2008){Stone}, {Gardiner}, {Teuben}, {Hawley}, \&
  {Simon}}]{2008ApJS..178..137S}
{Stone}, J.~M., {Gardiner}, T.~A., {Teuben}, P., {Hawley}, J.~F., \& {Simon},
  J.~B. 2008, \apjs, 178, 137, \dodoi{10.1086/588755}

\bibitem[{{Sun} {et~al.}(2020){Sun}, {Leroy}, {Ostriker}, {Hughes},
  {Rosolowsky}, {Schruba}, {Schinnerer}, {Blanc}, {Faesi}, {Kruijssen},
  {Meidt}, {Utomo}, {Bigiel}, {Bolatto}, {Chevance}, {Chiang}, {Dale},
  {Emsellem}, {Glover}, {Grasha}, {Henshaw}, {Herrera}, {Jimenez-Donaire},
  {Lee}, {Pety}, {Querejeta}, {Saito}, {Sandstrom}, \&
  {Usero}}]{2020ApJ...892..148S}
{Sun}, J., {Leroy}, A.~K., {Ostriker}, E.~C., {et~al.} 2020, \apj, 892, 148,
  \dodoi{10.3847/1538-4357/ab781c}

\bibitem[{{Sun} {et~al.}(2023){Sun}, {Leroy}, {Ostriker}, {Meidt},
  {Rosolowsky}, {Schinnerer}, {Wilson}, {Utomo}, {Belfiore}, {Blanc},
  {Emsellem}, {Faesi}, {Groves}, {Hughes}, {Koch}, {Kreckel}, {Liu}, {Pan},
  {Pety}, {Querejeta}, {Razza}, {Saito}, {Sardone}, {Usero}, {Williams},
  {Bigiel}, {Bolatto}, {Chevance}, {Dale}, {Gensior}, {Glover}, {Grasha},
  {Henshaw}, {Jim{\'e}nez-Donaire}, {Klessen}, {Kruijssen}, {Murphy},
  {Neumann}, {Teng}, \& {Thilker}}]{2023ApJ...945L..19S}
---. 2023, \apjl, 945, L19, \dodoi{10.3847/2041-8213/acbd9c}

\bibitem[{{Sutherland} \& {Dopita}(1993)}]{1993ApJS...88..253S}
{Sutherland}, R.~S., \& {Dopita}, M.~A. 1993, \apjs, 88, 253,
  \dodoi{10.1086/191823}

\bibitem[{{Tan} \& {Fielding}(2024)}]{2023MNRAS.tmp.3640T}
{Tan}, B., \& {Fielding}, D.~B. 2024, \mnras, 527, 9683,
  \dodoi{10.1093/mnras/stad3793}

\bibitem[{{Thornton} {et~al.}(1998){Thornton}, {Gaudlitz}, {Janka}, \&
  {Steinmetz}}]{1998ApJ...500...95T}
{Thornton}, K., {Gaudlitz}, M., {Janka}, H.~T., \& {Steinmetz}, M. 1998, \apj,
  500, 95, \dodoi{10.1086/305704}

\bibitem[{{Tre{\ss}} {et~al.}(2021){Tre{\ss}}, {Sormani}, {Smith}, {Glover},
  {Klessen}, {Mac Low}, {Clark}, \& {Duarte-Cabral}}]{2021MNRAS.505.5438T}
{Tre{\ss}}, R.~G., {Sormani}, M.~C., {Smith}, R.~J., {et~al.} 2021, \mnras,
  505, 5438, \dodoi{10.1093/mnras/stab1683}

\bibitem[{{van der Velden}(2020)}]{CMasher}
{van der Velden}, E. 2020, The Journal of Open Source Software, 5, 2004,
  \dodoi{10.21105/joss.02004}

\bibitem[{{van der Walt} {et~al.}(2011){van der Walt}, {Colbert}, \&
  {Varoquaux}}]{vanderWalt2011}
{van der Walt}, S., {Colbert}, S.~C., \& {Varoquaux}, G. 2011, Computing in
  Science and Engineering, 13, 22, \dodoi{10.1109/MCSE.2011.37}

\bibitem[{{Vandenbroucke} \& {Wood}(2018)}]{2018A&C....23...40V}
{Vandenbroucke}, B., \& {Wood}, K. 2018, Astronomy and Computing, 23, 40,
  \dodoi{10.1016/j.ascom.2018.02.005}

\bibitem[{{Virtanen} {et~al.}(2020){Virtanen}, {Gommers}, {Oliphant},
  {Haberland}, {Reddy}, {Cournapeau}, {Burovski}, {Peterson}, {Weckesser},
  {Bright}, {van der Walt}, {Brett}, {Wilson}, {Millman}, {Mayorov}, {Nelson},
  {Jones}, {Kern}, {Larson}, {Carey}, {Polat}, {Feng}, {Moore}, {Vand erPlas},
  {Laxalde}, {Perktold}, {Cimrman}, {Henriksen}, {Quintero}, {Harris},
  {Archibald}, {Ribeiro}, {Pedregosa}, {van Mulbregt}, \& {SciPy 1. 0
  Contributors}}]{2020SciPy-NMeth}
{Virtanen}, P., {Gommers}, R., {Oliphant}, T.~E., {et~al.} 2020, Nature
  Methods, 17, 261, \dodoi{10.1038/s41592-019-0686-2}

\bibitem[{{Vogelsberger} {et~al.}(2020){Vogelsberger}, {Marinacci}, {Torrey},
  \& {Puchwein}}]{2020NatRP...2...42V}
{Vogelsberger}, M., {Marinacci}, F., {Torrey}, P., \& {Puchwein}, E. 2020,
  Nature Reviews Physics, 2, 42, \dodoi{10.1038/s42254-019-0127-2}

\bibitem[{{Walch} \& {Naab}(2015)}]{2015MNRAS.451.2757W}
{Walch}, S., \& {Naab}, T. 2015, \mnras, 451, 2757,
  \dodoi{10.1093/mnras/stv1155}

\bibitem[{{Walch} {et~al.}(2015){Walch}, {Girichidis}, {Naab}, {Gatto},
  {Glover}, {W{\"u}nsch}, {Klessen}, {Clark}, {Peters}, {Derigs}, \&
  {Baczynski}}]{2015MNRAS.454..238W}
{Walch}, S., {Girichidis}, P., {Naab}, T., {et~al.} 2015, \mnras, 454, 238,
  \dodoi{10.1093/mnras/stv1975}

\bibitem[{{Watson}(1972)}]{1972ApJ...176..103W}
{Watson}, W.~D. 1972, \apj, 176, 103, \dodoi{10.1086/151613}

\bibitem[{{Weingartner} \& {Draine}(2001{\natexlab{a}})}]{2001ApJS..134..263W}
{Weingartner}, J.~C., \& {Draine}, B.~T. 2001{\natexlab{a}}, \apjs, 134, 263,
  \dodoi{10.1086/320852}

\bibitem[{{Weingartner} \& {Draine}(2001{\natexlab{b}})}]{2001ApJ...548..296W}
---. 2001{\natexlab{b}}, \apj, 548, 296, \dodoi{10.1086/318651}

\bibitem[{{Wiersma} {et~al.}(2009){Wiersma}, {Schaye}, \&
  {Smith}}]{2009MNRAS.393...99W}
{Wiersma}, R. P.~C., {Schaye}, J., \& {Smith}, B.~D. 2009, \mnras, 393, 99,
  \dodoi{10.1111/j.1365-2966.2008.14191.x}

\bibitem[{{Wolfire} {et~al.}(1995){Wolfire}, {Hollenbach}, {McKee}, {Tielens},
  \& {Bakes}}]{1995ApJ...443..152W}
{Wolfire}, M.~G., {Hollenbach}, D., {McKee}, C.~F., {Tielens}, A.~G.~G.~M., \&
  {Bakes}, E.~L.~O. 1995, \apj, 443, 152, \dodoi{10.1086/175510}

\bibitem[{{Wolfire} {et~al.}(2003){Wolfire}, {McKee}, {Hollenbach}, \&
  {Tielens}}]{2003ApJ...587..278W}
{Wolfire}, M.~G., {McKee}, C.~F., {Hollenbach}, D., \& {Tielens}, A.~G.~G.~M.
  2003, \apj, 587, 278, \dodoi{10.1086/368016}

\bibitem[{{Wolfire} {et~al.}(2022){Wolfire}, {Vallini}, \&
  {Chevance}}]{2022ARA&A..60..247W}
{Wolfire}, M.~G., {Vallini}, L., \& {Chevance}, M. 2022, \araa, 60, 247,
  \dodoi{10.1146/annurev-astro-052920-010254}

\bibitem[{{Zweibel}(2013)}]{2013PhPl...20e5501Z}
{Zweibel}, E.~G. 2013, Physics of Plasmas, 20, 055501,
  \dodoi{10.1063/1.4807033}

\end{thebibliography}
\end{document}